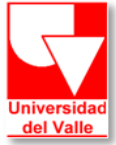

# GENERATING MUSIC AND GENERATIVE ART FROM BRAIN ACTIVITY
Undergraduate Thesis Report

Ricardo Andres Diaz Rincon
ricardo.andres.diaz@correounivalle.edu.co

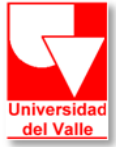

# GENERATING MUSIC AND GENERATIVE ART FROM BRAIN ACTIVITY
Undergraduate Thesis Report

Ricardo Andres Diaz Rincon
ricardo.andres.diaz@correounivalle.edu.co

Degree Project Submitted in Partial Fulfillment of the Requirements for the Degree of Systems Engineer

Supervisor
Paola J. Rodríguez C. PhD
paola.rodriguez@correounivalle.edu.co

Co-Supervisor
Javier M. Reyes V. PhD
javier.reyes@correounivalle.edu.co

Facultad de Ingeniería
Escuela de Ingeniería de Sistemas y Computación
Programa Académico de Ingeniería de Sistemas
Cali, May 15, 2020

# Contents

# List of Figures

# List of Tables

# Acknowledgements

I would like to thank my mom for her love, patience, affection, and support. Thanks to my family, friends, and professors for their constant feedback, and dedication. This research work couldn't have been done without them.

After 5 years, education and research budget cuts, 3 student strikes for more than 4 months each, and thousands of marches I still can't believe I had to finish up my Bachelor during a global pandemic.

# Contributions

Journal Papers:

- **2020. Diaz Rincon, R. A.** Towards Brain Computer Interfaces and Generative Art. GASATHJ (Generative Art Science and Technology Hard Journal). https://www.gasathj.com/tiki-read_article.php?articleId=76.

- **2019. Diaz Rincon, R. A.**, Reyes Vera, J. M., & Rodriguez C, P. J. Generando música a través de la Actividad Cerebral. Brazilian Journal of Development, 5(6), 5375–5388. https://doi.org/10.34117/bjdv5n6-077.

Conference Papers:

- **2019. Diaz Rincon, R. A.**, Reyes Vera, J. M., & Rodriguez C, P. J. An approach to Generative Art from Brain Computer Interfaces. In Celestino Soddu (Ed.), XXII Generative Art Conference - GA2019 (pp. 332–343). Retrieved from http://www.generativeart.com/GA2019_web/50_RicardoDiaz_168x240.pdf.

- **2018. Diaz Rincon, R. A.**, Reyes Vera, J. M., & Rodriguez C, P. J. Generando música a través de la Actividad Cerebral. Nuevas Ideas en Informática Educativa, 14, 600–605. http://www.tise.cl/Volumen14/TISE2018/600.pdf.

Conference Presentations:

- **2019**. XXII Generative Art Conference. Etruscan National Museum of Villa Giulia. Rome, Italy. An Approach to Generative Art from Brain Computer Interfaces.

- **2018**. TISE 2018. University of Brasilia. Brasília, Brazil. Generando música a partir de la actividad cerebral.

Software:
- GitHub: https://github.com/Ricardo0621/BrainBeats.

# Abstract

Nowadays, technological advances have influenced all human activities, creating new dynamics and ways of communication. In this context, some artists have incorporated these advances in their creative process, giving rise to unique aesthetic expressions referred to in the literature as Generative Art, which is characterized by assigning part of the creative process to a system that acts with certain autonomy (Galanter, 2003).

This research work introduces a computational system for creating generative art using a Brain-Computer Interface (BCI) which portrays the user's brain activity in a digital artwork. In this way, the user takes an active role in the creative process.

In aims of showing that the proposed system materializes in an artistic piece the user's mental states by means of a visual and sound representation, several tests are carried out to ensure the reliability of the BCI device sent data.

The generated artwork uses brain signals and concepts of geometry, color and spatial location to give complexity to the autonomous construction. As an added value, the visual and auditory production is accompanied by an olfactory and kinesthetic component which complements the art pieces providing a multimodal communication character.

# CHAPTER 1. PROBLEM DESCRIPTION

## 1.1. Approach and problem statement

Years of research and hard work by researchers, scientists, and engineers have brought humans closer to one of the most fascinating organs of living beings: The brain.

The relationship of human beings with science and technology has created new types of interaction and with the emergence of new technologies, interaction paradigms are evolving. Thus, nowadays brain activity recording devices are used to explain what happens in human beings and the brain while different tasks in our daily life are performed and how we react to different stimuli.

Over the years the brain activity — associated with daily life processes — has been extensively studied. Brain activity is as unique as Music and Art, which are not perceived in the same way on different occasions, but change regarding the experiences and mood of the person who experiences them. It is therefore believed that brain activity is as unique as the human being who generates it, however, it may change from the task performed or the mental states of the person under study.

Nonetheless, little has been said about brain activity and Brain-Computer Interfaces outside contexts rather than psychology and neuroscience. Therefore, this research work aims to explore the use of Brain-Computer Interfaces (BCI[1]) in a different area of knowledge such as art, and particularly its application in Music and Generative Art.

Thus, the question then arises: ***Which functionalities should a software application incorporate in a way that allows to generate music and generative art from brain activity?***

## 1.2. Justification

This research work is relevant since it allows building a bridge between areas of knowledge that have been widely studied: Brain-Computer Interfaces, music and art.

Thus, the research carried out will allow not only to make use of the skills acquired throughout the undergraduate studies, but also to gain new ones by diving into other disciplines. In the same way, it allows — once some of the basics of music theory are understood — to apply them to develop a tool that makes use of an EEG[2] to generate music from brain activity in the context of Generative Art.

Similarly, thanks to the absence of limitations regarding the users who will use the EEG, it is possible to present this research work as a tool that provides to those who use it an immersive artistic experience by visualizing geometric patterns, shapes, and music created from user interaction.

---

[1] A Brain-Computer Interface (BCI) is a hardware and software communications system that permits cerebral activity alone to control computers or external devices. (Nicolas-Alonso & Gomez-Gil, 2012).
[2] EEG measures mostly the currents that flow during synaptic excitations of the dendrites of many pyramidal neurons in the cerebral cortex (Teplan, 2002).

### 1.3. Objectives

**Main Objective**

To explore the use of Brain-Computer Interfaces (BCI) to generate Music and Generative Art from brain activity.

**Specific Objectives**

- Select an EEG device from those available in the market and identify the variables required for creating music and generative art.
- Development of the necessary algorithms for creating a piece that combines music and generative art.
- Design the functional prototype of an interactive installation using the BCI device and the selected algorithms.
- Apply hardware tests and static tests to the developed software component.

### 1.4. Expected results

| Specific Objectives | Expected Product(s) | Documented at: |
|---|---|---|
| Select an EEG device from those available in the market and identify the variables required for creating music and generative art. | Document with: Description of EEG devices, advantages, disadvantages and selection justification. | CHAPTER 3. BCI DEVICE SELECTION |
| Development of the necessary algorithms for creating a piece that combines music and generative art | Development and/or adaptation of the algorithms to be used, source code and executable. | 4.2. Musical component 4.3. Generative Art component |
| Design the functional prototype of an interactive installation using the BCI device and the selected algorithms. | Interactive installation design description. | CHAPTER 5. CEREBRUM: INTERACTIVE INSTALLATION DESIGN |
| Apply hardware tests and static tests to the developed software component. | Testing design and execution. | CHAPTER 6. DESIGN AND TESTS EXECUTION |

*Table 1. Expected Results.*

### 1.5. Scope and delimitation

This research work will be limited to the development of a prototype that, using a specific EEG, allows creating music and generative art taking as input brain activity, understanding music as *organized sound* (GOLDMAN, 1961) and not as the subjective appreciation or perception of the listener.

In this regard, this research work is not intended to produce different musical rhythms or genres by user interaction, nor to carry out a study on how music affects the brain.

# CHAPTER 2. FRAME OF REFERENCE

The frame of reference is presented below:

## 2.1 Theoretical framework

The theoretical framework of this proposal includes Brain-Computer Interfaces, Music and Generative Art.

### 2.1.1 Brain-Computer Interfaces (BCI)

Brain-Computer Interfaces are hardware and software communication systems whose purpose is to help users interact with the external environment by predicting their intentions based on data related to their brain activity.

These types of systems have been fundamentally studied and used as assistance tools for people with reduced mobility because they do not involve the use of muscular channels for user interaction (Nicolas-Alonso & Gomez-Gil, 2012).

A Brain-Computer Interface is a hardware and software system that can recognize a set of patterns in the brain's signals by following four consecutive stages: Signal Acquisition, Signal Pre-processing, Feature Extraction and Classification.

The first stage captures neural signals using algorithms to reduce the external noise. The second stage prepares the data for further processing. The feature extraction stage identifies the information, discriminating against that which has been recorded in the brain. The classification stage allows brain waves to be divided into different frequency ranges.

BCI systems allow the brain to communicate with external mechanical devices and involve important aspects such as voluntary control of electroencephalographic signals, synchronization of brain rhythms and measurement, interpretation, and classification of neural activity. The correct performance of the BCI system depends on the control of the amplitude of the brain rhythms, so it has a direct impact on the two-dimensional manipulation of the connected devices.

A person's movements — real or imaginary — significantly involve brain activity, resulting in different responses. The stages of a BCI system are shown in Figure 1.

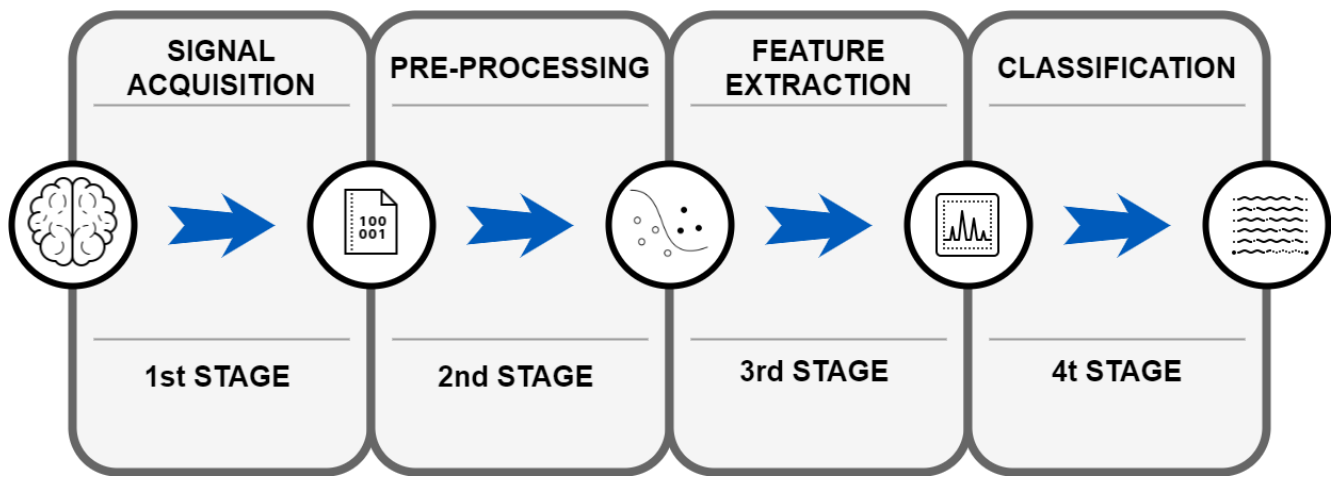

*Figure 1. BCI system stages.*

As shown in Figure 1, brain signals are acquired and pre-processed to extract, identify, collect information, classify it, organize the data, and transfer them into understandable commands for any connected device, such as a wheelchair or a computer (Nicolas-Alonso & Gomez-Gil, 2012). This approach shows that signal extraction involves processing that allows manipulation of an external device or application. Once the brain signals are mapped into a characteristic vector, the next step is the extraction of this information. Brain signals are mixed with other signals from an infinite set of activities that overlap in time and space. Additionally, the signal is not stationary and may be distorted by the artifacts used to pick it up.

### 2.1.2 Brain waves classification

Brain waves are produced when the brain cells (neurons) are activated and produce local current flows that are translated into electrical impulses and changes. Encephalography or EEG primarily measures the currents flowing during the synaptic excitations of the dendrites of many pyramidal neurons in the cerebral cortex (Teplan, 2002). Brain patterns form sinusoidal graphs that commonly range from 0.5 to 100µV in amplitude, that is, almost 100 times less than ECG signals (electrocardiograms).

The Fourier transform allows these raw signals to be recorded and amplified to obtain a higher volume of information. Brain waves are measured in cycles per second (Hz), the higher the number of Hz, the higher the frequency or brain activity. The first approach to brain waves was made by the German Hans Berger in 1924 (Jung & Berger, 1979).

Between 1930 and 1940 brain waves were classified into 5 groups summarized in Table 2 (Jung & Berger, 1979).

| Waves | Ranges | Mental states |
|---|---|---|
| Delta (δ) | < 4 Hz | Unconsciousness, deep sleep. |
| Theta (θ) | 4Hz – 7Hz | Relaxation, intuition, creativity, remembrance, Imagination. |
| Alpha (α) | 8Hz – 12Hz | Mental effort, sleepless relaxation, stillness, awareness. |
| Low Beta (β) | 12Hz – 15Hz | Relaxation and focus. |
| Mid Beta (β) | 16Hz – 20Hz | Thinking, self-consciousness. |
| High Beta (β) | 21Hz – 30Hz | Alert, agitation, disturbance. |
| Gamma (γ) | 30Hz – 100Hz | Motor functions and high mental activity. |

*Table 2. Brain waves classification.*

According to the above, a BCI device allows inference to be made about an individual's mental state and some of its motor functions.

In recent years, companies such as Emotiv, Neurosky, g.tec, and OpenBCI have been dedicated to manufacturing non-invasive BCI devices that allow brain signals acquisition and processing.

### 2.1.3 Music

The word *"music"* derives from the Greek mousike (μουσική) which means *"art of the muses"* (H. G. Liddell and R. Scott, 1843). According to (Allen, Fowler, & Fowler, 1990) music is *"the art of combining vocal, instrumental (or both) sounds to produce beauty of form, harmony, and expression of emotion"*. Also, according to modernist composer Edgard Varèse, music is defined as *"organized sound"* (GOLDMAN, 1961).

However, many authors have expressed different opinions about what is or is not considered music. Therefore, in order to understand the definition and language of music, it is necessary to become familiar with concepts such as: Tone (height), Duration, Intensity, and Timbre (Patterson, Gaudrain, & Walters, 2010). Figure 2 shows the relationship between the concepts mentioned above.

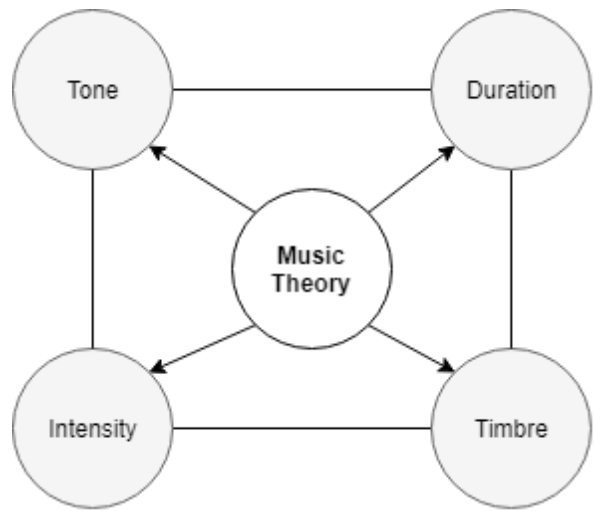

*Figure 2. Pillars of music theory.*

**Tone:** The tone is an essential characteristic that allows us to distinguish between high-pitched and low-pitched sounds (Plack & Oxenham, 2005). The frequency of each sound — usually measured in Hertz (Hz) — indicates the number of sound waves per second and allows identifying the musical note to which it corresponds.

**Duration:** In music, the duration is the time in which the vibrations produced by a sound are maintained, that is, the period or time interval in which a specific note sounds (Benward & Saker, 2008). This is represented graphically by musical figures assigned to different sounds, where the whole note (semibreve) is the reference unit and each subdivision (musical figure) lasts in time the half of the previous note.

**Intensity:** Also known as volume, is the property that allows human beings to identify how loud or soft a sound is perceived. Volume levels are measured in decibels (dB). The range of human hearing lies between 0 and 120 dB approximately, for this reason, sounds above the upper limit — such as that produced by an aircraft take-off — can cause irreversible damage to hearing. While frequency is set by the length of the sound waves, intensity is determined by its height (Michael Hewitt, 2008).

**Timbre:** The timbre is an intrinsic property that allows the human ear to differentiate between sounds emitted by multiple sources, even when they do not belong to the same category. For example: The sound emitted by a guitar and an electric bass or the same musical note played by different instruments. Each of the above definitions constitute the structure and foundation of what we know as music.

### 2.1.4 Generative Art

One of the first approaches to a unified definition of what is Generative Art was proposed by Philip Galanter in 2003, who in his paper: *"What is Generative Art? Complexity Theory as a Context for Art Theory"* states that it refers to any artistic practice where the artist uses a system that possesses *"A set of rules of natural language, a computer program, a machine or other procedural invention, which is put into motion with a degree of autonomy contributing or resulting in a complete work of art."* (Galanter, 2003).

Technological advances and new forms of interaction allowed Generative Art to gain renown. Over the years, artists such as Georg Nees, Vera Molnár, Lilian Schwartz, John Maeda, and Casey Reas positioned Generative Art on the global scene, adding multiple elements, shapes, and figures within their pieces of art.

A new definition of Generative Art was coined in 2011 by Matt Pearson, author of the book Generative Art: A practical guide using Processing. In this book the author states that *"Generative art is neither programming nor art, in its conventional sense. It's both and neither of these things. Programming is an interface between man and machine; it is a clean, logical discipline, with clearly defined objectives. Art is an emotional, highly subjective, and challenging subject. Generative art is the meeting place between the two; it is the discipline of taking strict, cold, logical processes and subverting them to create illogical, unpredictable, and expressive results"*.

Thus, Generative Art often refers to algorithmic art (computer-generated art determined algorithmically) which is usually created from chemistry systems, biology, mechanics, robotics, intelligent materials, manual randomization, mathematics, data mapping, symmetry, mosaics and more.

Generative art is often expressed through interactive installations which are meant to be art forms that involve the viewer in a way that allows the art to achieve its purpose. Some installations achieve this by allowing the spectator or visitor to walk around them. Others do so by asking the artist or viewers to become part of the artwork.

**2.2 State of the Art**

Below are the works previously developed in the areas of interest of this research work.

### 2.2.1 Generative Art

**Schotter (Gravel) – Georg Nees:** One of the first and best-known pieces of generative art. Schotter begins with a standard row of 12 squares and gradually increases the magnitude of randomness in the rotation and placement of the squares as he moves through the rows. The art piece was created on a Siemens-System 4004 and is shown in Figure 3.

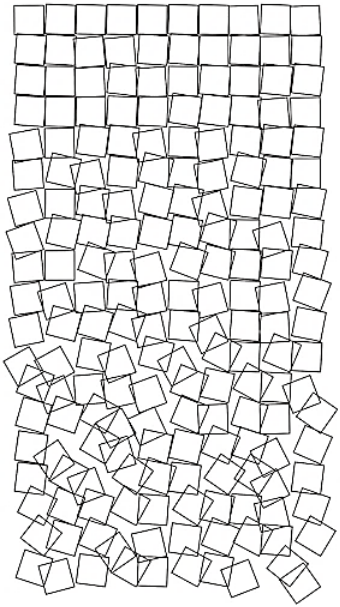

*Figure 3. Schotter (Gravel) – Georg Nees.*

**Head – Lilian Schwartz:** After conceiving the graphic Head (1968), she sketched it on graph paper. Using the sketch as model, she redrew the image with alphanumeric characters designating levels of gray. Eventually, the image was output onto black-and-white microfilm. The final image was created with the aid of traditional silkscreen techniques.

One screen was for the yellow background. A positive of Head was used for the red foreground. Graph paper is still used where an artist wishes to determine coordinates in advance of typing them in, particularly where the project involves three dimensions.

However, an artist can now compose a final color image on the screen and then use a color separation program prior to filming. Each frame then contains only those elements of the image that will be printed or silkscreened in a selected color or black as shown in Figure 4.

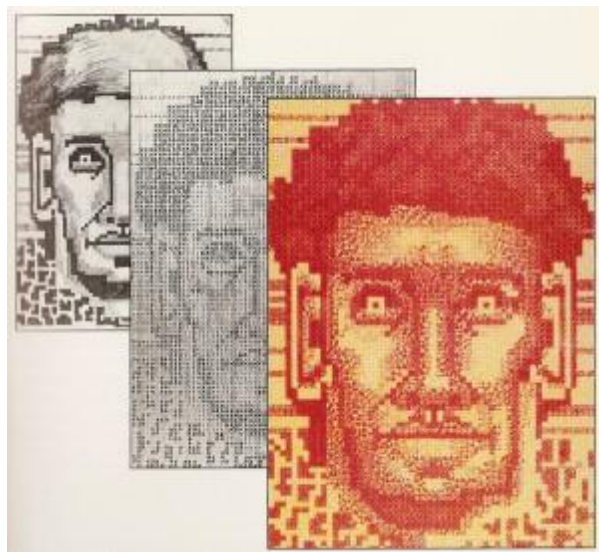

*Figure 4. Head – Lilian Schwartz.*

**AI Infinity – John Maeda:** It is a piece of generative art created in 1993 by John Maeda, former president of the Rhode Island School of Design and author of the book "*Laws of Simplicity*".

This was created with the help of the NeXT computer and was the first art piece the author coded using the PostScript language after a long time away from programming. The art piece comprises a collection of strokes made with the help of an infinite loop that stopped at the 10,000th iteration as shown in Figure 5.

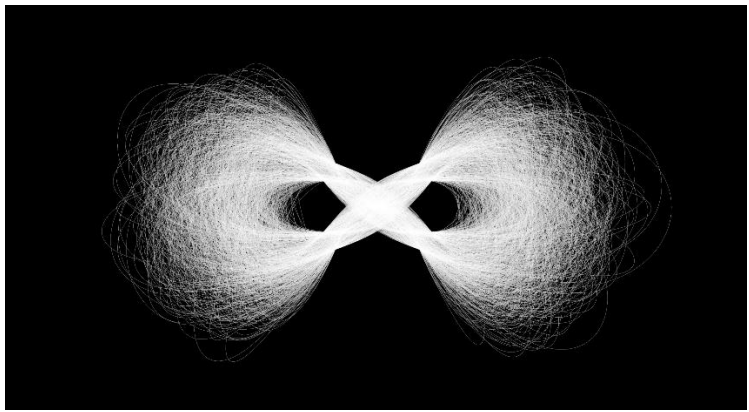

*Figure 5. AI Infinity – John Maeda.*

**I am Portraits – Sergio Albiac:** This project aims to add some literary and philosophical meaning to the average selfie. To do this, the artist takes photos of participants while asking them to describe themselves.

This is then converted into text using a Web Speech API. For this project the artist combined the custom code and semantic analysis with machine learning networks he trained to copy his style. This is because he didn't want the portraits to mimic other artistic styles like those in Google's Deep Dream neural networks do, but to maintain his own individual aesthetic as shown in Figure 6.

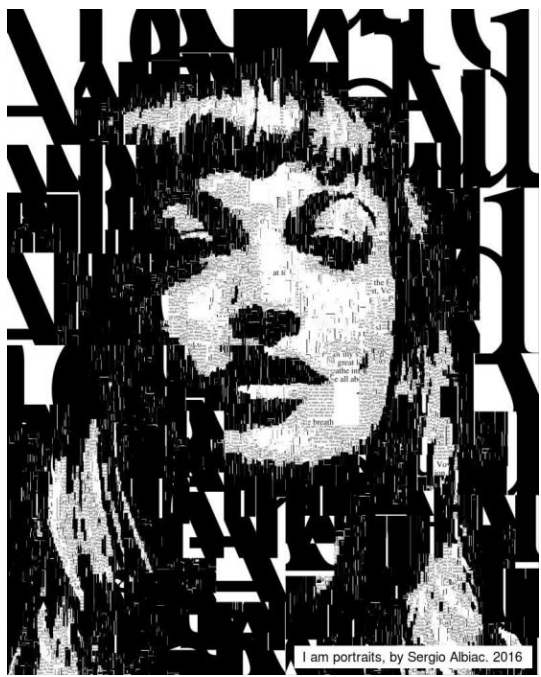

*Figure 6. I am Portraits – Sergio Albiac.*

## 2.2.2 BCI and Music

**Music for Solo Performer:** Created and produced in 1965 by the American composer Alvin Lucier, who is praised for his valuable contributions to the world of sound, by creating installations and experimental music that allows exploring the physical properties of sound.

Alvin Lucier was the first musician to use brain waves recordings as an input on the music scene. His creation *"Music for Solo Performer"* is also the first work in history that uses brain waves to generate sound (Eduardo R. Miranda, Magee, Wilson, Eaton, & Palaniappan, 2011).

It was composed during the winter of 1964–1965, with the technical assistance of physicist Edmond M Dewand. This work used electrodes to capture and/or detect the presence of alpha waves in the meditation process, whose waves would later be used to vibrate (make sound) percussion instruments around the installation (Lucier, 1976).

The first live appearance took place on May 5th, 1965, at the Rose Art Museum of Brandeis University (USA) with the participation of the composer John Cage (Straebel & Thoben, 2014). After this, it has been performed by Lucier on multiple occasions in individual concerts on tours around Europe and the United States with the Sonic Arts Union, a collective of experimental musicians.

The main goal of this work was proposed by the scientist Edmond M Dewand, who by 1965 was investigating the alpha waves: Dewan asked Lucier if he would be interested in using his equipment to detect alpha waves and thus, use them in a piece of music. Lucier, who was not composing at that time, took the opportunity.

While thinking about what to do with brain waves in musical terms, he realized that alpha waves are so low (about 8–12 cycles per second) that it would be easier to think of them as rhythms and create a piece for percussion. Alpha waves are produced only with the eyes closed, in a relaxed state of mind, without any activity. With this in mind, Lucier decided to *"take a dangerous course that involves sitting on stage and trying to produce alpha waves, live, in front of the audience"*. This means that once the electrodes are connected to his head, the artist must sit on the stage doing nothing.

Eventually, his brain will enter the alpha state. Brain waves are constantly collected with electrodes and amplified with a brain wave amplifier. A filter allows only alpha waves to pass through. The signal is then divided into several channels and each of them is amplified and sent to a speaker. The speaker cones follow the alpha rhythm and make the percussion instruments sound, either by hitting them directly or by air movement. One or two assistants control the volume of the individual channels and by doing so they determine the musical form of the piece.

Music for Solo Performer is considered one of the precursors of what is known today as experimental music. It was also one of the first works in which music (sound) was generated without the direct intervention of a musical instrument. This was a turning point in Lucier's career since it allowed him to find his musical language. This established at least three important elements in his work: The physicality of sound, unintentionality, and the use of non-musical instruments (preferably test equipment) in the performance.

Lucier's work not only encouraged the study of brain waves but also inspired the creation of new instruments capable of amplifying them, being such the case of John David Fullman, who was a student of Lucier and founder of Magic Boxes co. who created the Alpha Amp I in 1974 for experimentation in the activation of LED lights with brain waves. A later version of this device, the Alpha Amp II – created in 1975 – was later purchased and used by Lucier for multiple performances (Straebel & Thoben, 2014).

Although the work was visionary for its time, it had some disadvantages that are mentioned below:

- **Number of people:** In fact, the artist was not alone since he was dependent on two other people to control the overall sound volume and assist him during the performance.

- **Signal acquisition:** By 1965, very large and heavy equipment had to be used which in case of failure could compromise the physical integrity of the person operating them and the performer.

- **Scenery:** The staging of the play uses a variety of percussion instruments. While this means that the sound generated is much richer, it also involves the preparation and tuning of each instrument separately, which requires a considerable amount of time.

Below is the list of instruments involved in Lucier's work according to (Straebel & Thoben, 2014).

- 1 large mixer, 16 outputs
- 16 high-quality speakers, at least 1200 diameter
- 2 bass drums
- 2 snare drums
- 2 timpani
- 2 cymbals
- 2 triangles
- 1 tambourine
- 1 large tam tam
- 1 small gong
- 1 grand piano
- 1 cardboard box large enough to contain a speaker
- 1 metal trash can large enough to contain a speaker
- 1 CD player

**Eunoia:** Eunoia uses an EEG-Headset (Mindwave) that allows real-time acquisition of the artist's brain waves for her performance. The name is derived from the Greek words *"good"* and *"mind"*, which are translated as *"beautiful thought"*. The headset physically manifests the artist's current mental state within pools of water, which offer a direct visualization of thought and emotion.

During the performance, Lisa Park used the headset to measure her brain activity so that it could be translated into sound waves and then transmitted via Bluetooth to produce sounds through Reaktor and Max/MSP. Standing in the middle of five 24-inch flat plates containing a thin layer of water, the sounds produced reflect the intensity of the artist's thoughts and emotions in real time. NeuroSky's (Mindwave) EEG could monitor several common brain waves frequencies, including alpha, beta, delta, gamma and theta (Park, 2012).

The customized source code allowed the volume, pitch, and panning of the sound associated with the attention and meditation values to be calibrated. The plates work in a similar way to those used in cymatics, the study of visible sound. Modulations become safe as they reverberate in water, causing the drops to jump in unpredictable formations. This dance depends on the speed, volume, panning, and pitch of the wobbly noises. The attributes increase in conjunction with the chaos of brain activity and emotional intensity.

The result is a display that delivers data in a convincing but discreet way. Park hoped that by calming her thoughts, she could mute the speakers and prevent the water from moving.

The author admitted this was feasible when she was alone, but the performance made her nervous, which affected her inner calm. The aim of her work is *"To use technology to capture and evoke human emotions rather than alienation and to create art that empowers the public through awareness of them"*. For the Eunoia project, Park told Slate magazine that the goal of the art piece was *"To exert control over your mind, to make your mind so calm and relaxed that there is nothing for the EEG sensor to read"*. (Victoria Woollaston, 2014).

Without a doubt, Lisa Park uses her work to implicitly point out that the spectrum of human emotions, from placid to overwhelming, must be embraced and harnessed as the source of our creativity, vitality, and individuality. Instead of building bridges over troubled waters, Park's work encourages us to splash around in them. In essence, Park's evocative installation reveals the exquisite and fleeting beauty of a feeling, no matter where it is on the pleasure-pain scale.

**Disadvantages**
Some of the disadvantages found in this work were:

- **Lack of developer tools:** While Mindwave represents a convenient and easy-to-use option, it does not have good support for the application's development, so the programmer must search for libraries and documentation on the internet to get the most out of the device.

- **Limitations on the data provided:** Neurosky's device returns values corresponding to each of the brain waves, and the user's attention and relaxation levels. However, on one hand, it does not allow to know how the previous values were calculated and on the other hand it does not allow to make inferences about other mental states presented by the user, as some of its competitors in the market.

- **Bluetooth connection problems:** Once paired with the computer the device has a persistent connection. However, the difficulty lies in making the initial pairing, since, when problems are found, the device and Bluetooth driver on the computer must be installed and uninstalled causing frustration in the users who purchased the product and the developers.

**Activating Memory:** It is a project developed by Brazilian composer Eduardo Reck Miranda from the University of Plymouth in the United Kingdom, which uses Brain-Computer Interfaces so that patients with physical diversity or reduced mobility can create music.

Activating Memory is an innovative experimental composition for 8 performers. The ensemble comprises two parts: The first is a string quartet, which will play a score that is modified in real-time regarding the user's choice. The second is a BCMI (Brain-Computer Music Interface) quartet composed by users with reduced mobility and/or locked-in syndrome (pseudocoma).

Each of the 4 users who integrate the BCMI quartet chooses which segment to play next through a BCI device that allows getting signals derived from brain activity. The device is the g.SAHARAsys manufactured by g.tec (Guger Technologies) which is a brain cap that covers the entire scalp. It contains dry 8-pin electrodes, made of a special gold alloy that is long enough to reach the skin through the hair. The gold alloy and the 8 pins reduce the impedance between the skin and the electrodes in the visual's cortex region (the back of the head), from where the brain signal is recorded.

Once the signals are obtained, a neurological phenomenon known as SSVEP (Steady State Visually Evoked Potential) is used to identify the signals detected by the EEG that are produced as natural responses to visual stimulation at different frequencies (Eduardo Reck Miranda, 2013). For this reason, the brain data is recorded from the visual cortex.

Each user is asked to direct their gaze to four flashing icons located in the corners of the screen which flash at different frequencies (usually between 8 and 16Hz). When this happens, or even when the person focuses on a point, it is possible to detect in the users' EEG signals the frequency of the icon they are looking at by using techniques such as basic spectral analysis (Cheng, Gao, Gao, & Xu, 2002).

Filters were used for noise removal to ensure the accuracy of the recordings. Each of the icons corresponds to the segment of the piece of music that is played in real time (Eduardo Reck Miranda, 2013). This way, users may choose which part is played next.

Throughout the development of the project, two artificial intelligence algorithms were considered: The first allows obtaining EEG signals that will be used in the musical process. The second allows composing music based on the user's choice.

Patients with reduced mobility or physical disabilities cannot make music because the desired action – playing a musical instrument – requires movement of the limbs.
Therefore, the main goal of Activating Memory is to allow people with physical diversity to make music. The BCMI system allows them to play with signals detected from their brain activity.

The contributions of Miranda's work are valuable both socially and academically. On one hand, there is the fact that patients with reduced mobility can create music on their own without the dependence of a musical instrument (Eduardo R. Miranda et al., 2011). Most of them have suffered devastating accidents that have left them paraplegic, so it is satisfying to know that even in their condition, human beings can make music thanks to the scientific advances of our era. On the other hand, academic advances are found. Some of the most notable are the use of Brain-Computer Interfaces, artificial intelligence algorithms for recording and processing EEG signals, and the use of SSVEP signals for the development of the project, which is an area of neurology and neuroscience that is still under development.

Eduardo Miranda has spent over 10 years in these research areas. His interest in systems that enhance human creativity led him to create the ICCMR (Interdisciplinary Centre for Computer Music Research) at the University of Plymouth in the United Kingdom, which is a pioneer in a field he has called *"Musical Neurology"* (Eduardo Reck Miranda, 2013) thus, the development of this project is the result of much effort and dedication in this emerging area of knowledge.

Activating Memory has been performed at multiple international music festivals, including the London Symphony Orchestra's participation in the *MusicTech Fest* at St Luke's (London), the 11th International Symposium on Computer Generated Music Research in Plymouth and the Mainly Mozart Festival in San Diego (USA).

The piece won a special mention by the jury of the Medicine Unboxed Creative Prize in 2014 and was selected for the BASCA (British Academy of Songwriters, Composers & Authors) Composers Award in 2016.

**Disadvantages**

In contact via email the author stated that although the system works very well *"The only disadvantage is that the number of things that it can control is very limited. It is not as sophisticated as a real musical instrument. A lot of research is needed in order to make this type of system more flexible."*

### 2.2.3 Art Installations

**Plasma Reflection – Danny Rose Studio:** A generative and interactive artwork that makes it possible for visitors to see their reflection in the fourth state of matter: Plasma. Plasma Reflection is designed like a distorting mirror that would turn the matter of those who are facing it.

On the other side of the mirror, matter is ionized. The silhouettes of those who stand or move in front of Plasma Reflection are diffused in space and are subject to the turbulence generated by the magnetic field.

**As Above So Below – Nelson Ramon:** Wall projection mapping of Generative Visuals involving a complex but subtle particle system. The system reacted to the audience presence. Using a Kinect sensor, it detected the visitors' position and influenced the behavior of the particles. Enabling collaboration between deterministic machine rules and the randomness found in our real world, Nelson Ramon projected generative visuals embedded in this installation.

This project involves design, projection Mapping, implementation, and execution of generative algorithms in the Lingo (Adobe Shockwave) programming. The installation was originally created by Jessica Angel who used different materials and techniques to sculpt and fabricate life-size Polyhedrons props destined to be placed and complete the visual adaptation of the exhibit space.

**The Act of Seeing – Riar Rizaldi:** The artist sits in front of a digital projector wearing electrodes that measure electrical potential from his eye muscle movements. These electrooculographic signals (EOGs) are then generated into prepared visuals using software written by the author, determining the video's colors and frame rate.

The visuals are then projected back onto his body, with the light triggering his eye muscles, creating a feedback loop of audiovisual cinema. Also included in the feedback loop is audio generated eye muscle movements. To create "*The Act of Seeing*", Rizaldi built the software in Processing. He then connected an Arduino to an Electrooculogram circuit board, which he routed to electrode patches attached to his face.

The author was inspired by Alvin Lucier's piece "Music for Solo Performer" where the artist amplified his brainwaves to create sounds from moving objects on the percussion set. He was also influenced by James Whitney's generative computer art films, and the expanded cinema performances of Junichi Okuyama, who interacted with projector beams and flickers.

**Share your Unicorn – Chris Walter:** A permanent art installation at the Futurium Museum in Berlin that uses a BCI to record the user's thoughts which are then graphically drawn by a machine using lines of different lengths (Walter, 2019).

In the museum, the installation is located in a part of the *"Denkraums"* (thinking room) where everything revolves around BCIs. The "world of thoughts" of the user is measured by an EEG with a headband at three points. The algorithm creates a variable that captures the user's tension and relaxation levels. This same variable is transmitted from the headband via Bluetooth to a drawing machine which is visible behind a glass pane.

When it receives the signals, it sets itself in motion and draws the "world of thoughts" translated by the algorithm. When the user is relaxed, the pencil lowers in the middle and draws long lines and the machine glows blue. If the user is restless, the machine paints only short lines and turns green. Through the overlapping of the "painted thoughts" of the visitors', pictures are created.

# CHAPTER 3. BCI DEVICE SELECTION

The following is a review of some of the most widely used BCI devices on a commercial and academic level.

## 3.1. Mindwave Mobile

It is a non-invasive BCI manufactured by Neurosky, which allows the acquisition and classification of EEG signals. It is paired with the computer through a Bluetooth link and acquires signals through passive biosensors connected to an electrode that makes contact with the prefrontal area, above the eye (FP1 position according to the 10-20 system). It also has a reference terminal that has to be connected to the left earlobe (position A1) as shown in Figure 7.

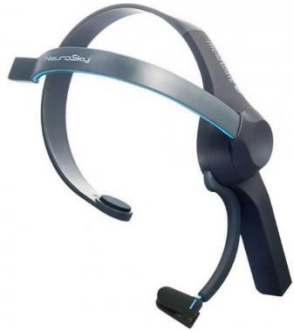

*Figure 7. Mindwave device.*

**Mindwave Exploration and Feature Identification**

Mindwave's App Central offers several applications. One of the best is MyndPlayer, a compendium of games developed in flash. MindPlayer consists of the following parts:

- **Main Panel:** Where the user can select the games of its preference. On the right-hand side, there are a series of buttons that will allow the user to choose, pause, forward and rewind the current game (cinematic). On the left side, there are the volume control and the device's signal meter. The main panel is shown in Figure 8.

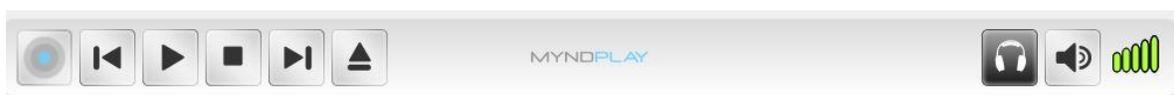

*Figure 8. Main Panel.*

- **Signal Panel:** This panel shows the user's attention and relaxation levels and its current state as well as a real-time graph showing the EEG raw signal, as seen in Figure 9.

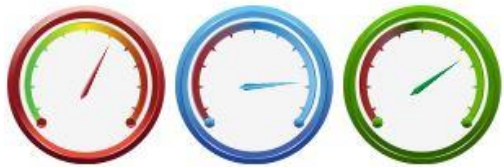

*Figure 9. Levels: attention (red), relaxation (blue) y user (green).*

The exploration and feature identification begins with the game "*Archery Life*" whose goal is for the user to score 30 points. To achieve this, the player must be focused or relaxed, depending on the situation. The course of the game is shown in Figure 10.

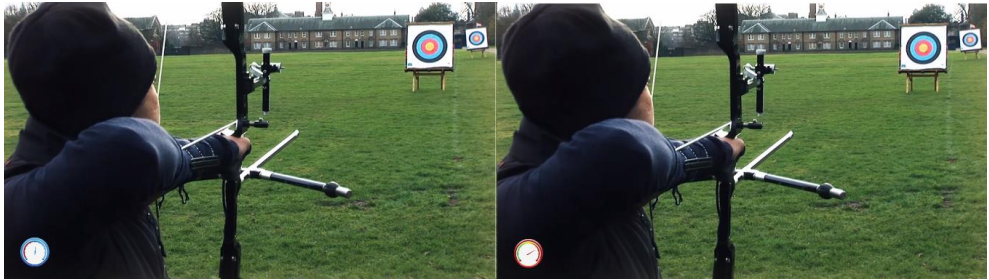

*Figure 10. Attention and relaxation levels in the game.*

Higher levels of attention and/or relaxation will have a positive effect on the player's score, which will be determined by the accuracy of the arrow. Normal levels of attention will allow the user to reach 6–8 points and optimal levels 9–10 points. If the highest score is to be obtained, the user must reach 10 points in each of the 3 stages (attention, relaxation, attention).

The second game "*Bullet Dodger*" puts the user in the skin of a person being chased by a boss of the Irish mafia. It presents a set of stressful situations, where the player's ability to stay calm or focused at any given time will be tested. The first challenge the player will face is to enter the facility where much of the game will take place. Here it will face a bodyguard who encourages the player to fight. The user will only pass the test if his/her level of attention is the right one, as shown in Figure 11.

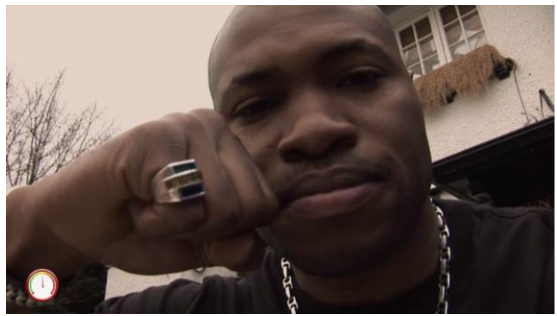

*Figure 11. Bullet Dodger: First challenge.*

The second challenge takes place inside the enclosure and it is more complicated than the first one. Here, a woman will intimidate the player and try to scare him/her with a short sharp weapon, as shown in Figure 12. This challenge is designed to measure the user's level of relaxation and will be completed only if its level is high enough.

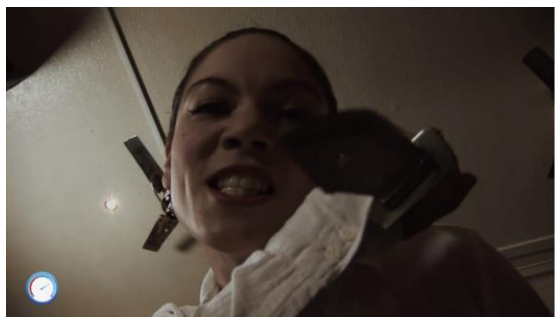

*Figure 12. Bullet Dodger: Second challenge.*

In the third and final challenge, the user will face the final boss. The interaction between the two involves a game of Russian roulette in which each one fails repeatedly until it is the boss' final turn.

Again, the player will be immersed in a stressful scenario, so it will be important to keep a high attention level to win the game, as shown in Figure 13.

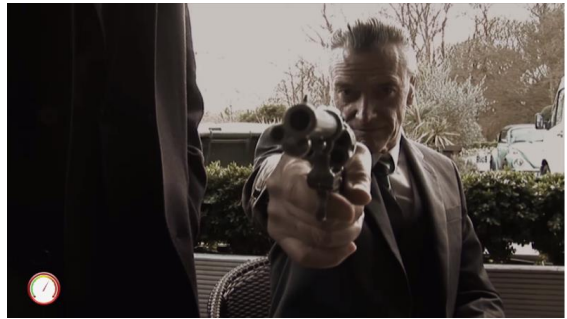

*Figure 13. Bullet Dodger: Final challenge.*

As in previous occasions, the game displays the attention/relaxation meter to get the user's brain data. Here, if the user's attention level is low, the game will end in an unfortunate outcome for the player. If the player's attention level is high, he or she will dodge the bullet and run away, ending the game. This is one of the most exciting games of the whole franchise of applications developed by Neurosky, because it offers different scenarios to those the player would face in daily life. Also, the interrelationships between the game's characters, dialogues, and music make the experience even more real and put on pressure on the player which is translated into an effort to complete the game successfully.

The third game is called *"Paramynd: Paranormal Mynd"*. Here, the player personifies a mentalist/exorcist whose mental powers will be the key to freeing a young woman from her "evil oppression". Paranormal or supernatural situations are common in the game, and the cinematics are charged with flashes, gloomy music, and strange sounds worthy of a horror movie. The game's warning is clear: "Remain focused or die", which is why it presents situations where things get out of control. Unlike other games, the time for acquiring the player's mental states is not limited to a few seconds, but the periods are much longer since the player is exposed to changes in the game's course that could scare him/her. The final stage of the game is shown in Figure 14.

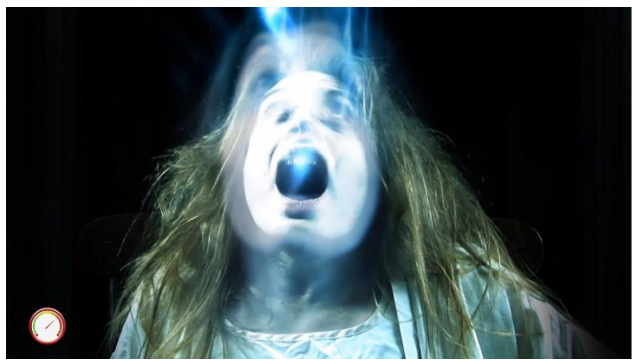

*Figure 14. Paramynd: Final stage.*

By keeping the required level of attention, the game ends with the release of the oppressed young woman. Because the game contains no scoring or point accumulation, it is very easy to win if the attention levels are slightly above normal or somewhat high.

The fourth and final game is called *"Parkour Heroes: Running all over the city"*. Here, the player will become a practitioner of this discipline, who must show that has sufficient skills to become a *"Parkour Hero"*. The perfect opportunity arises when a woman's cell phone is stolen by a person who practices this sport. The player embarks on the adventure and takes an obstacle course through the city to stop the thief who also knows Parkour. Performing acrobatics during the game is an attractive feature that makes it visually enjoyable and exciting for the player.

In addition, the third person's camera provides an experience that allows the player to enjoy the cinematics and acrobatics of the stunt performer as shown in Figure 15.

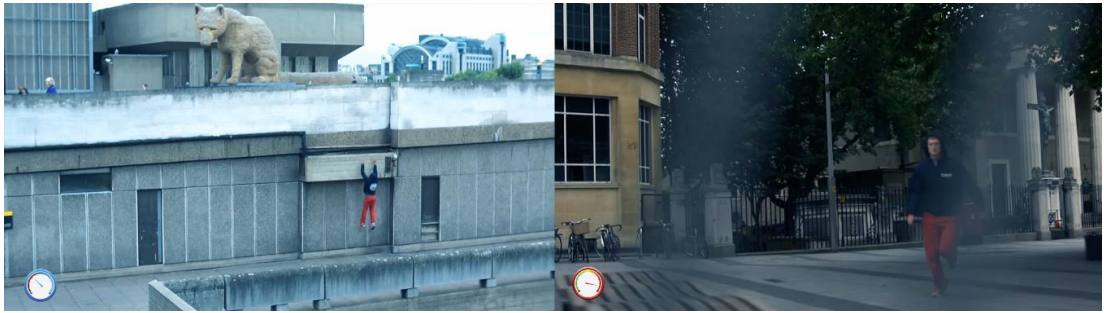

*Figure 15. Scenarios throughout the game.*

The locations where the game was filmed also add to its appeal, however, the real difficulty lies in the fact that, again, the player's attention/relaxation levels are measured, but unlike previous games, "*ParkourHeroes*" makes jumps between measurements, that is, if in the previous cinematic the attention level was measured, in the next one the relaxation level will be measured and both are opposite and inversely proportional states for most people: The more attention, the less relaxation and vice versa, which implies a challenge for the person who wants to win the game.

In conclusion, each of the games displayed here has the following characteristics:

1. Measuring the user's attention/relaxation levels.
2. Evaluating the user's mental states between cinematics.
3. Multiple number of scenarios throughout the game.
4. Use of stories in the games.

Table 3 shows the characteristics found in more detail:

| Game | Story | Scenarios | Score | How to win | Changes in mental states | Main mental state |
|---|---|---|---|---|---|---|
| *Archery Life* | No | 2 | Yes | Score 30 points | Relaxation Attention | Neutral |
| *Bullet Dodger* | Yes | 3 | No | Complete the game | Attention Relaxation Attention | Attention |
| *Paramynd* | Yes | 2 | No | Complete the game | Attention Attention | Attention |
| *Parkour Heroes* | Yes | 6 | No | Complete the game | Attention Attention Relaxation Attention Attention Relaxation | Attention |

*Table 3. Games' characteristics.*

From the table above it can be seen that attention is the main mental state in most games and that those that are most difficult for the user are the ones that have a combination of attention and relaxation levels. The number of scenarios also adds complexity to the game, because the more stages to be completed and the more diverse their characteristics, the greater the variation in the player's mental states, so there will be a greater challenge when it comes to completing the games.

Thus, most of the games provided by Neurosky measure either attention or relaxation, however, finding a state of complementarity in which both cooperate and the presence of one does not diminish or make null the other is a real challenge for any person.

**Advantages:**

- **Consumer-oriented:** The Brainwave Starter Kit features a set of interactive apps and games that allow the end consumer to have fun while learning about its brain waves. Among these applications, some are dedicated to meditation and exercise.
- **Affordable:** Mindwave is currently one of the most affordable BCI devices on the market.
- **User-friendly:** The provided GUI allows the user to visualize, in real-time, the values their brain waves and associated moods.

**Disadvantages:**

- **Lack of support:** Neurosky doesn't offer much support or free tools for developers, only applications for brain waves' visualization, and games. In this regard, most of the improvements, innovations, and contributions to repositories, libraries and source code available are made by developers. This slows down considerably the progress expected when developing software.
- **Number of channels:** As a device used to bring the end consumer closer to Neuroscience, Mindwave Mobile has a single channel, which limits the raw EEG data and the processing that can be done from this information.
- **Bluetooth connection/pairing:** Mindwave has some communication problems with the Serial (COM) port, this causes the Bluetooth connection to be interrupted repeatedly, so it is necessary to install/uninstall the device driver on multiple occasions.

### 3.2. Emotiv EPOC Model 1.0

EPOC is one of the signature products of EMOTIV Inc., a privately held bioinformatics and technology company founded in Australia in 2011, headquartered in San Francisco, California, that manufactures portable encephalography (EEG) products, including neuroheadsets, SDKs, software, mobile applications and data products. The EPOC is an EEG designed for scalable and conceptual research of the human brain and advanced Brain-Computer Interface applications. The EPOC provides an easy-to-use design, while allowing access to high-quality raw EEG data. The device has two types of connections: The first is a Bluetooth to PC and mobile devices. The second is via a proprietary USB receiver, which allows the headset to be configured. Unlike Mindwave, the EPOC has nine axes of motion sensors, a rechargeable battery for up to 12 hours, and saline or dry electrodes. The EPOC is shown in Figure 16.

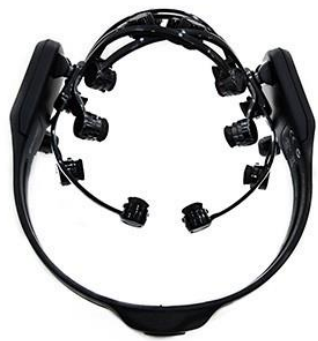

*Figure 16. EMOTIV EPOC.*

It also has 14 channels: AF3, F7, F3, FC5, T7, P7, O1, O2, P8, T8, FC6, F4, F8 and AF4 according to the 10–20 system which determines how the electrodes should be placed on the scalp when encephalograms are performed, as shown in Figure 17.

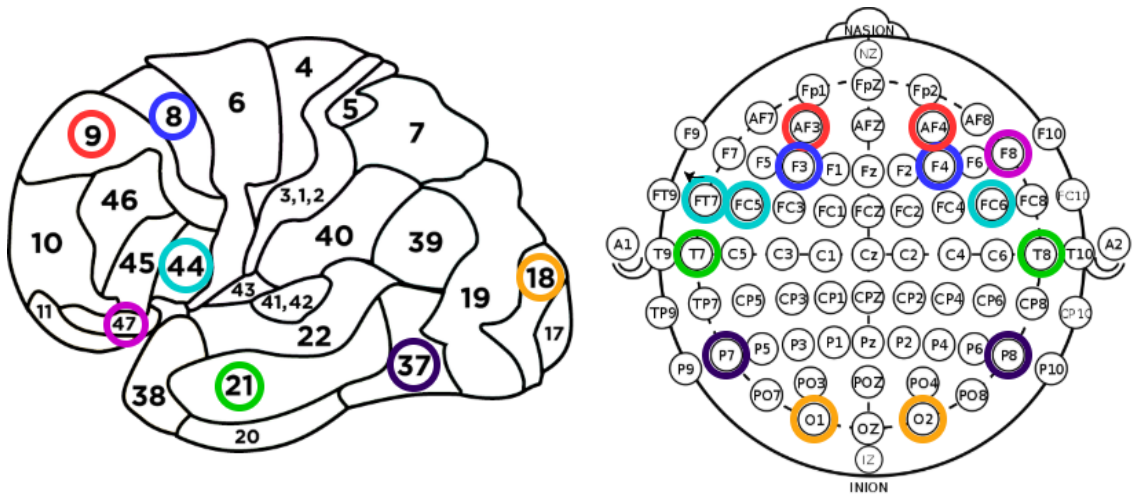

*Figure 17. EMOTIV EPOC: Electrodes' placement (10 – 20 system) and Brodmann areas.*

**Advantages:**

- **Tools for development:** One of the major advantages is the set of tools that EMOTIV offers for signal acquisition and development of mobile and desktop applications. This speeds up the software development process, since developers do not need to expend additional efforts and time looking for support, libraries or source code in another website.

- **Support:** The support that EMOTIV offers to the developers involves three parts: The first is the documentation, which is where a description of the methods used, data types, concepts, workflows, and error messages the user can experience is found. The second is a set of examples that show how to call the API from C++ using the Qt framework. It also contains the source code and a guide for its use. The last part is the knowledge base, which contains detailed information about possible problems with reading and/or accessing EEG information and technical problems with the device, its use, licenses, supported operating systems, and programming languages.

- **Visualization tools:** EMOTIV allows users to visualize, through color-coded projections, information about their brain waves (theta, alpha, beta, gamma, delta) in real-time, which is ideal for presentations, demonstrations or educational uses.

**Disadvantages:**

- **Preparation time with the user:** Since EPOC is a more specialized device suitable for neurological research, a considerable amount of time is required for the user's training and preparation. This means more time spent preparing the device, applying the saline solution to each of the electrodes, making sure that the electrodes are placed in the area to which they correspond according to the 10-20 system, among other activities.

- **Saline solution electrodes:** This is one of the major disadvantages of this version of the EPOC. Using these electrodes means that they must be well attached to the user's scalp, which involves using a considerable amount of saline solution. This causes discomfort for the user, not only because the sensation of using non-dry electrodes attached to the scalp is unpleasant, but also because when the device is removed, the hair remains attached to it.

Users with a lot of hair have been reluctant to use the EEG more than twice since it is very uncomfortable for them. Also, the use of saline electrodes might limit the pool of potential users who can use the device for two reasons: The first is that much of the users' hair gets attached to the EEG. The second reason is that the greater the length of hair, the lower the quality of the data, making the device more susceptible to errors caused by the impedance between the electrodes and the scalp. For this reason, when working with EEGs that use saline electrodes, it is ideal to choose users with thin or no hair.

- **Constant use of spare parts**: This is a direct consequence of the previous item. Using EEG devices with saline electrodes implies, on repeated occasions, the purchase of replacement electrodes and pads due to their quick deterioration or the impossibility of using them again because of the amount of hair attached.

### 3.3. Ultracortex "Mark IV" EEG Headset

The Mark IV is, along with the Cyton, one of the best-selling products by OpenBCI, an American open-source company founded in 2013 dedicated to the manufacturing and sale of low-cost, high-quality biosensor hardware. These devices are used to measure and record the electrical activity produced by the brain (EEG), muscles (EMG), and heart (ECG). The Ultracortex is an open-source, printable 3D headset designed to work with any OpenBCI board. It can record and/or monitor brain activity at a research level and receive EEG signals, however, it is not designed for transcranial stimulation. The Ultracortex can sample up to 16 EEG channels from 35 different electrode locations as shown in Figure 18.

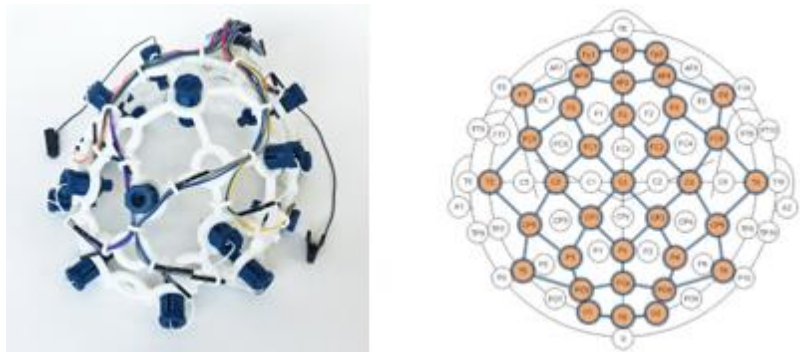

*Figure 18. Ultracortex Mark IV and electrodes' placement (10 – 20 system).*

This headset is easily adjustable and uses dry electrodes that take less than 30 seconds to position and adjust for data acquisition.

**Advantages:**

- **Number of channels and electrodes:** Compared to the devices and companies mentioned in this chapter (Neurosky and EMOTIV), OpenBCI offers the largest number of channels and electrodes (16 and 35 respectively). This means excellent news for researchers, since this amount of electrodes allows to cover a greater part of the cranial surface, which results in greater coverage and getting better quality raw signals.

- **Dry electrodes:** The use of this type of electrodes represents a huge advantage in terms of usability and time since the user does not require any prior preparation and the time spent placing the electrodes does not exceed 30 seconds. Besides, it is very easy for any patient to use the headset, since the amount of hair will not be a barrier and the hair will not be glued to the electrodes, as with saline and some gel electrodes.

- **Support and development tools:** Since OpenBCI is an open-source company, one of the main advantages to acquire its services is the multiplatform support and development tools. That is why the GUI for data visualization is available in Windows, Mac, and Linux. Also, software development can be done in Processing (Java), JavaScript (Node.js), Arduino, and Python. The drivers for research tools are available in OpenViBE, MATLAB, MINE - Python, and BrainBay.

**Disadvantages:**

- **Board sold separately:** Unlike its competitors, OpenBCI decentralized the way the headsets communicate with the computer, which means that a board is needed to establish the connection. However, this is one of its biggest disadvantages: Whoever buys the device will also be required to buy the board to develop applications, receive data from the BCI and use the tools offered by the company.

Thus, the total cost of the entire research kit (which includes headset, plate, electrodes, and wires) goes up to approximately $950,000. Nonetheless, this price could be justified by the number of channels and electrodes which is much higher and substantially exceeds that of its direct competitors.

Table 4 shows some of the most commonly used BCI devices and their main differences:

| Characteristics/Device | Mindwave | OpenBCI Cyton | Emotiv Epoc | g.tec nautilus | Cognionics Mobile–128 |
|---|---|---|---|---|---|
| **Number of channels** | 1 | 16 | 14 | 64 | 128 |
| **Sampling rate** | 512Hz | 250Hz | 128Hz | 500Hz | 500–1000Hz |
| **Communication** | Bluetooth | Bluetooth/ Arduino radio | Wireless (USB) | Bluetooth | Wireless |
| **Operating time** | 8 hours | 24 hours | 12 hours | 10 hours (SD Card) | 8 hours (SD Card) |
| **Weight** | 90g | 260g | 125g | 360g | 460g |
| **On-board storage** | No | Yes | No | Yes | Yes |
| **Accelerometer** | No | Yes | Yes | Yes | Yes |
| **Medical certification** | No | No | No | No | No |
| **Price(USD)** | 99.99$ | 499.99$ | $799 | 5,127$ | 50,000$ |

*Table 4. Commonly used BCI devices.*

Considering the above table and the analysis presented, it was decided to use Mindwave as a device to carry out this research because of the following reasons:

The first is that Mindwave is a quality, non-invasive, and affordable BCI device.

The second is that it uses dry electrodes and not saline. This is relevant since, according to (Rodriguez-C, Jimenez, & Paterno, 2015) saline electrodes may cause interference in data acquisition since in users with abundant hair some of the sensors may lose contact with the scalp and also interfere negatively in the user experience due to the wet sensation of the electrode.

Finally, the third reason is that the main interest of this research work is to record the brain activity and to associate it to the user's mental states, which is the reason why it is enough to obtain the information available in the frontal lobe from where Mindwave acquires brain data.

# CHAPTER 4. SOFTWARE DEVELOPMENT

To guide this research work, which involves the development of a software and tangible component (Artistic Installation), the agile methodology SCRUM was adapted, so the artifacts Product Backlog, User Stories, and Sprint Release were implemented.

As a result of the elaboration of the Product Backlog (Figure 19) four modules were defined: 1) BCI, which allows the acquisition and processing of encephalographic signals. 2) Music, which collects the filtered signals from the BCI and transforms them into music using algorithms and data structures. 3) Generative Art, which using the brain data, applies algorithms to compose an image which reacts to the user's brain waves and mental states. 4) Artistic Installation, which involves the design of an Artistic Installation.

The product backlog is available at the following link: http://tiny.cc/PBacklog. 17 user stories were generated, an example of these is shown in Figure 20. The user stories compilation is available at the following link: http://tiny.cc/HUs.

## Product Backlog

| Module | Id | User story name | Priority | Estimated Points | Progress |
|---|---|---|---|---|---|
| BCI | 1 | Signal Acquisition (Data recording) | High | 5 | |
| BCI | 2 | Data pre-processing | High | 5 | |
| BCI | 3 | Data processing | High | 3 | |
| BCI | 4 | Data Usage | High | 3 | |
| Music | 5 | Sound exploration | Medium | 3 | |
| Music | 6 | Key selection | Medium | 4 | |
| Music | 7 | Musical structure creation | Medium | 5 | |
| Music | 8 | Brain waves assignation | Medium | 2 | |
| Music | 9 | Overall sound fluctuation | High | 3 | |
| Music | 10 | Adding expressiveness | High | 4 | |
| Generative Art | 11 | Geometric shapes exploration | High | 5 | |
| Generative Art | 12 | Geometric shapes selection | High | 4 | |
| Generative Art | 13 | Prototype developmet | Medium | 5 | |
| Generative Art | 14 | Generative prototype development | Low | 3 | |
| Generative Art | 15 | Using brain data | Low | 5 | |
| Artistic Installation | 16 | Spatial location | Medium | 2 | |
| Artistic Installation | 17 | Sensorial elements | Low | 3 | |

*Figure 19. Product Backlog.*

## User story

| | | | |
|---|---|---|---|
| Author | Ricardo Diaz | Module | BCI |
| User story name | Signal Acquisition (Data recording) | Id | 1 |
| | | Priority | High |
| Actor(s) | User | Iteration | 1 |
| Points | 5 | | |
| Description: The brain waves' raw data and mental states are required to be recorded for further use. | | | |
| Last modification | 12/11/2019 | | |

*Figure 20. User Story.*

### 4.1. Software Architecture

As shown in Figure 21, the software application consists of four components:

- EEG: Responsible for acquiring raw data from the device.
- BCI: In charge of pre-processing and processing the raw data and convert them into a graphical and sound output.
- MIDI: Allows using and constructing MIDI[3] data structures.
- Arduino: Allows controlling the light bulbs and the air humidifier.

The BCI component processes the EEG data and transforms them into:

- MIDI data sent to a virtual port, whose sound is played back by a DAW[4].
- Graphical elements which change regarding the user's brain waves.
- Values that allow the color of the light bulbs to be modified and the scented airflow to be triggered in the interactive installation.

It can also be observed that three libraries are used as well:

- MindSetProcessing: Allows the acquisition of raw data from the BCI device.
- ProMIDI: For handling and constructing MIDI data.
- Serial: For writing and reading data on serial ports.

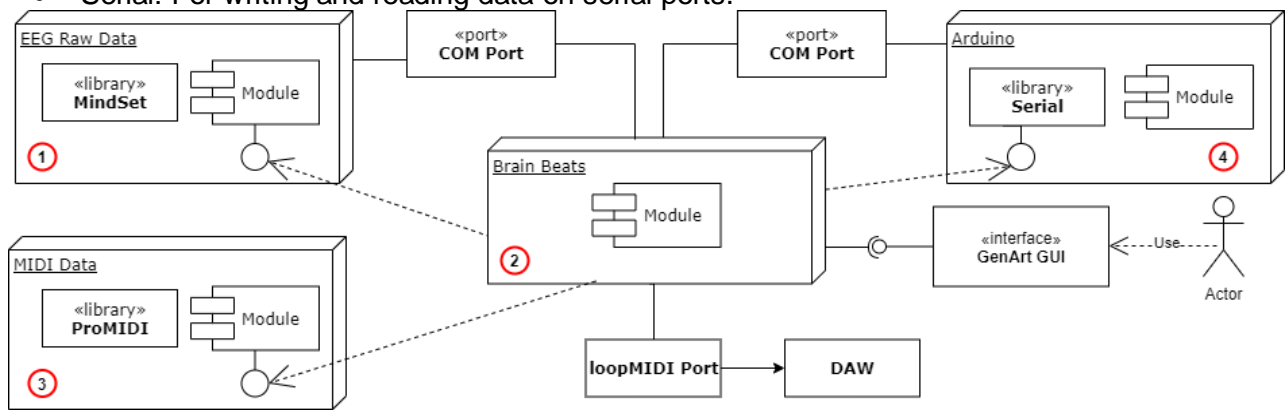

*Figure 21. Software Architecture.*

Processing was selected as programming language and environment because of these reasons:

- It is one of the most widely used programming languages in the context of projects related to Visual Arts.
- It allows an easy integration with Java/JavaScript and MIDI data handling.
- The ease it provides for the deployment of information, the use of graphic and multimedia libraries and the developing of user interfaces.

Additionally, code was developed in Arduino to control the air humidifier, whose development process is specified in Section 5.1. It is to be noted that Processing, Arduino, and the libraries and tools used are free and Open Source technologies.

The following is a detailed description of the above components:

---

[3] MIDI is a technical standard that describes a communications protocol, digital interface, and electrical connectors that connect a wide variety of electronic musical instruments, computers, and related audio devices for playing, editing, and recording music.

[4] A digital audio workstation (DAW) is an electronic device or application software used for recording, editing and producing audio files.

### 4.2. Musical component

**Sound Exploration**

After defining the BCI device, programming languages and libraries, a sound exploration was made, that is, the available choices regarding the potential musical notes that would give structure to the overall sound were inquired.

In this sense, each musical note has an associated MIDI representation, which is defined by an integer value, as shown in Figure 22.

| Note | Octave |   |   |   |   |   |   |   |   |   |
|------|-----|----|----|----|----|----|----|----|-----|-----|
|      | -1  | 0  | 1  | 2  | 3  | 4  | 5  | 6  | 7   | 8   | 9 |
| C    | 0   | 12 | 24 | 36 | 48 | 60 | 72 | 84 | 96  | 108 | 120 |
| C#   | 1   | 13 | 25 | 37 | 49 | 61 | 73 | 85 | 97  | 109 | 121 |
| D    | 2   | 14 | 26 | 38 | 50 | 62 | 74 | 86 | 98  | 110 | 122 |
| D#   | 3   | 15 | 27 | 39 | 51 | 63 | 75 | 87 | 99  | 111 | 123 |
| E    | 4   | 16 | 28 | 40 | 52 | 64 | 76 | 88 | 100 | 112 | 124 |
| F    | 5   | 17 | 29 | 41 | 53 | 65 | 77 | 89 | 101 | 113 | 125 |
| F#   | 6   | 18 | 30 | 42 | 54 | 66 | 78 | 90 | 102 | 114 | 126 |
| G    | 7   | 19 | 31 | 43 | 55 | 67 | 79 | 91 | 103 | 115 | 127 |
| G#   | 8   | 20 | 32 | 44 | 56 | 68 | 80 | 92 | 104 | 116 |     |
| A    | 9   | 21 | 33 | 45 | 57 | 69 | 81 | 93 | 105 | 117 |     |
| A#   | 10  | 22 | 34 | 46 | 58 | 70 | 82 | 94 | 106 | 118 |     |
| B    | 11  | 23 | 35 | 47 | 59 | 71 | 83 | 95 | 107 | 119 |     |

*Figure 22. MIDI note values chart.*

**Key selection**

From the set of musical notes, those in a defined key were selected. Similarly, during the development of this research work major chords were selected since usually, on an emotional level, major chords are related to bright and happy sounds and minor chords to darker sounds.

In this way, G Major (or the key of G) was selected, with the pitches G, A, B, C, D, E, F# as shown in Figure 23.

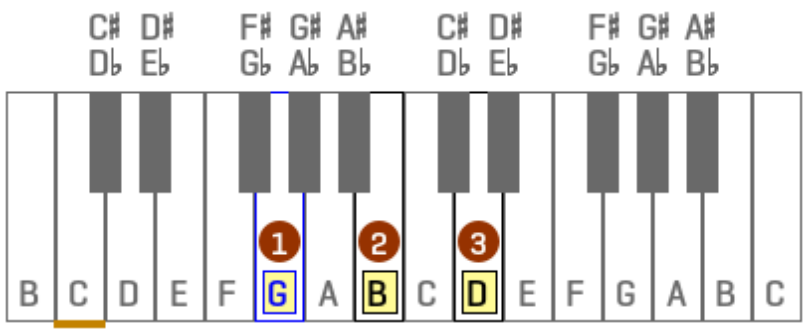

*Figure 23. G major chord.*

After choosing the key, the notes that form the triad are selected, in other words, the notes that form the chord, as shown in Figure 24.

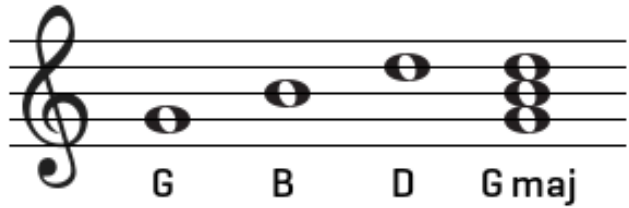

*Figure 24. G major triad.*

The selected notes are then declared as variables, which are then sent on a MIDI channel. Figure 25 shows the notes of the G major triad (G, B and D) located in octaves 2–5 and their MIDI representation.

```
int G4 = 67;
int B4 = 71;
int D5 = 74;
int G5Flat = 78;
int G2 = 43;
int D3 = 50;
int G3 = 55;
```

*Figure 25. Musical notes used.*

**Musical structure creation**

After having the notes declared, a specific length and tempo are assigned to each one of them. In this way, a musical structure is created, consisting of notes in the same scale with a given duration.

**Brain waves assignment**

The notes are then assigned to each brain wave as shown in Table 5.

| Musical note | Brainwave |
|---|---|
| G4 (G, fourth octave) | Theta |
| B4 (B, fourth octave) | Alpha |
| D5 (D, fifth octave) | Beta |
| G5Flat (G Flat, fifth octave) | Gamma |
| G2 (G, second octave) | Delta |
| D3 (D, third octave) | High Alpha |
| G3 (G, third octave) | AVG Gamma (Low + High gamma) |

*Table 5. Musical notes and their corresponding assigned brainwave.*

A matrix of integer values (two-dimensional array) is used as a musical pattern which includes a sequence of notes.

Each number, from zero to seven, represents a musical note: C, D, E, F, G, A, B, and C. This will determine the occurrence of a note in a sequence of 8 (the number of musical notes in a scale). The brain waves' values will determine which pattern is played.

**Overall sound fluctuation**

In order to increase expressiveness, attention, and relaxation levels are used to modify the overall sound: High levels of attention will increase the sound's intensity while high levels of relaxation will decrease it. The fluctuation of both levels (which are not complementary at first) and the user's emotions will be a characteristic feature of the music generated by each person (Diaz Rincon, Reyes Vera, & Rodriguez C, 2019b).

**Adding expressiveness**

While Processing allows the developer to use the computer's MIDI output to play back the sent data, using a DAW (Digital Audio Workstation) allows to choose between several musical instruments and audio effects.

However, for this to be possible, it is necessary to create a link between the application and the DAW, so a virtual MIDI port is used to allow communication between the two.

Once the MIDI port is running, it is possible to establish communication between the application and the DAW. Figure 26 shows the communication scheme.

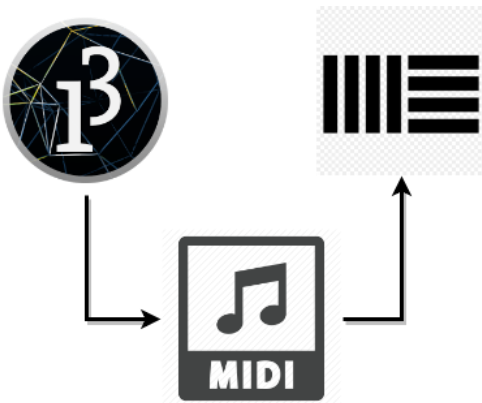

*Figure 26. Communication scheme.*

When a new note is sent, the musical structure is refreshed and its previous appearance is removed, then a new structure that includes the note is built up. The duration and MIDI values are then defined. Next, for each note, a given brain wave is used to select which musical pattern is played next.

Figure 27 shows the process of creating music, described above.

*Figure 27. Source Code: Music Generation.*

### 4.3. Generative Art component

**Proposed system: Preliminary stages**

The integration of the generative art component in this research work is described below: Throughout the exploration, tests were carried out with different colors and geometric figures, so that at each stage significant improvements could be seen. In this way, the final proposal shows an evolution in the algorithm's robustness and the art pieces created as a result of the user interaction.

The following is a description of the experimental process: Processing was used as a programming environment to create generative art. Similarly, an exploration of the most relevant geometric forms that are more easily identifiable in nature and daily life was carried out to promote the naturalness of the generated art pieces. Some of the geometrical figures chosen were hexagons, triangles, and circles. However, the exploration was carried out with circles because the low computational complexity required to draw them (Chen & Sundaram, 2005) allows including different characteristics into the final piece.

The first approach came from a mathematical function whose curves have the shape of a flower petal. This function is called *"Rhodonea Curve"* or *"Rose curve"* and was named by Italian mathematician Guido Grandi between 1723 and 1728. It allows drawing roses from polar coordinates. Figure 28 describes the function expressed in polar coordinates and its representation in parametric equations:

$$r = \cos(k\theta)$$
$$x = \cos(k\theta)\cos(\theta)$$
$$y = \cos(k\theta)\sin(\theta)$$

*Figure 28. Polar coordinates and parametric equations.*

If k is a half-integer (1/2, 3/2, 5/2), the curve will be rose-shaped with 4k petals. For example: n=7, d=2, k= n/d = 3.5, as θ changes from 0 to 4π. In other words, the rose will have a greater or lesser number of petals. Figure 29 shows the curves defined by different values of n and d.

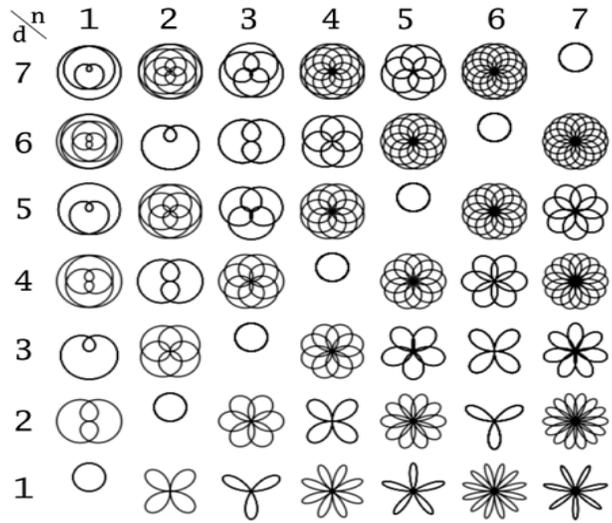

*Figure 29. Rose Curves with different values of n y d.*

The second approach continued under the premise of using circles for constructing the art pieces. In this regard, color and location were used as characteristics of generative art, so a variable associated with color and a pair of coordinates for the placement of the figure were introduced as random elements. The circles are formed by lines starting from a given position. The algorithm makes a for loop that goes from zero to 360 (degrees) to draw a circle.

Within it, a straight line is drawn whose $X_1$ and $Y_2$ coordinates are computed using random numbers between 50–150 and 150–360 respectively. The $X_2$ and $Y_1$ coordinates are both zero. In this way, the center of the circle changes and its internal points are defined, as can be seen in Figure 30.

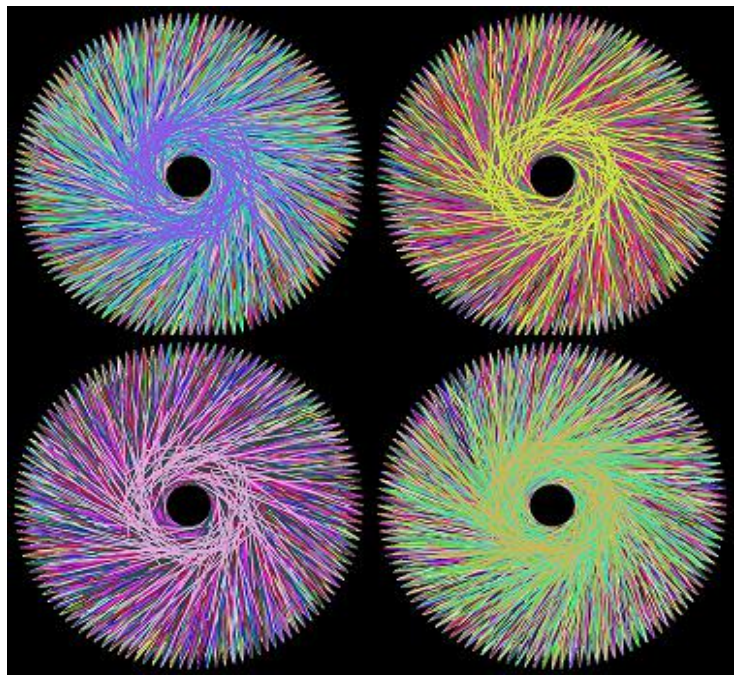

*Figure 30. Multicolor circles generated randomly.*

**Proposed system description**

Regarding generative art, the third approach gathers the ideas introduced in previous stages, however, instead of randomizing the whole algorithm, it takes as input the brain waves' values acquired by user interaction in order to generate art from brain activity.

Figure 31 shows the user interaction and some of the art pieces created by its brain data.

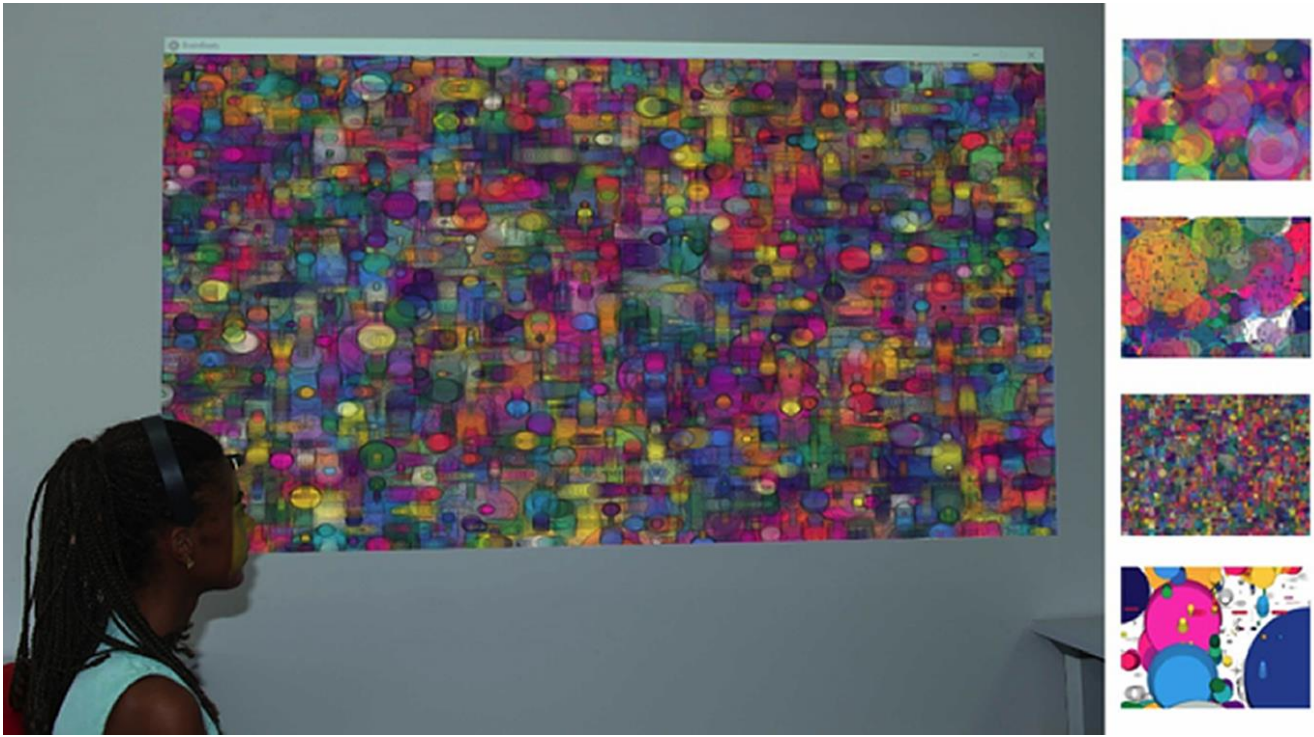

*Figure 31. User interaction.*

The way the application works is explained below:

It starts by defining a variable that will measure time in milliseconds. This variable will be used to provide continuity, and movement to the final generated art pieces. A seed is also defined for the random elements of the algorithm. For this approach, the values of the five brain waves are used: Delta, Theta, Alpha, Beta, and Gamma.

The raw data and brain signals recorded by the EEG are pre-processed and filtered so the brain data can be used by the algorithm.

Regarding Alpha waves, two types are distinguished: The normal ones, which are in a range of 8Hz–12Hz and High Alpha waves between 10Hz–12Hz.

Following the concept of using circles, two models are proposed:

The first is a training set formed by Delta, Theta, and Alpha waves that will control the amount of circles in the art piece. This set will be called C1. The second set (C2), formed by High Alpha, Beta, and Gamma waves will be used to control the radius of each circle.

An additional variable (named Z) that measures the screen width and divides it by the brain waves' values of the C2 set is used to graph the generated circles within the display area.

Then, the algorithm iterates from zero to the brain waves' values in C1. Three new variables are then introduced: The first two are the X and Y coordinates, whose values will be the product of multiplying the Z variable by a random number between zero and the value of the brain waves at C1, plus one.

The third variable is in charge of the displacement of each circle in the figure. This variable is the product of multiplying the next factors: The time variable (introduced at the beginning), a random number between 0.1 and 1, the number 60, and a random number between zero and two.

The mentioned values were set after testing and observing that they made possible the perception of movement. Each circle moves along one of the two axis following a straight line whose sinusoidal displacement and wavelength change randomly.

The color of each circle and its interpolation with the next color are chosen in the same way. For this purpose, the user's attention and relaxation values are used to select randomly the opacity of each circle in the art piece, varying in a range that has the user's attention level as the lower bound and the relaxation level as the upper bound.

To provide variability and expressiveness in the final art piece, it was necessary to determine which pair of brain waves from sets C1 and C2 would be selected. The training sets are shown in Table 6.

| *C1* (Number of circles) | *C2* (Radius) |
|---|---|
| Alpha | High Alpha |
| Theta | Beta |
| Delta | Gamma |

*Table 6. Training sets.*

Thus, tests were carried out with 6 users, whose generated art pieces showed divergent behaviors. Each user had different mental states and brain activity.

As can be seen in the figures below, the brain waves of each user allows creating art pieces with different morphological characteristics. The user's brain data used in the algorithm are expressed uniquely in the generated art pieces.

Similarly, each of the figures displayed shows the uniqueness of each of the art pieces. The geometric shapes that are formed contain a series of basic design concepts such as translation, superposition, and gradation of: shape, size, color, and scale.

The change in the opacity of the geometric figures combined with the scale gradation generates a controlled chaos effect. Additionally, the visual weight and the cognitive load of the generated art pieces are modified as the interaction process lasts longer.

Below, the different art pieces generated are explained in detail, taking into account the brain waves and the users who participated in the tests.

**User 1: Alpha – High Alpha**

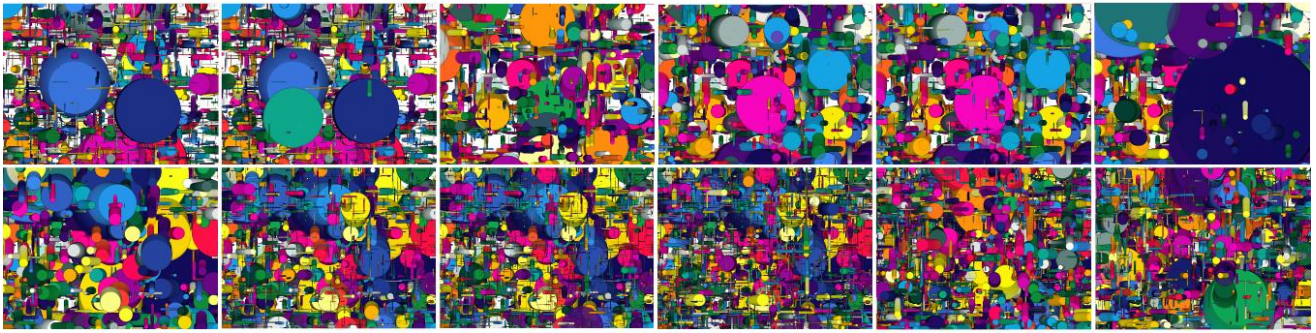

*Figure 32. User 1 (Alpha – High Alpha).*

As seen in the results, a grid of superimposed circular shapes with size grading is created. The randomness of the position of the figures ends up being recurrent in the sense that a certain order can be seen in the X and Y axis, as well as the distribution of the filler colors in the whole composition.

**User 2: Theta – Beta**

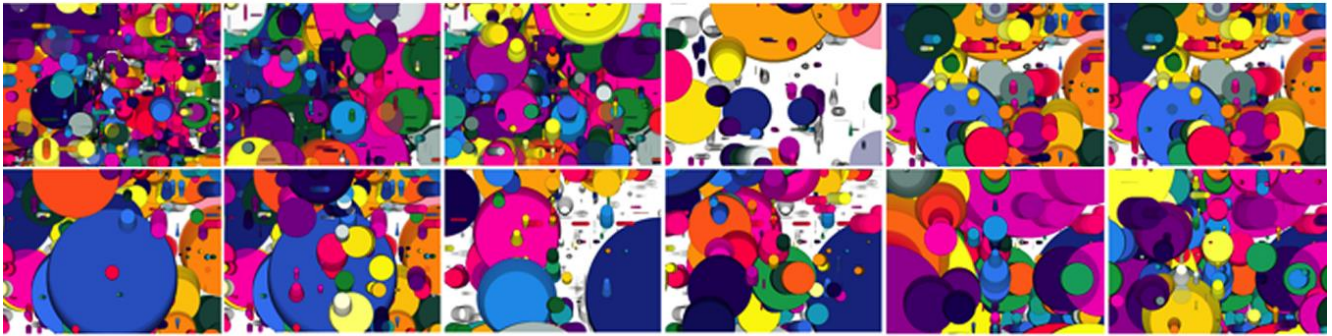

*Figure 33. User 2 (Theta – Beta).*

Unlike the others, these brain waves do not saturate the image composition, leaving blank spaces that allow breathing. The figures tend to group in certain positions, which generates a visual imbalance in the results.

**User 3: Delta – Gamma**

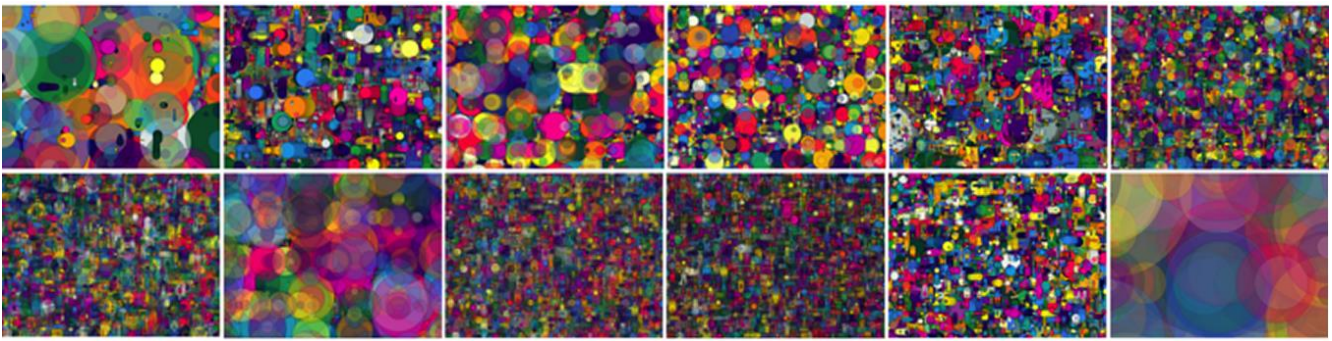

*Figure 34. User 3 (Delta – Gamma).*

With these brain waves the figures are massively grouped along the whole composition, this generates a grid of overlapping circular shapes that change their opacity.

Regarding the users' mental states, it was found that each one experienced different sensations before the interaction with the BCI device.

Thus, user one was sleepy at the time of the interaction, user two was attentive-expectant and the user three was relaxed.

In this sense, not only the brain waves of each individual provide variability to the generated art pieces, but also their mental states are a great input in the construction of the images. Based on the information provided by the previous figures it was decided to select Delta – Gamma waves, since they are the pair that brings up more expressiveness and variability (Diaz Rincon, Reyes Vera, & Rodriguez C, 2019a).

This decision is also supported by the fact that these brain waves are the ones in a lower and higher frequency range respectively, as shown in the study conducted by (Abo-Zahhad, Ahmed, & Abbas, 2015) and Figure 35.

The above is reflected in the fact that having fewer circles that change rapidly in size provides greater contrast, variability, and motion in the art pieces.

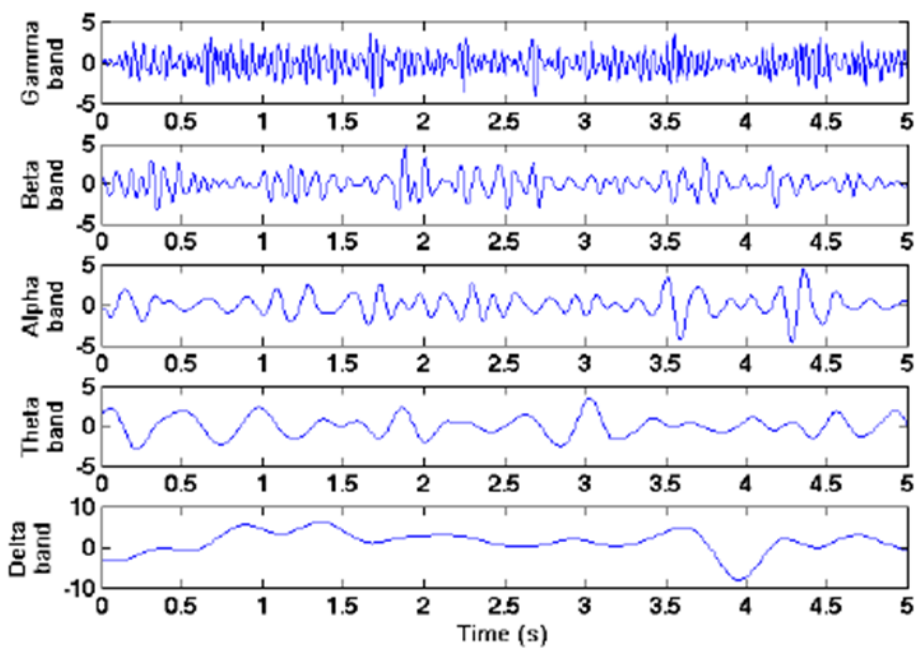

*Figure 35. EEG brain frequency chart.*

The Figures below show tests performed on three different users aiming to obtain the values of their Delta – Gamma waves to produce generative art.

**User 4**

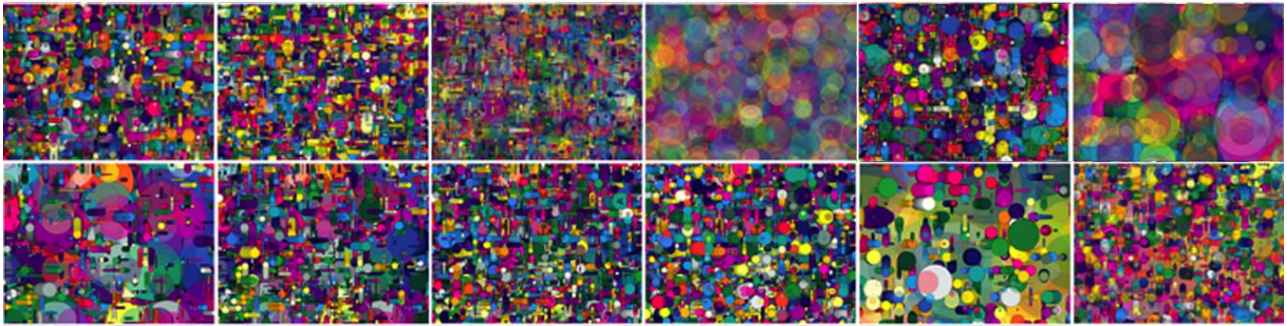

*Figure 36. User 4.*

Here, it can be seen how with these brain waves it is possible to perceive color grading, shape translation, and figures interweaving. It is important to see how the figure also presents a change in opacity which generates a diffuse image.

**User 5**

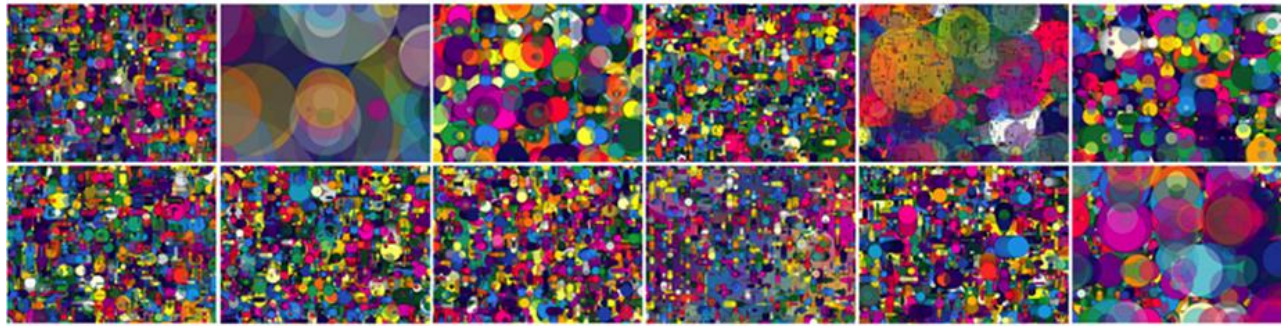

*Figure 37. User 5.*

The reticular organization provided by these brain waves is clearer, since the size of the figures is smaller and contributes to a more logical order in the arrangement of its elements. There are certain moments in which opacity, in conjunction with scale, blurs the composition; however, they tend to be homogeneous.

**User 6**

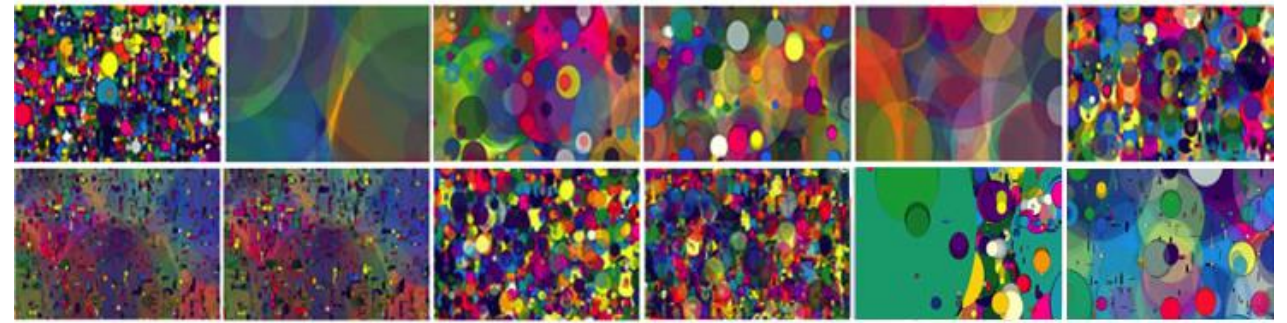

*Figure 38. User 6.*

These brain waves tend to form lattices of circular forms, however, there are certain moments where there is variation in opacity and scale that generate visual asymmetry concerning the sizes of some figures.

The tests performed and images shown above were executed in the same time frame for each user (one minute). This allows getting a greater accuracy in the data and images presented.

Regarding the tests carried out with users whose brain waves were different, it was found that the variability of these and the profile of the users allowed to get diverse results, as shown in Figures 32–34.

Similarly, each of the user-generated art pieces had different images, color patterns, and compositions, even if for each of them, the tests were carried out using the same brain waves, as evidenced in Figures 36–38.

This generates a composition with depth, whose randomness provides unique art pieces. Thus, despite using the same waves and capture times, the results of the generated art pieces are diverse.

The amalgamation of each of the above elements and their interrelationship provides the necessary inputs for creating a generative art piece that not only relies on the brain activity of a user but also fluctuates with it. This means that none of the images generated by the users are the same, since their mental states and/or brain activity change constantly. The generated art pieces are shown in the following link: http://tiny.cc/ImagenesGeneradas.

Figure 39 shows a fragment of the application's source code. The execution is available at the following link: http://tiny.cc/LivePerformanceRoma.

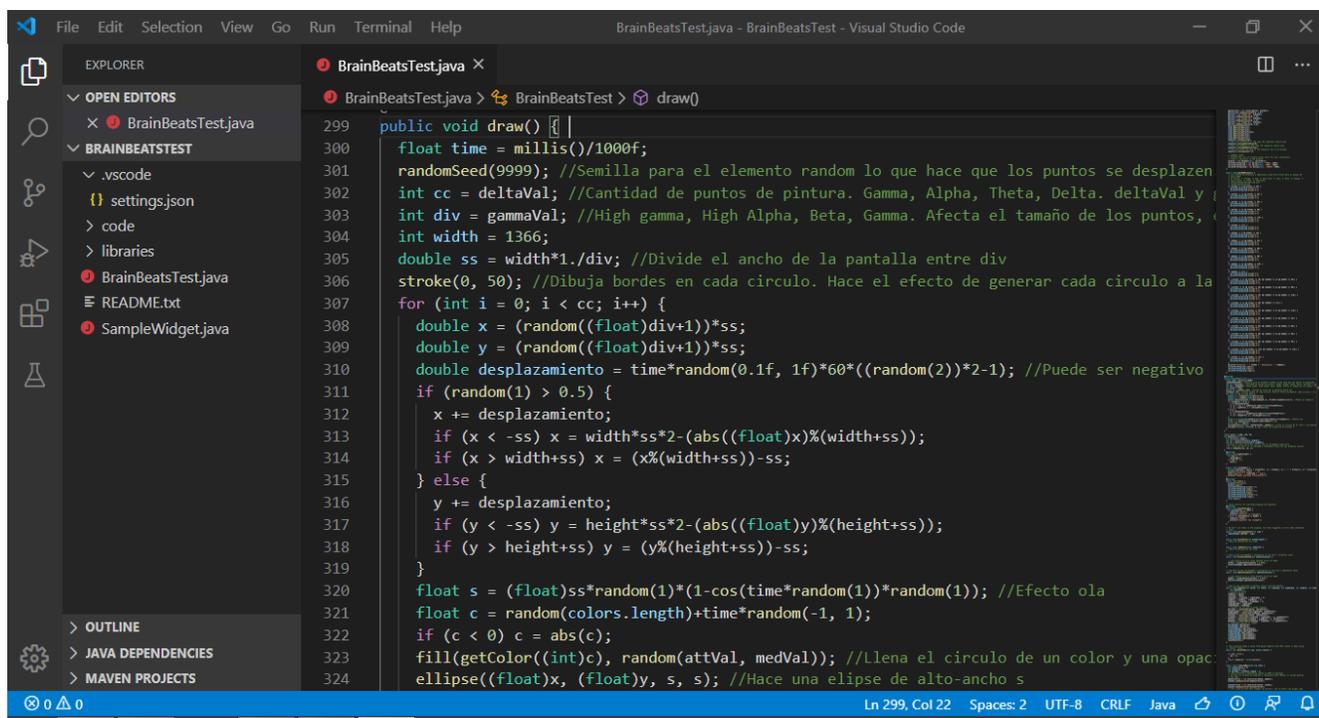

*Figure 39. Source code: Generative art.*

# CHAPTER 5. CEREBRUM: INTERACTIVE INSTALLATION DESIGN

In this section, design refers to the adequacy, categorization, and characterization of the elements that constitute the interactive installation.

The main goal is to produce in the spectators (users) a multisensory experience with the elements arranged in the installation to achieve inclusion among the different users that interact with the installation.

Thus, to provide an immersive and inclusive experience, it was decided to add a set of light bulbs, air currents, and scents that allow visually impaired users to enjoy the installation. For this reason, it was named: Cerebrum: Inclusive Interactive Installation.

Thus, the installation contains the following items:

1. **Computer:** This element provides the logic functioning and is in charge of the proper management of all the resources and components. It is camouflaged and its access is exclusively to the developer.
2. **BCI device:** This element is located on a Styrofoam–mannequin head in a position that suggests the user how it should be used.
3. **Speakers:** They allow listening to the sound that is being generated. In order to offer a surrounding effect, they are located at different distances from the spectator, around the area.
4. **Humidifier light bulbs:** These are light bulbs that change color synchronously based on changes in the brain waves and mental states of the user. In addition, they generate scented air currents regulated from the users' brain activity. This element is relevant for the visually impaired users, since it represents one of its main ways of interacting with the installation.
5. **Chair with armrest:** In addition to serving as a reference point for the sighted user, it is also a reference point for the visually impaired users to locate themselves properly. This element has a hidden access, where the computer and other electronic components (wires, Arduinos, actuators, sensors, among others) will be located. In this way, it becomes a container that holds the elements that are essential to the installation, but imperceptible to the user.
6. **Element of visual projection:** It allows the sighted user to appreciate in a unique format his/her art creations. This format has the characteristic that it is not a flat surface but a sequence of pyramids in a spiral-shaped scale. This is done with the purpose of offering a deep sensation while projecting the art pieces created by the interaction between spectators and the art installation.
7. **Video-beam in back-projection:** It is strategically located at the back of the viewer, so that their shadow does not cover the generated art pieces.
8. **Physical space:** The installation is designed to be projected in a facility of 3m length, 4m width, and 2m height. Although small variations can be made, what is required is that acoustic and lighting control can exist. For this reason, in the facility arranged for the installation the passage of light from the outside has been interrupted.
9. **Accessibility:** The floor has a rough-textured path that highlights the route. This path has a reflective paint and high relief which benefits any user, providing them with a guide regarding the route they should follow to interact with the installation.

Once these elements have been defined, a brief description of the conceptual model (use) of the installation by a spectator is made.

Use sequence:

1. **Before entering:** The exterior of the facility provides graphical instructions on how to use the BCI device. There are also Braille instructions strategically located to allow visually impaired users to read.
2. **Entrance:** The place will be in darkness since the projector is in repose and the exterior light is regulated. The attendees will be guided through the rough-textured path in reflective painting so they can reach the chair arranged for them.
3. **Once seated:** On the right-hand side there is a Styrofoam mannequin head that shows how to place the BCI device in the spectator's head. For the visually impaired users there are Braille instructions located on the chair and a high relief diagram that will allow them to understand how to use the device and where is it located.
4. **When the interaction starts:** Both types of users will activate the image, sound, light bulbs, and scents through their brain waves, which are recorded by the BCI device and processed by the developed algorithm. The interaction will be parallel, which will allow the perception from sight, touch, and smell. In this way, not only the mental states influence the creation of the art pieces, but also the users' senses.
5. **Interaction time:** In the tests carried out in this research work, it was found that the adequate time to avoid cognitive user fatigue should not exceed 5 minutes.

The front and lateral projections on the visual element are shown in the following links: http://tiny.cc/FrontalProjection and http://tiny.cc/LateralProjection.

Figure 40 shows the installation drawing and the arrangement of the listed elements.

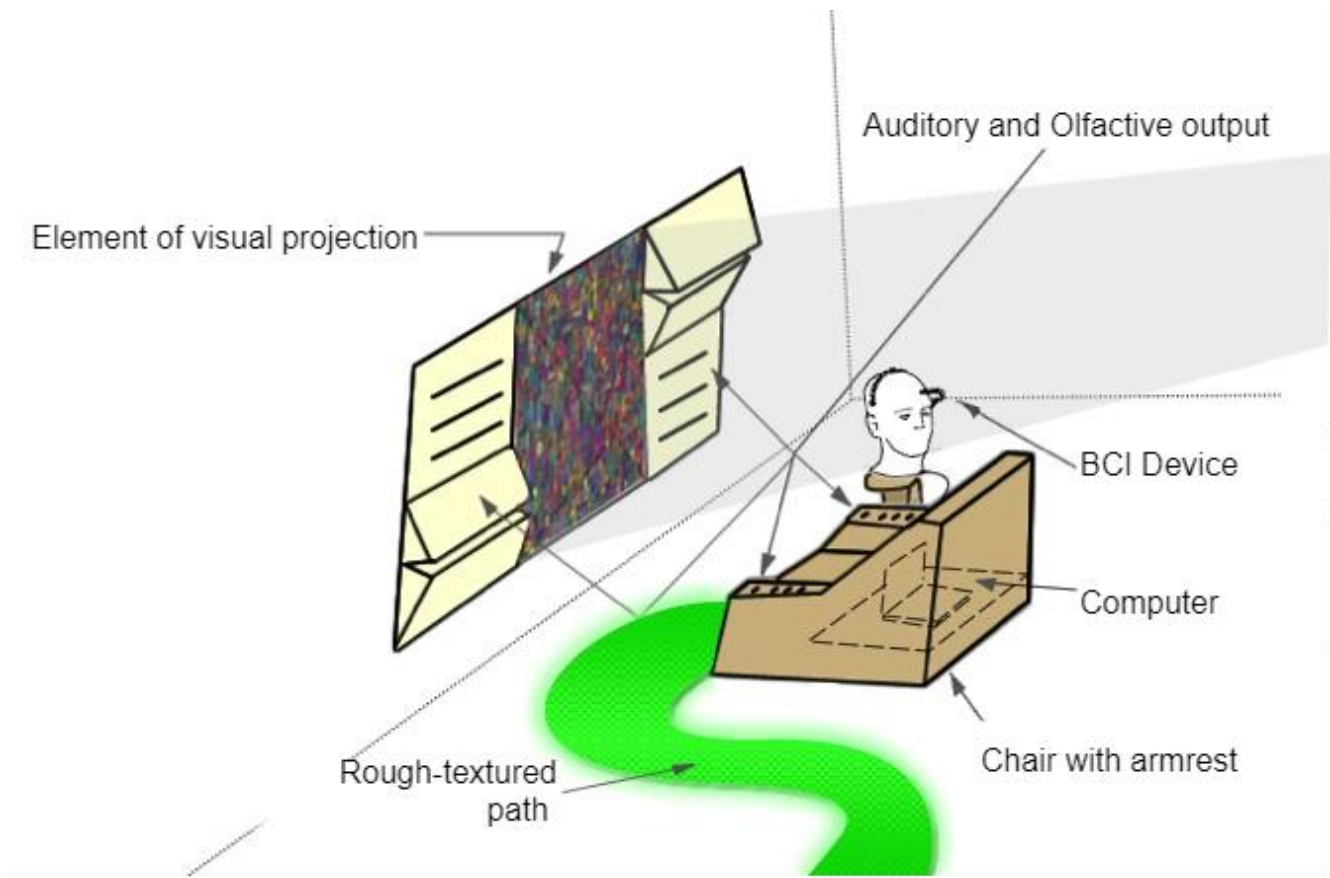

*Figure 40. Installation drawing.*

## 5.1. Electronic component

To make the artistic installation accessible to both sighted and visually impaired users it was decided to include 3 air humidifiers located in a chair according to the installation drawing. Their function is to act as light-emitting and scented air currents. This way, in addition to the visual and auditory modalities the olfactory and kinesthetic modalities are included, thus allowing a Multimodal Artistic Installation to be proposed. These humidifiers were intervened to be able to connect them to an Arduino and control the light bulbs and spray using the developed algorithm. Figure 41 shows the design of the electronic circuit in charge of performing the actions indicated by the microprocessor.

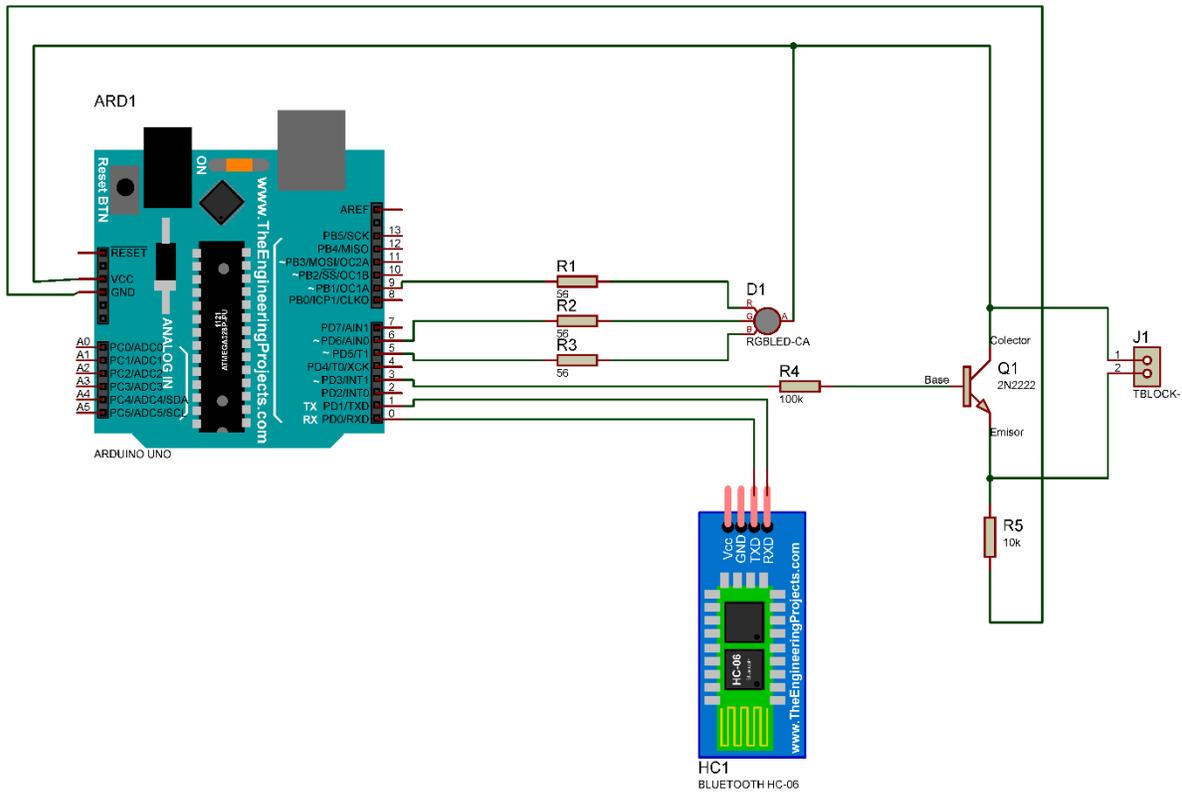

*Figure 41. Control circuit diagram.*

From the above figure it can be seen that the electronic circuit consists of two parts: Light-color and Spray control. To control the color of the light bulbs, three 56 ohm resistors and the outputs 5, 6, and 9 of the Arduino which correspond to PWM (pulse width modulation) outputs were used. With these outputs the voltage is changed, so that it is less than or equal to 5V. This is achieved by writing in the source code a number between 0–255, corresponding to the range of RGB colors.

The light bulbs used by the humidifier are RGB common anode, which means that when selecting a color 255 must be subtracted from each component. Resistors of 56 ohms were used to limit the current flow and voltage drop (between 2V and 3.2V) on the humidifier's LEDs. It is important to point out that if these resistors are not used, the LEDs could burn out, since 5V could reach them directly from the Arduino's outputs.

Figure 42 shows the electrical circuit for light control.

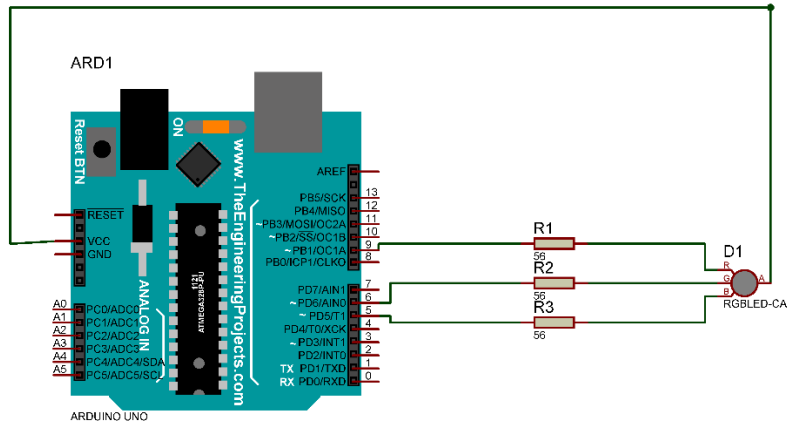

*Figure 42. Electrical circuit for RGB light control.*

To control the spray, a circuit that works like a button or switch was designed. Doing so, it is possible to use the Arduino to change the on/off status of the spray.

To achieve this, a 2N2222A transistor and resistors with values Rb = 100Kohm and RE = 10Kohm were used so that the transistor could work in strong saturation, in other words, to allow the passage of the maximum current. This ensures the output between the collector and the transistor emitter to work as a switch controlled by the Arduino output (pin 3), as shown in Figure 43.

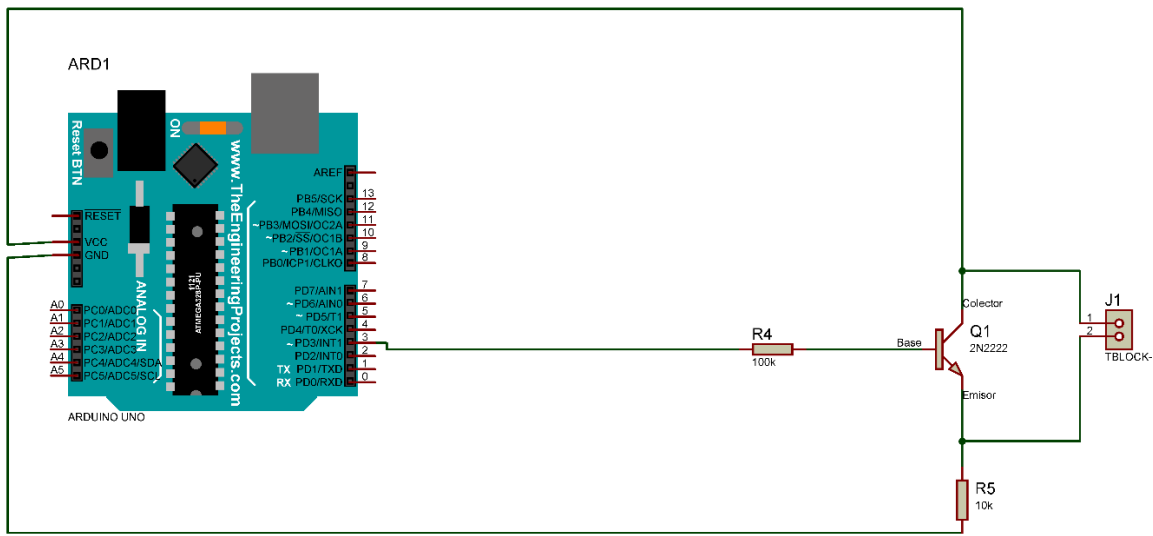

*Figure 43. Electrical circuit diagram with transistor 2N2222A, in strong saturation.*

Finally, a HC06 Bluetooth module was used in order to handle the communication between the Arduino and the application, where the TX and RX pins of the module were connected with their opposites in the Arduino, as shown in Figure 44.

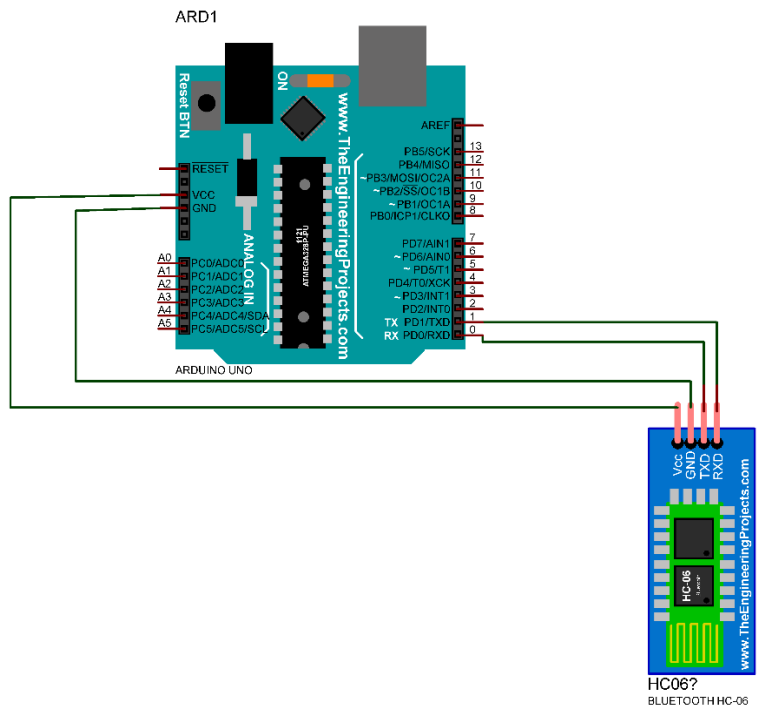

*Figure 44. Connection of the Bluetooth module in the Arduino.*

In order to integrate the light bulbs into the prototype, a printed circuit was implemented.

By soldering the individual components to the boards, it is possible to present the configuration of the humidifiers in a better and simpler way. Similarly, this process also makes it possible to reduce the number of wires in the installation.

Finally, the integration and connection between the application, the humidifier, the Arduino, and the board will make it possible to obtain as a result what is shown in Figure 45.

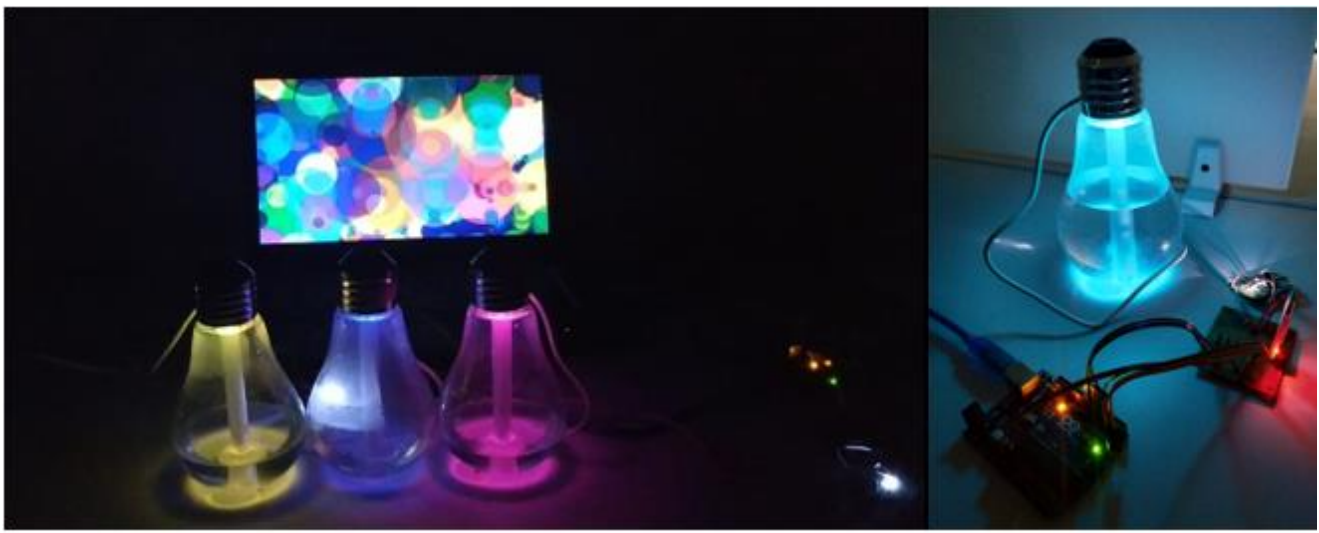

*Figure 45. Integration between components.*

The source code written in Arduino is explained below: Three integer variables with the values 6 (Red), 5 (Green), and 9 (Blue) are declared for controlling the RGB light colors. A variable for spray control is also declared with the integer value 3. Each of these values corresponds to the Arduino outputs or pins that will be used. Then, in the Arduino's setup function, the transmission rate for serial data is set to 9600, the color-related pins are set as analog outputs and the spray as a digital output.

Then the simulation of a digital pulse is explained. To do this, a logical one must be sent to the port, which means that 5V is sent for 200 milliseconds (switch closed). Then a logical zero must be sent to the port, which means that 0V is sent for the same period (switch open). The above is intended to simulate the button states (on/off) that allow to control the spray by sending the character "S".

Then the color variations in the light bulbs are defined by sending different characters. To do this, the color's RGB values are sent to the analog pins (those related to the light bulbs). Since the light bulbs are common anode, it is necessary to subtract 255 from each value. However, this is easily solved by using an auxiliary function, so the exact RGB color values can be sent to the output.

Finally, a delay of 200 milliseconds is set so that the change between colors is not too fast and can be observed by the human eye. In this regard, 16 colors were defined, represented by numbers from 0 to 9 and letters from A to F. Consequently, the change between colors represents the variation in the user's mental states, specifically, in its attention and relaxation levels which due to the use of MIDI data, were transformed so that they were in a range between 0 and 127 and not between 0 and 100 as defined by the manufacturer of the BCI device.

Thus, six ranges are defined: [0, 20], [21, 40], [41, 60], [61, 90], [91, 110] and [111, 127]. Since all the values in the proposed range were considered, the color of the light bulbs will change each time the attention and relaxation levels are in one of the 6 intervals.

In this regard, to distinguish more easily the mental states, the first bulb shows when the attention levels are in the aforementioned intervals, the second one shows the relaxation levels and the third one shows when both mental states are in the same interval.

Figure 46 shows a fragment of the code written in Arduino and its use in the main application. The functioning of the light bulbs, spray and their integration with the main application is available in the following links: http://tiny.cc/InstallationLights1 and http://tiny.cc/InstallationLights2.

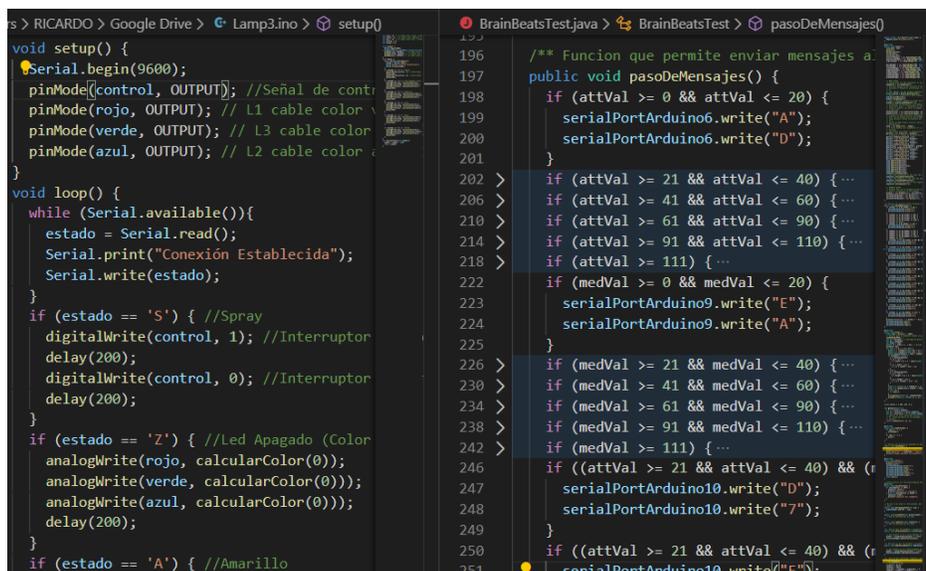

*Figure 46. Arduino source code.*

# CHAPTER 6. DESIGN AND TESTS EXECUTION

To guarantee some basic elements of quality in the developed solution, hardware tests (BCI device), software tests (static tests), and two usability tests were carried out. It is important to clarify that, due to the national decree of mandatory isolation, the installation could not be completed and the usability tests could not be executed.

However, both a heuristic test and a direct observation test were designed and included because of the contribution that adapting usability evaluation heuristics can have in the context of generative art.

As for the software tests, only static tests were chosen since the devices and development equipment were left inside the research laboratory, and it was not possible to access them.

## 6.1. BCI testing and Signal Processing

The purpose of the tests is to verify the reliability of the data sent by the BCI. This is relevant since this research work transforms brain data into an audio and visual output. For this reason, it is necessary to pre-process, process, and analyze the data making sure they correspond to the users' brain activity and not defects on the device's electrode or other sources of noise.

To test the proper functionality of the device OpenViBE[5] was used. OpenViBE is a software platform for neuroscience research that allows designing, testing and using BCI devices. It can also be used to acquire, filter, process, classify and display brain signals in real-time. OpenViBE offers a set of tools that allows the use of different algorithms to perform processing and classification of encephalographic signals (Renard et al., 2010).

The designed tests are divided into three parts: The first consists of verifying the connection between the BCI and the computer. The second consists of testing the device's electrode and identifying cases or scenarios that may create noise. The third and fourth tests consist of applying different filters to both the electrode and encephalographic signals.

**First test:** Verify the connection between the BCI and the computer.

**Objective:** OpenViBE will verify that the BCI device and the computer are successfully connected.

**Software Resources:** OpenViBE Acquisition Server. OpenViBE Designer. Operating System: Windows or Linux.

**Hardware Resources:** Mindwave (BCI). Memory: Minimum 2 GB RAM. Bluetooth link. Processor: Any with a speed higher than 2.0 GHz.

**Task script:** With the device placed on the user's head, make sure the electrode is positioned in the FP1 position (above the left eyebrow) according to the 10 – 20 system and making contact with the forehead.

In this way, recordings of the lateral prefrontal cortex (LPFC) and Brodmann 10 area will be acquired. The device's reference terminal is then connected to the left earlobe, this is, the position A1 in the 10 – 20 system, as shown in Figure 47.

---

[5] http://openvibe.inria.fr/

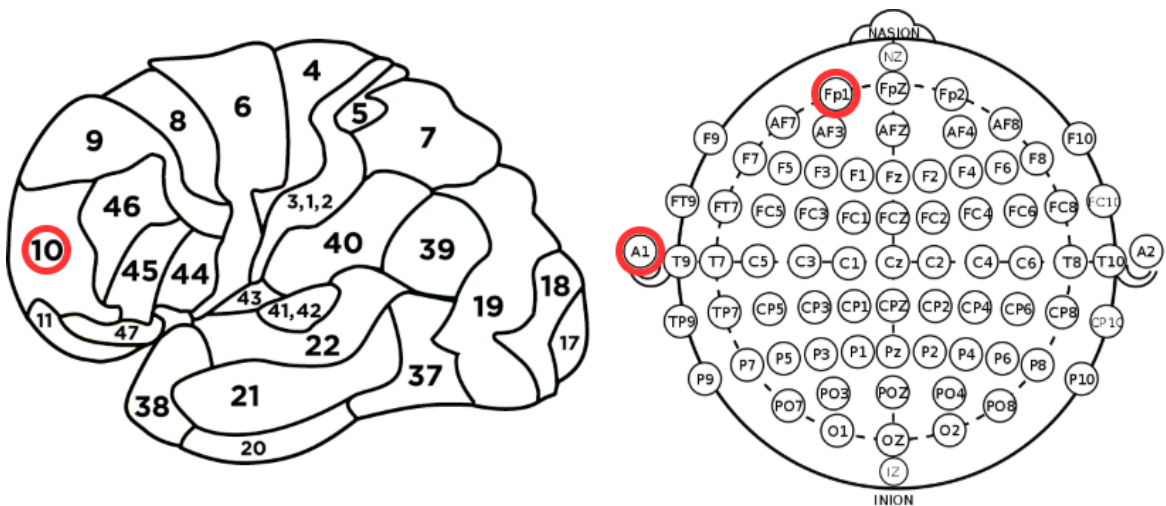

*Figure 47. Mindwave: Electrodes placement (10 – 20 system) and Brodmann areas.*

Once the electrode and the device are correctly positioned, the OpenViBE Acquisition Server is launched. In the new window, the name of the device (Neurosky Mindset) is searched for. Likewise, the sample rate sent by blocks is set to 32 since this value allows a good data transmission speed regarding the device's electrode, attention and relaxation levels. In this way, it is possible to obtain reliable data. Figure 48 shows the configuration.

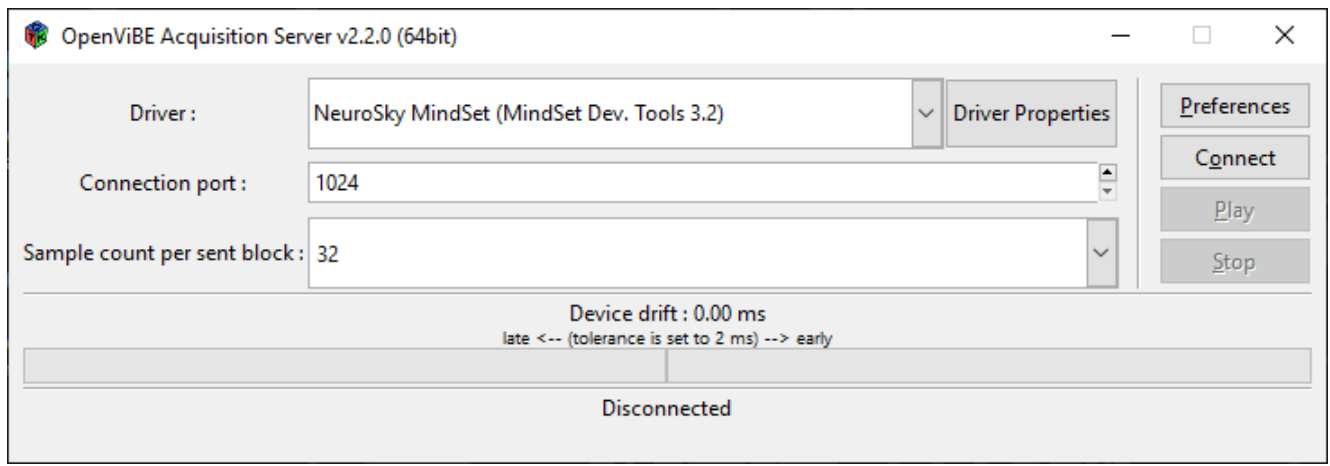

*Figure 48. BCI device ready to be used.*

Then click on the preferences button to set the BCI device configuration. The configuration panel allows to select the age, gender, sampling rate and to choose which values to display, including the eSense values (attention and relaxation levels), brainwaves' values and blink strength.

Since we are only interested in the eSense values at the moment, the user must click on this option and then apply the configuration as shown in Figure 49.

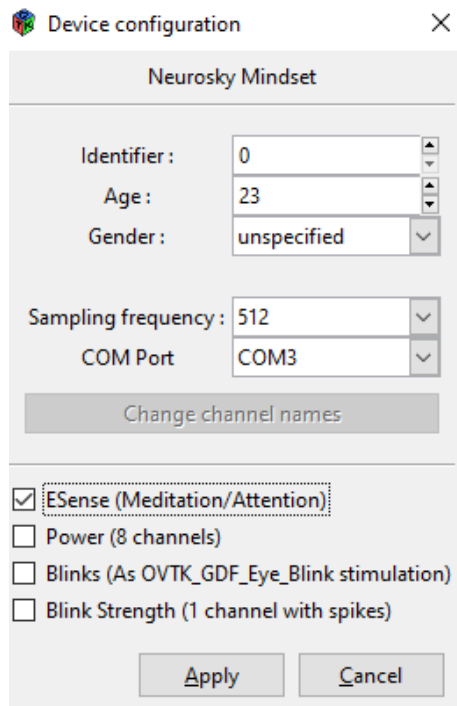

*Figure 49. Device configuration.*

Once the configuration is complete, make sure the BCI device is loaded, turn it on and click on connect, as shown in Figure 50.

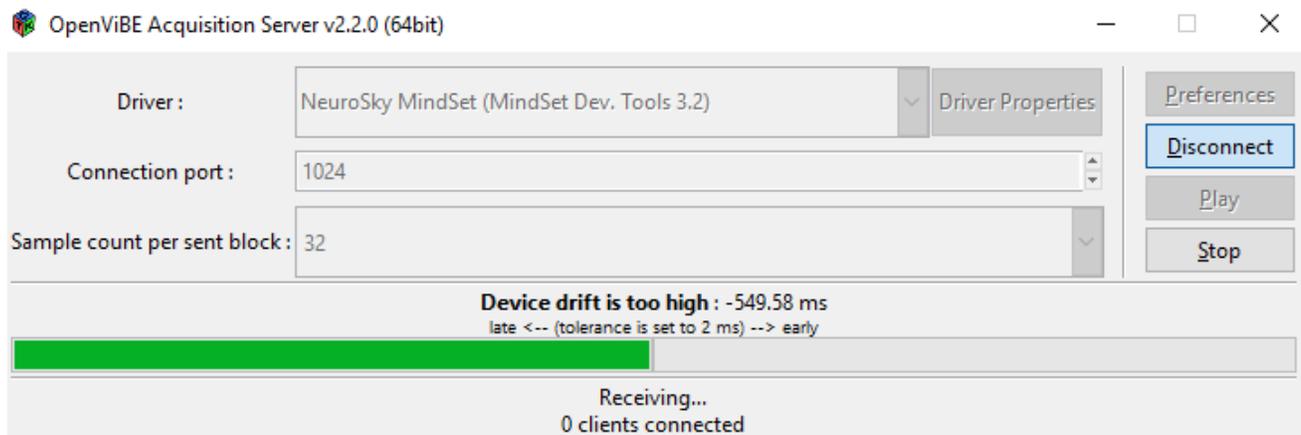

*Figure 50. OpenViBE Acquisition Server running.*

**Test execution results:** Once the test has been executed, it can be seen the successful connection with the serial port in which the brain data is transmitted via Bluetooth. Also, the signal quality shows low noise levels (less than 12.5%) as can be seen at the end of Figure 51.

Similarly, the figure shows how the connection to ThinkGear is made, which works as a background-running process on the computer in charge for directing the device data from the serial port to an open network socket.

*Figure 51. BCI device and OpenViBE connection.*

When the device is disconnected, the signal is low or the electrode does not make contact with the forehead, the software automatically detects the anomaly, as shown in Figure 52.

*Figure 52. Low signal.*

**Second test:** Testing the device's electrode and identifying cases or scenarios that may create noise.

**Objective:** The OpenViBE Acquisition Server and the OpenViBE Designer are used to verify that the data between the BCI device and the computer are sent properly. Also, cases or scenarios that may create noise are identified.

**Software Resources:** OpenViBE Acquisition Server. OpenViBE Designer. Operating System: Windows or Linux.

**Hardware Resources:** Mindwave (BCI). Memory: Minimum 2 GB RAM. Bluetooth link. Processor: Any with a speed higher than 2.0 GHz.

**Task script:** With the BCI device properly charged, placed and turned on and the OpenViBE Acquisition Server running, open the OpenViBE Designer and configure the desired scenario.

To do this, search and select in the search box the Acquisition client which is a generic network-based signal acquisition client that allows data acquisition from the BCI device. Also, search for the Signal Display box, which displays in a graph the acquired signal. Finally, the colors of the terminals are connected to each other. The design of the scenario is shown in Figure 53.

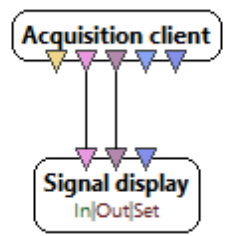

*Figure 53. Signal display.*

**Test execution results:** Once the test has been executed, the fluctuation of the electrode's raw signal and the different values it reaches can be seen.

During repeated tests in this scenario and by selecting Blink Strength in the device setup (Figure 49) it was found that blink and blink strength have a large impact on the validation of EEG signals, affecting the fluctuation of the received values and creating spikes in the signal.

These spikes would be relevant if there were under study the number of times a person blinks when reading a book while in a certain mental state or when studying eye blink characterization from frontal EEG electrodes using source separation and pattern recognition algorithms (Roy, Charbonnier, & Bonnet, 2014). However, this does not correspond to the studies carried out in this research work, for that reason, it is necessary to avoid the blink detection since it represents a source of noise.

The spikes produced by blinking can be seen in Figure 54.

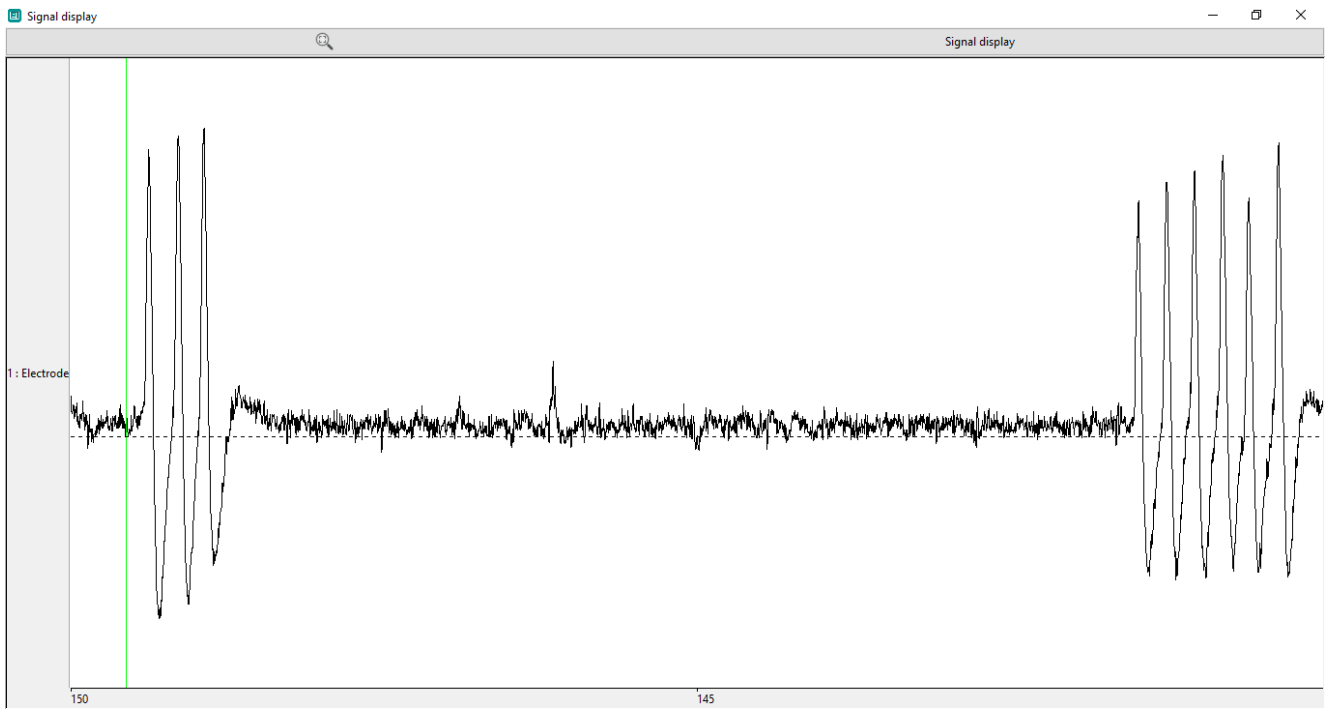

*Figure 54. Raw signal and spikes.*

**Third test:** Applying filters to the device's electrode.

**Objective:** Using the OpenViBE Acquisition Server and the OpenViBE Designer some filters and methods will be applied to the raw electrode signal. This is of vital importance since applying digital filters allows blocking or suppressing certain frequency components in the signal, passing the original signal minus these components to the output.

Unlike analog filters, digital filters work by applying mathematical operations to the signal (Kropotov, 2009). Thus, signal filtering allows for affecting and/or modifying the shape or time structure of EEG signals. In addition to this, by applying filters it is possible to verify the functioning of the electrode when acquiring encephalographic signals and their quality.

**Software Resources:** OpenViBE Acquisition Server. OpenViBE Designer. Operating System: Windows or Linux.

**Hardware Resources:** Mindwave (BCI). Memory: Minimum 2 GB RAM. Bluetooth link. Processor: Any with a speed higher than 2.0 GHz.

**Task script:** Four scenarios described below will be used to apply the filters to the electrode:

**Scenario 1:** With the BCI device properly charged, placed and turned on, and the OpenViBE Acquisition Server running, open the OpenViBE Designer and configure the desired scenario.

To do this, the Acquisition client is searched for, selected and connected to two terminals: The first is Signal Display, renamed as Raw Data. The second is Signal Average, which shows an approximation to the average of the raw signal. Then, this terminal is connected to Signal Display which will be renamed as Average Display. The scenario design is shown in Figure 55.

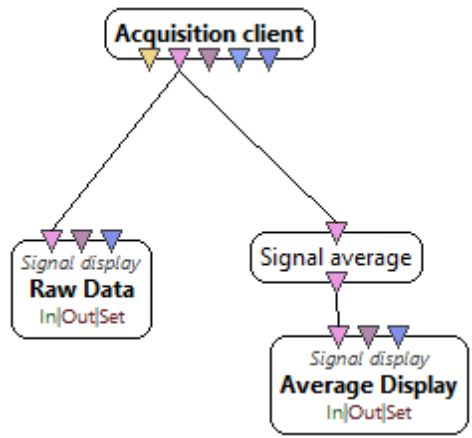

*Figure 55. Scenario 1: Signal average.*

**Scenario 1. Test execution results:** The first filter shows how the signal averaging can be calculated and plotted from the raw signal. Figure 56 shows on the left side the raw signal and on the right side the average of the same signal. This filter allows approximating the signal making it finer and using only the average of the values that have been common or frequent during signal recording.

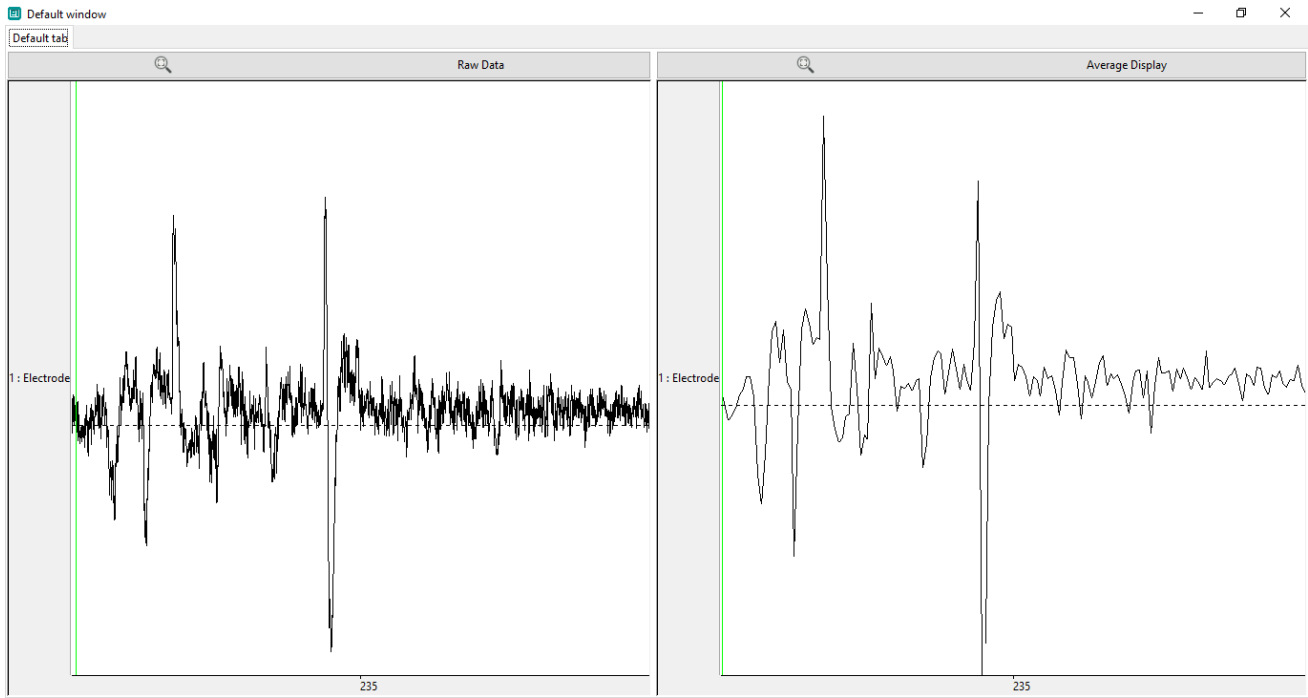

*Figure 56. Signal average.*

**Scenario 2:** With the BCI device properly charged, placed and turned on, and the OpenViBE Acquisition Server running, open the OpenViBE Designer and configure the desired scenario.

To do this, the Acquisition client is searched, selected and connected to Time Based Epoching which divides the EEG signal into a time interval where each epoch serves as a reference point from which time is measured. From there, two connections are made: The first goes to Signal Average with its respective Signal Display renamed as Average Display Epoch where epoch's average signal can be seen. The second connection goes to Signal Display, renamed as Raw Data Epoch which shows the raw signal captured during the epoch.

Then, from the Acquisition client, the Signal Average is connected to its respective Signal Display renamed as Average Display. Next, the Acquisition client is connected to Signal Display, renamed as Raw Data. Thus, this scenario shows the average and raw signals recorded by epochs and without them. Figure 57 shows the configuration.

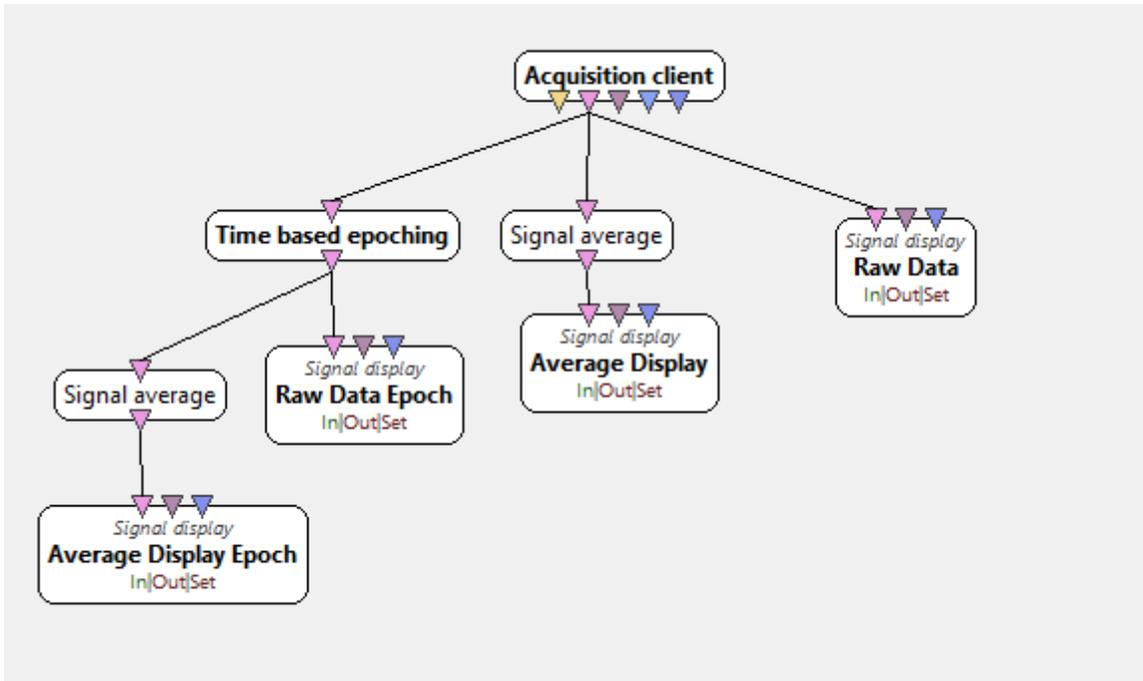

*Figure 57. Scenario 2: Time based epoching*

Since each epoch is configurable, these will be defined in intervals of 0.5 and 2 seconds. To do this, double-click on Time Based Epoching and type in the values, as shown in Figure 58.

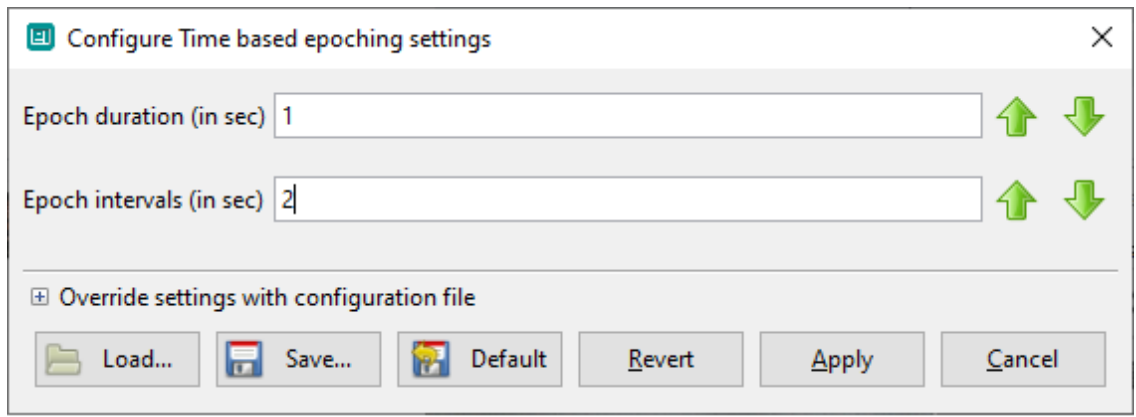

*Figure 58. Epoch configuration.*

Since the execution of this scenario involves several windows, the Window Manager will be used. This is a module dedicated to handling 2D and 3D graphic windows in OpenViBE scenarios. To use it, click on the Window Manager button and once the window is displayed, select and drag each of the elements accompanied by a magnifying glass (corresponding to each Signal Display in the scenario). The way the results are displayed is described below.

On the left side, from top to bottom, the raw signals are shown. Thus, the upper left-hand side shows the raw signal recorded by epochs (at certain time intervals) and the lower left-hand side shows the raw signal.

Similarly, the right side shows the average of the signal. Thus, the upper right-hand side shows the signal average per epochs and the lower right-hand side shows the signal average. Figure 59 shows the arrangement of the windows.

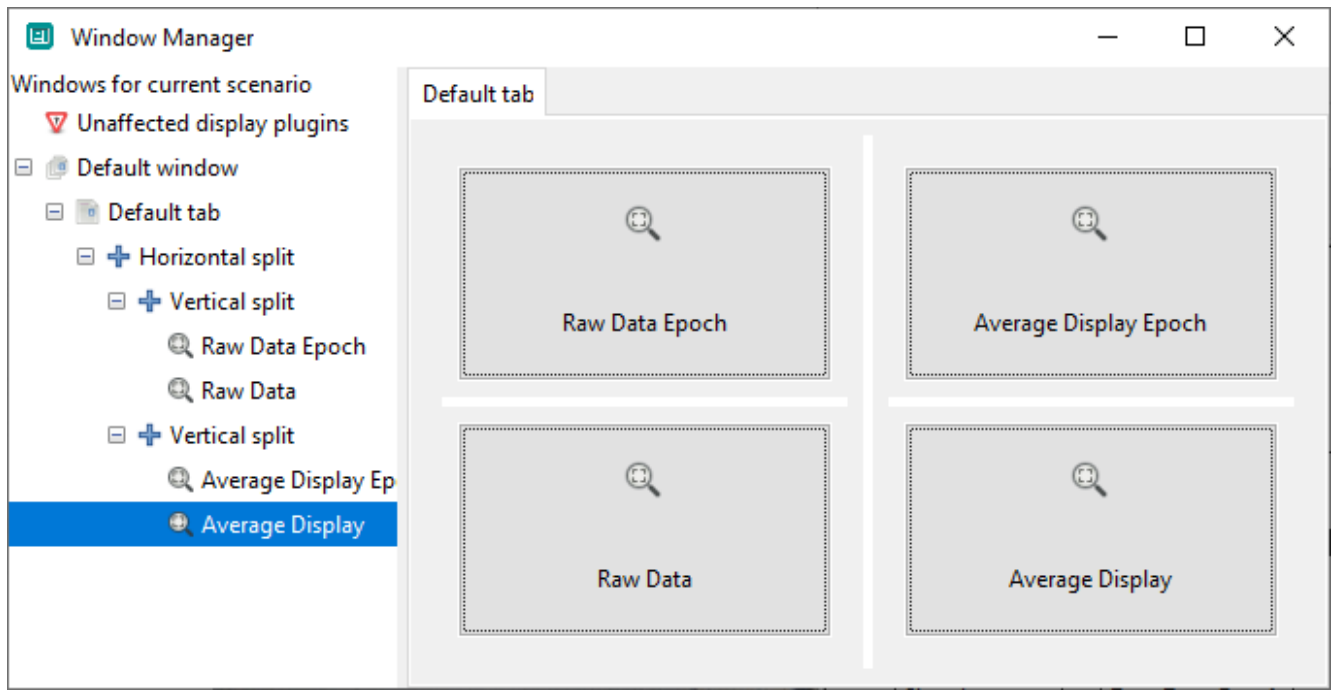

Figure 59. Windows configuration.

**Scenario 2. Test execution results:** The execution of this scenario is shown in Figures 60 and 61, with epochs of 0.5 and 2 seconds respectively.

Epochs are commonly used when analyzing a particular stimulus or mental state. In this way, the proposed scenario involved monitoring a user with eyes closed in order to analyze the values produced by the electrode while in a meditative state.

Figure 60 shows how the use of epochs significantly changes the result of the received signal. In this way, in the upper part of the figure, which shows the data captured by epochs, it can be seen how the signal is more balanced and consistent.

Let's look at this in detail: On the left side, from top to bottom, the raw data by epochs and the raw data without epochs are shown. Here it is possible to see the difference between the variation of the signals, going from regular and constant in the upper part, to chaotic and irregular in the lower part. The averages of the raw signals shown on the right side also remark this.

Accordingly, the right-hand side shows, from top to bottom, the average of the raw signal processed by epochs without them (in normal time periods). Here it becomes much easier to notice how the use of epochs, when analyzing a particular mental state or stimuli allows the recorded signal to be much more fluid and easy to process, since it does not present too many alterations or changes.

The same analysis is done for epochs in 2-second time periods as shown in Figure 61.

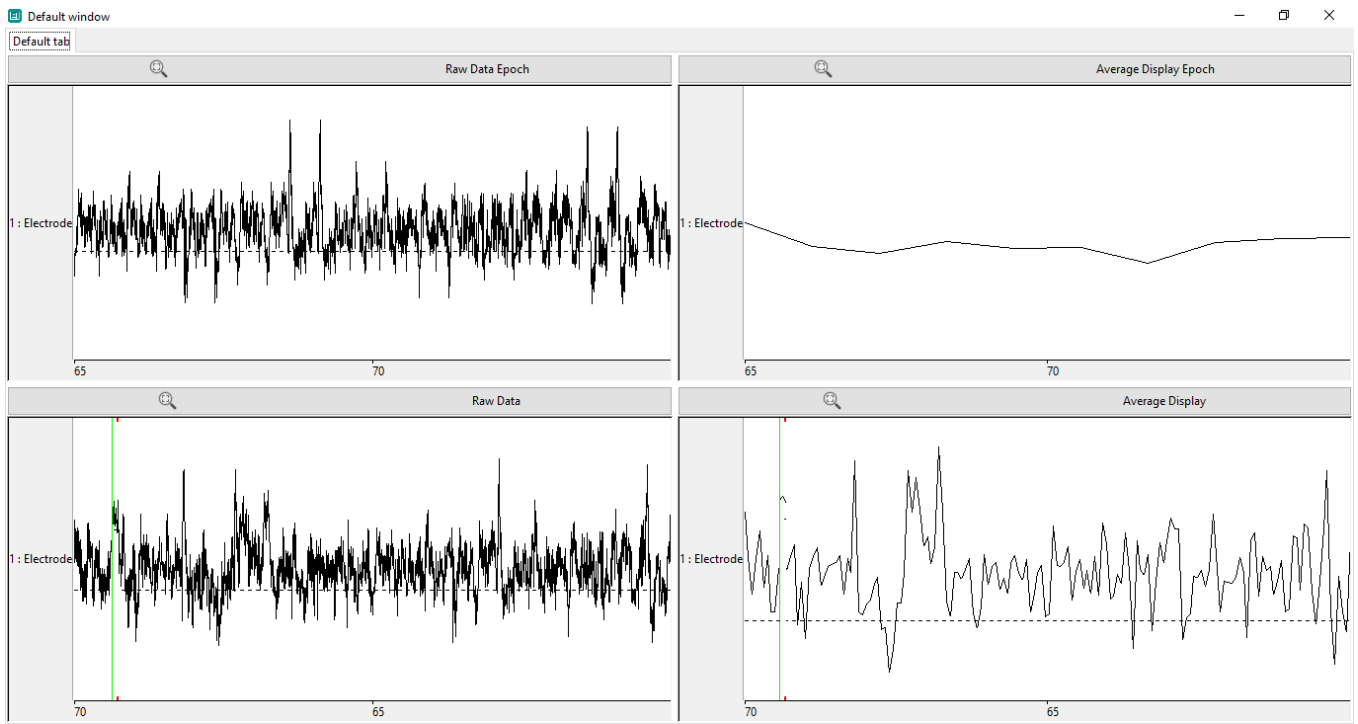

*Figure 60. Epoch at 0.5 second intervals*

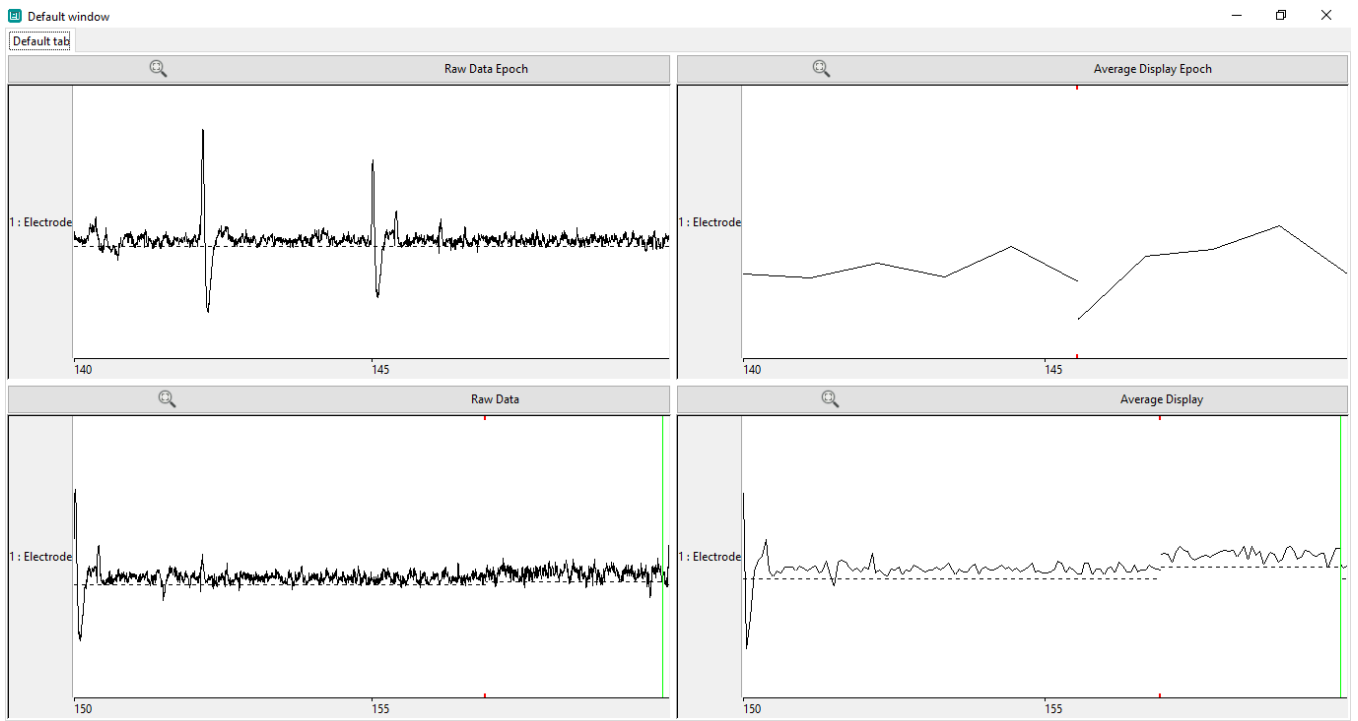

*Figure 61. Epoch at 2 second intervals*

**Scenario 3:** With the BCI device properly charged, placed and turned on, and the OpenViBE Acquisition Server running, open the OpenViBE Designer and configure the desired scenario.

To do this, select the Acquisition client and connect it to Channel Selector, which allows selecting the channels to which the filters will be applied.

From here three terminals are connected: The first is Signal Display, renamed as Raw Data, the second is Signal Decimation, renamed as Signal Decimation (16), which is a filter or process whose goal is to reduce the sampling rate which will be later connected to its respective Signal Display renamed as Decimated Signal (16). The third terminal is another Signal Decimation but this time with a value of 32, connected to its respective Signal Display.

In practice, Signal Decimation usually involves a low-pass filtering of the signal and then removing some of its samples. The decimation factor is simply the ratio of the input rate to the output rate. In this scenario, the sample reduction rates are 18 and 32. The proposed scenario is shown in Figure 62.

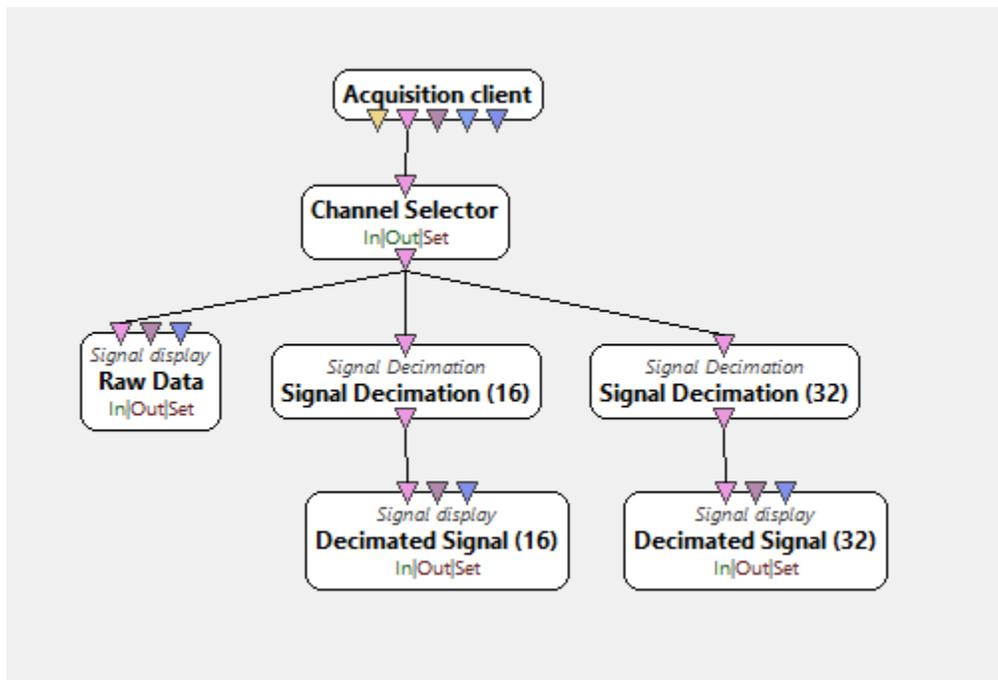

*Figure 62. Scenario 3: Signal decimation*

**Scenario 3. Test execution results:** The execution of this scenario is shown in Figure 63, where the raw signal is displayed first, followed by a signal decimated by factors of 18 and 32 respectively.

From the figure it can be seen that the following graphs become progressively finer and finer thanks to filtering. The most immediate reason for its application is to reduce the sampling rate at the output of a system so that another system operating at a lower sampling rate can receive the signal as input.

However, a major motivation for applying this filter is to reduce the cost of processing: The calculation and/or memory needed to implement a Digital Signal Processing system is usually proportional to the sample rate, so the use of a lower sample rate usually results in a cheaper implementation.

In this sense, at the beginning of the testing section it was mentioned that 32 would be the sample count sent by blocks. Thus, this value allows a proper relationship between the data sent and display rates and greater fluidity in data transmission.

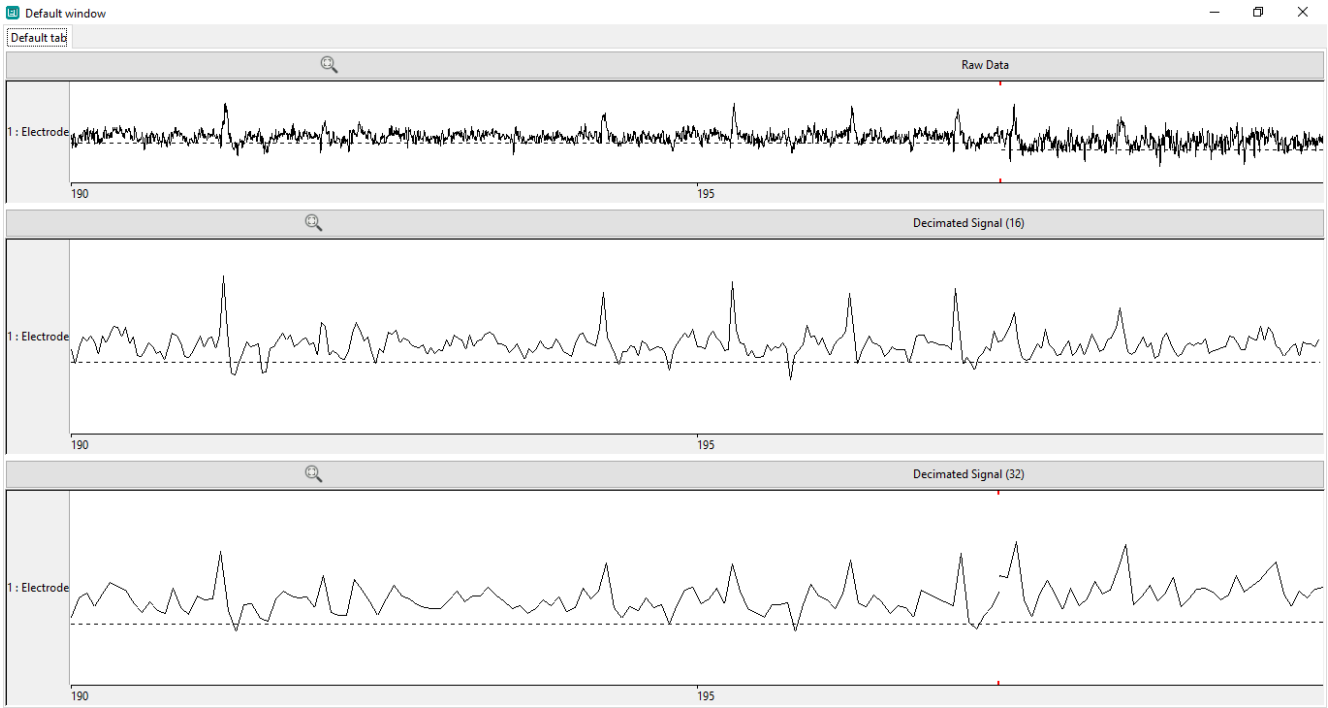

*Figure 63. Signal Decimation (Down sampling)*

**Scenario 4:** With the BCI device properly charged, placed and turned on, and the OpenViBE Acquisition Server running, open the OpenViBE Designer and configure the desired scenario.

To do this, the Acquisition client is selected and connected to Temporal Filter, which aims to remove or attenuate the frequencies within the raw signal that are not of interest. The temporal filter is then connected to Signal Display which allows the signal to be displayed. The scenario is shown in Figure 64.

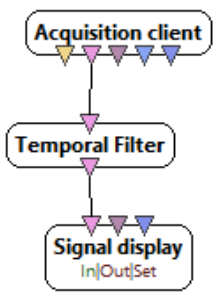

*Figure 64. Temporal filter.*

**Scenario 4. Test execution results:** The temporal filter transforms a multichannel signal $X(n)$ so that each channel $yi(n)$ on $Y(n)$ depends only on the channel $xi(n)$. Using temporal filters in signal processing is relevant, since it allows substantial improvement of the signal-to-noise ratio (SNR). The difficulty is to decide which frequencies are of interest and which are noise. Figure 65 shows the application of a temporary filter. In this way, it is shown how the raw signal on the left is processed, filtered and normalized which results in the signal on the right where its spikes are more consistent improving the signal-to-noise ratio.

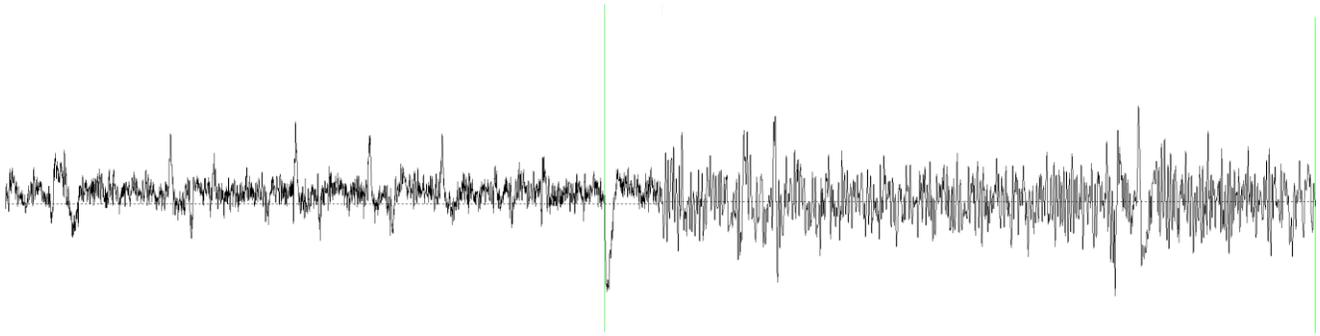

*Figure 65. Temporal filter.*

**Fourth test:** Applying filters to encephalographic signals.

**Objective:** Using the OpenViBE Acquisition Server and the OpenViBE Designer, filters and methods will be applied to the encephalographic signals to suppress and/or clean the signal output of certain unwanted components in the signal frequencies.

**Software Resources:** OpenViBE Acquisition Server. OpenViBE Designer. Operating System: Windows or Linux.

**Hardware Resources:** Mindwave (BCI). Memory: Minimum 2 GB RAM. Bluetooth link. Processor: Any with a speed higher than 2.0 GHz.

**Task script:** Since this time it's not just the electrode that's being tested, it is necessary to set the device to obtain the attention, relaxation levels, and brain waves. Therefore, with the BCI device properly charged, placed, and turned on, the OpenViBE Acquisition Server is launched and the preferences button is clicked. Then select the Power box, which will allow the user to obtain the brain waves' measurements. Finally, click on Apply to save the configuration and then on connect to start. Figure 66 shows the configuration panel.

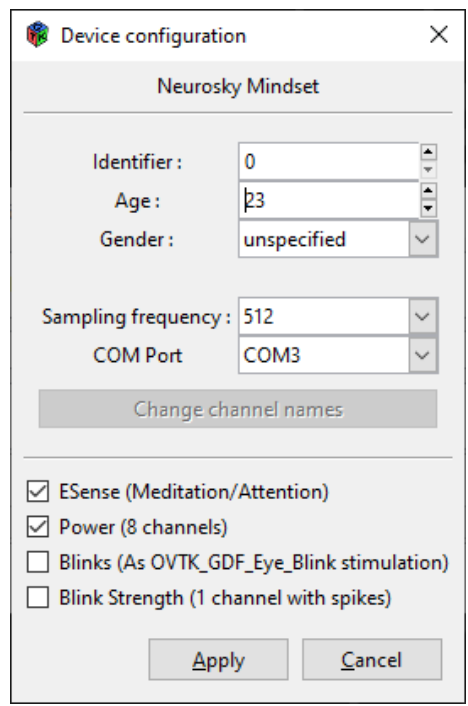

*Figure 66. Selecting brain waves for further filtering.*

Since the same scenarios mentioned above are used to run this test (See third test) the results obtained in each of them will be explained.

**Scenario 1. Test execution results:** This filter shows the average of the acquired signal. When applied, it can be seen how the raw signal on the left is smoothed when reconstructed by the common values obtained. Similarly, the signal presents more curves as a result of the approximation. Figure 67 shows the application of this filter.

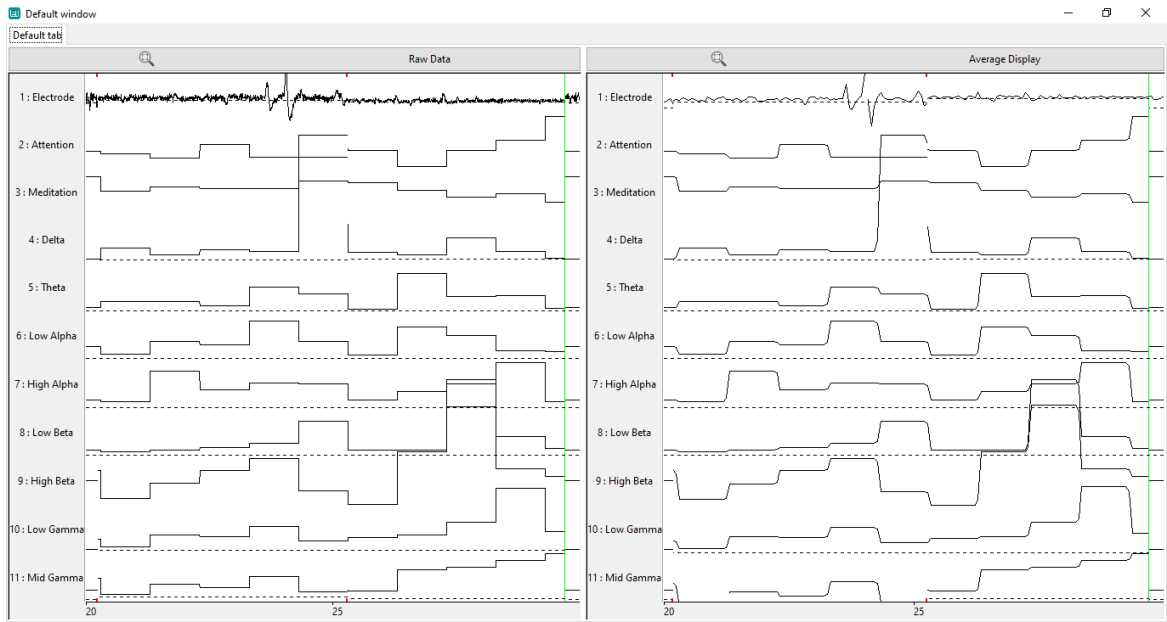

*Figure 67. Signal average (Brain waves).*

**Scenario 2. Test execution results:** Similar to the second scenario of the third test, the average signal and the raw signals captured by epochs and without them are shown. Thus, Figures 68 and 69 show how the use of epochs (0.5 and 2 seconds respectively) allows for more accurate data and significantly changes the spectrum of results obtained while measuring a stimulus or mental state.

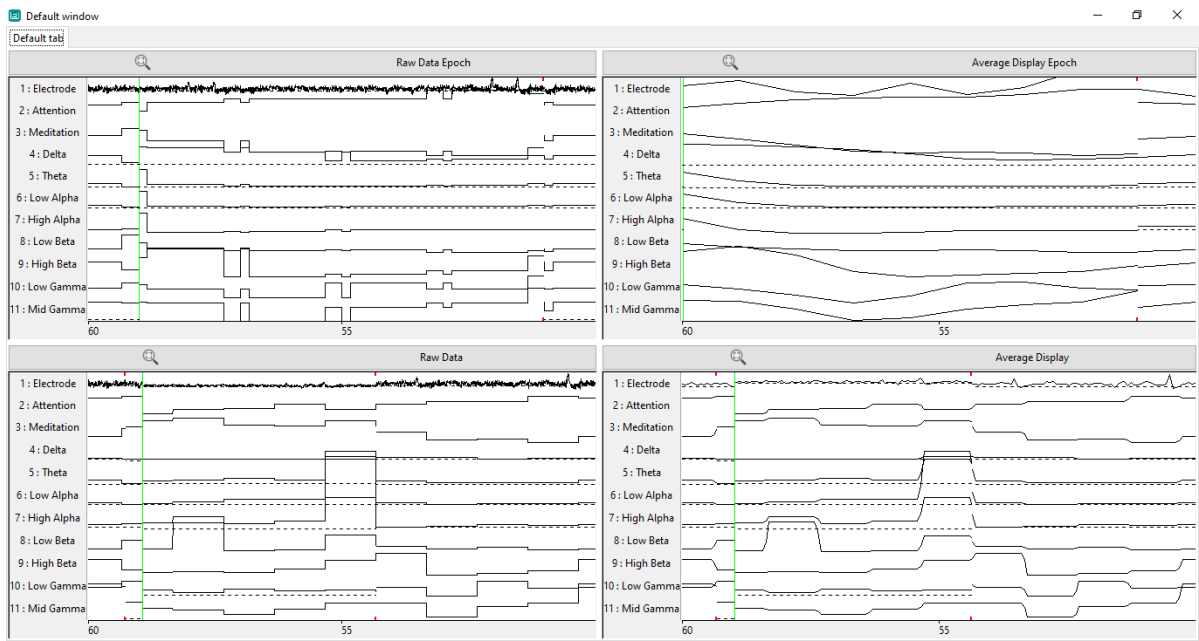

*Figure 68. Brainwave's epoch (0.5 seconds).*

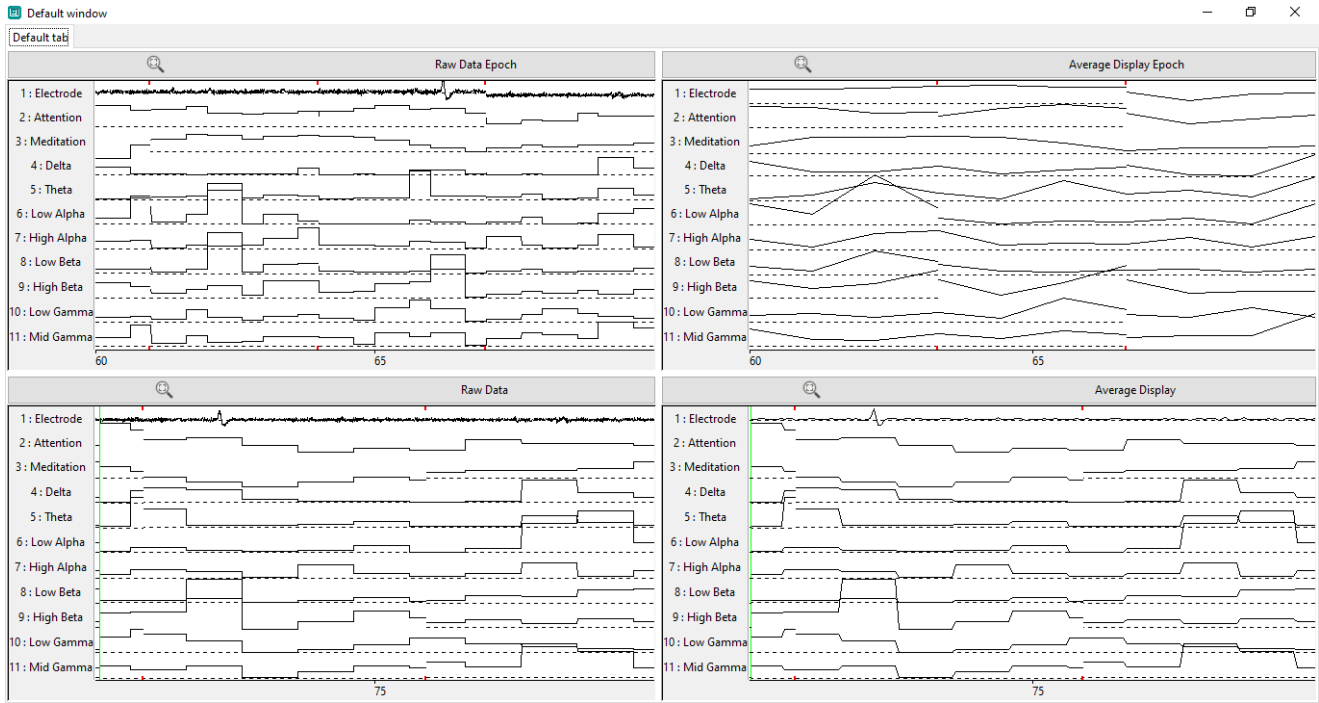

*Figure 69. Brainwave's epoch (2 seconds).*

**Scenario 3. Test execution results:** The third scenario shows the Signal Decimation, that is, the reduction in sampling rate for all 8 channels. This significantly reduces the cost of processing since the use of a lower sampling rate translates into a more economical implementation in terms of cost and computational complexity.

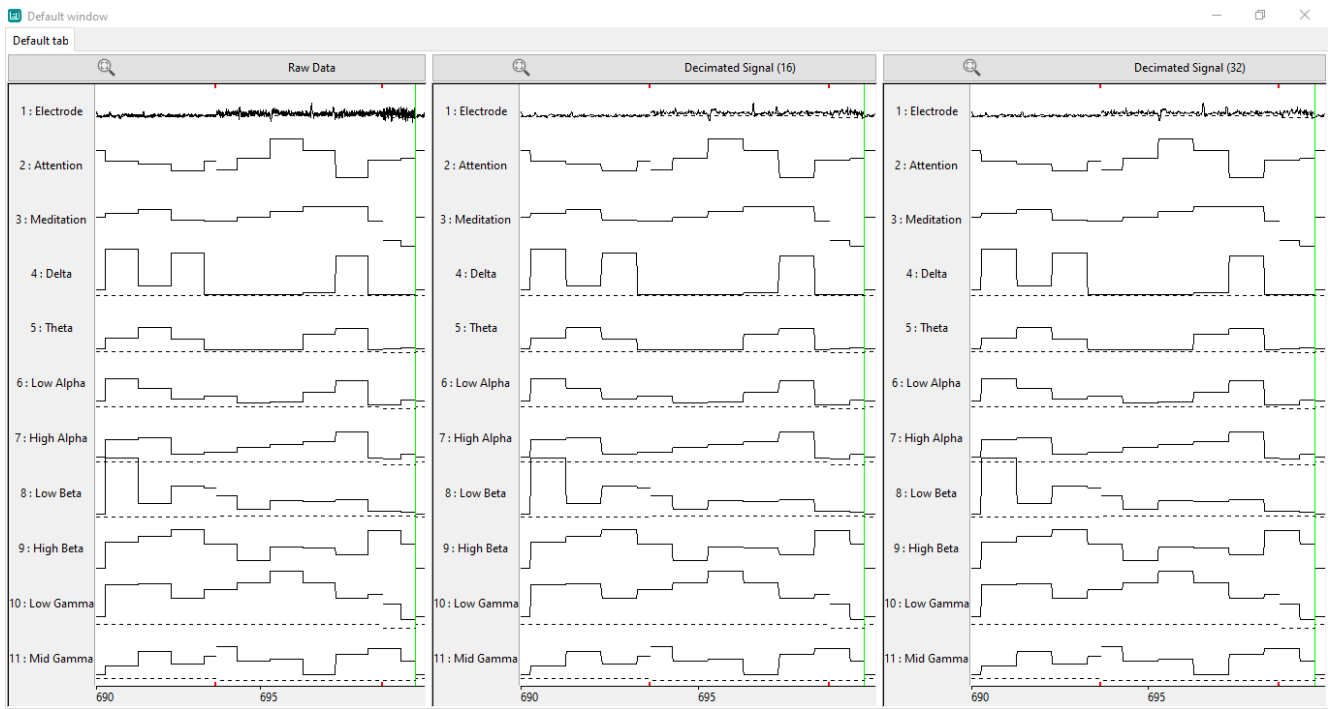

*Figure 70. Reduced sampling rate on all 8 channels.*

**Scenario 4:** As with the electrode testing a temporary filter was applied to the signal. However, for analysis purposes the above filter is not suitable as the variation in amplitude of each of the signals cannot be seen.

Therefore, it is required to use another element for this task. For this reason, it is necessary to connect the Temporal Filter to the Spectral Analysis box. Then it is connected to Stacked Bitmap (Horizontal) for data display.

Thus, this element will transform the signals present in the time domain to the frequency domain, and there it will be possible to observe the difference that occurs in every second of the signal amplitude. Figure 71 shows the proposed scenario.

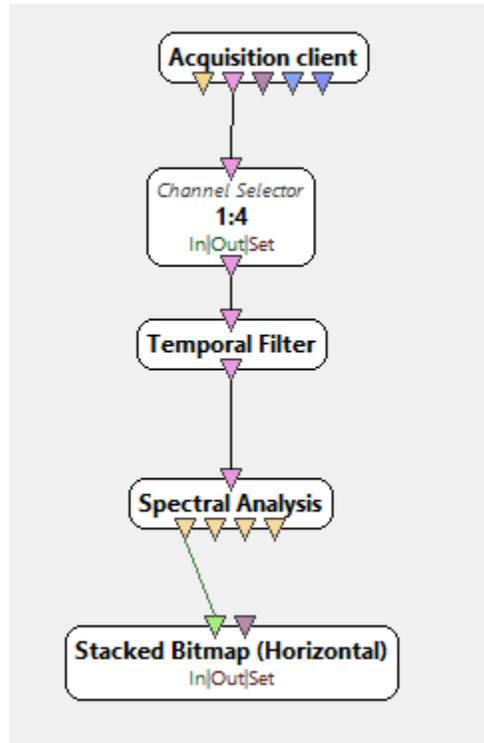

Figure 71. Spectral analysis.

**Scenario 4. Test execution results:** This scenario shows the need for a spectral analysis of the signal. Spectral analysis quantifies the amount of rhythmic (or oscillatory) activity at different frequencies in the EEG.

Based on numerous studies that reported a significant relationship between the EEG spectrum and human behavior, cognitive status or mental illness, spectral analysis is now accepted as one of the leading methods of analysis in the field of neuroscience (Im, 2018).

Spectral analysis requires applying a fast Fourier transform to the incoming signal (FFT). This allows to obtain the amplitude of the already filtered signal in the frequency domain.

Figure 72 shows the results of performing spectral analysis, where the electrode values and the user's attention and relaxation levels are shifted from the time domain to the frequency domain.

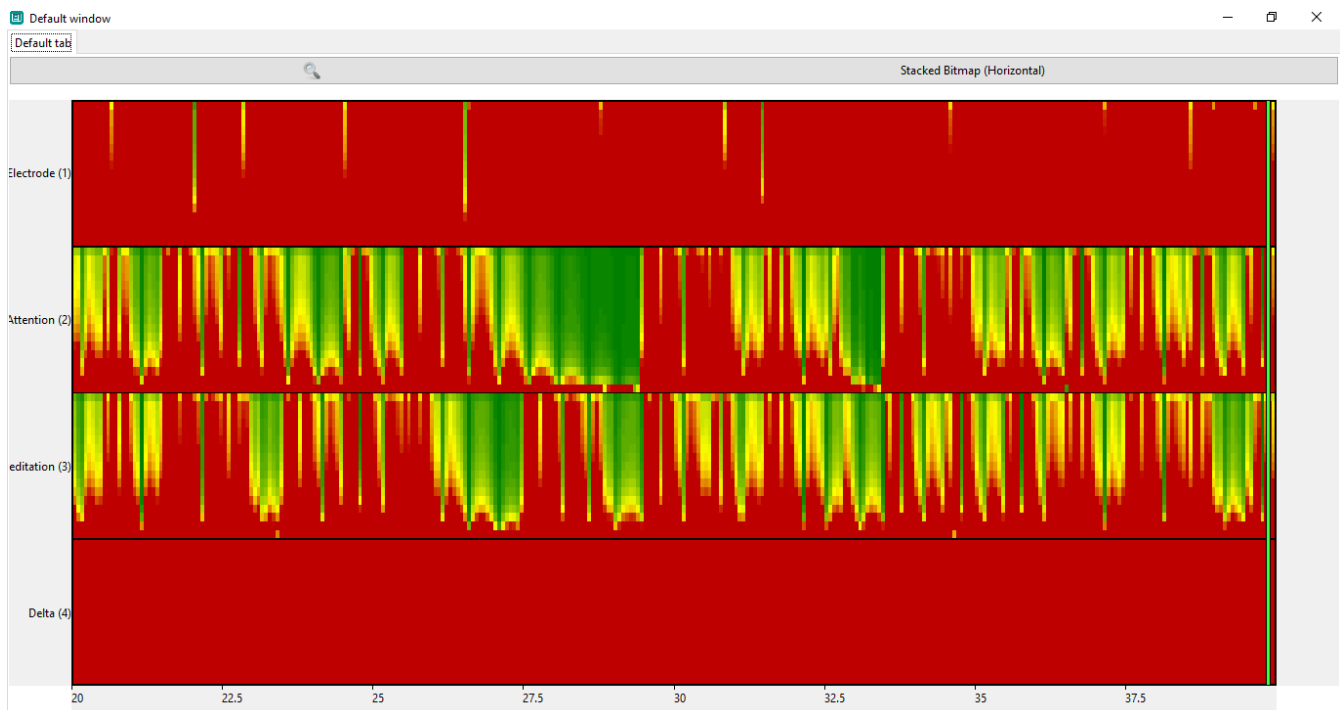

*Figure 72. Spectral analysis results.*

**Conclusions on BCI testing:**

In conclusion, applying filters and tests to the device's electrode and the encephalographic signals allowed us to understand and evidence in a better way which brain waves and mental states are more common in human beings. In this way, it was possible to identify the brain waves whose natural values would contribute more expressiveness and movement to the developed algorithms and the generated art pieces.

This allowed us to obtain a more general and objective vision insofar as it was possible to use the results of these tests to predict how users would react to certain stimuli and scenarios, such as closing their eyes, listening to music in a certain key at a certain tempo and smelling different scents.

In this regard, not only the mental states before the interaction were used, but also those generated by the sensations that music and art arouse during the interaction, becoming a cyclical system.

On the other hand, while carrying out tests with users, potential anomalies in the data delivered by the device were identified. For this reason, it was decided to filter the signal and process it before its use to ensure the software application to be executed with reliable data. In that sense, a correct interaction between the user and the application is guaranteed since the values received by the application have been previously processed.

It is important to highlight that some components that created noise were suppressed, for example, detection, stimulation, and blink strength since as explained above it generated undesired spikes in the signal.

By taking these measures and avoiding these scenarios a smoother interaction and a noticeable change in the fluctuation of the sound produced by the user's brain waves was observed. In this regard, the processed data allowed an easier integration of the EEG values with the MIDI data, as well as an improvement in the composition of the artistic production created by the user's mental states.

### 6.2. Static testing

To supplement this section dedicated to testing, it was decided to conduct static software tests. Static tests are performed on a component or system at the specification or implementation level without running the software.

There are two types: Review and Static code analysis. For this research work the second type has been selected. The static code analysis allows showing and/or evidencing bad programming practices in the code and segments that can be written in a better way.

To achieve the above, and to meet the required deliverables, automate the software development process and optimize writing and analysis of code in real-time, static code analyzers were used. These are tools that allow programmers to speed up the software development process and to correct errors related to programming style and code efficiency.

The selected tools are: SonarLint[6] which is an extension that can be found in most IDEs and is used to help developers to detect and fix problems related with coding.

Figure 73 shows how SonarLint displays error messages.

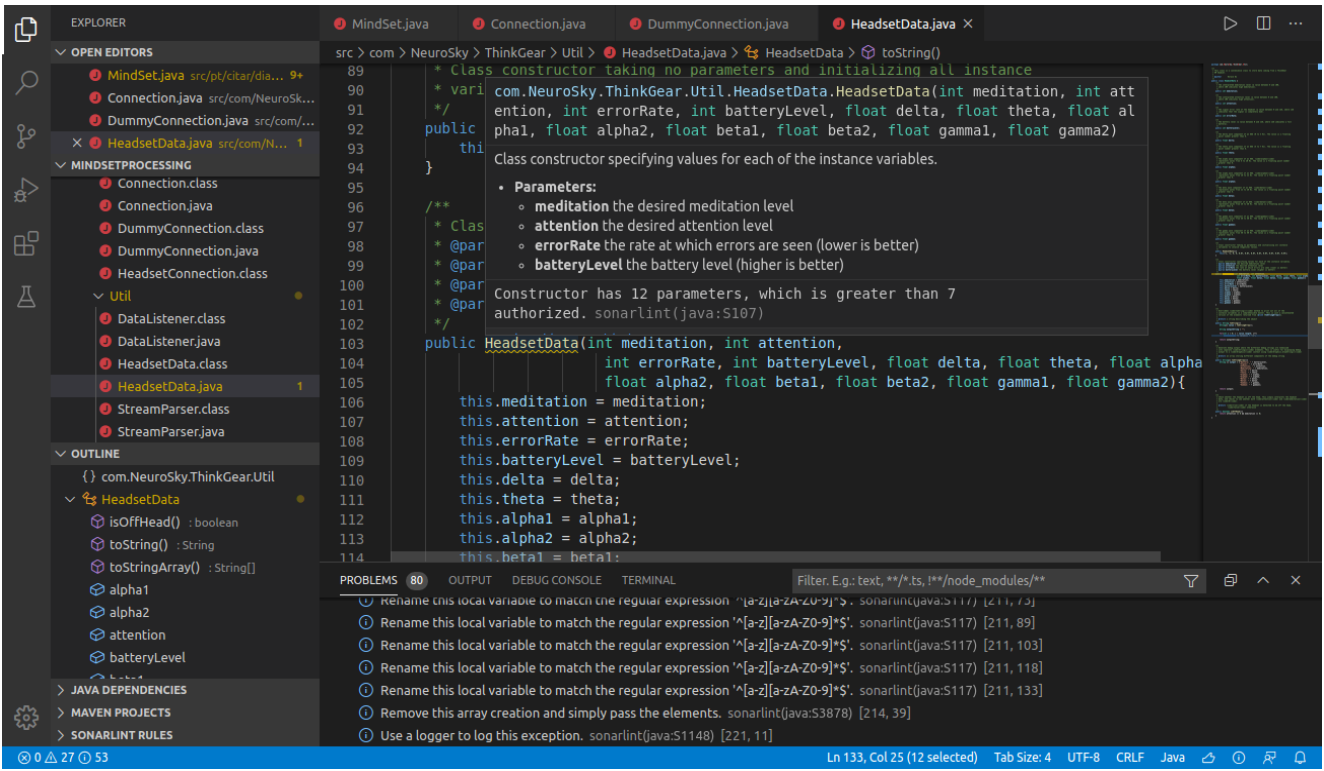

*Figure 73. SonarLint.*

SonarLint works like a Spell Checker, highlighting segments or lines of the code that can be written differently or optimized. SonarLint was used as a code analyzer because unlike many other tools, its execution does not require the use of a server. It runs easily in the IDE, as a plugin, avoiding re-coding. It also provides full descriptions of error messages, as shown in Figure 74.

---

[6] https://www.sonarlint.org/

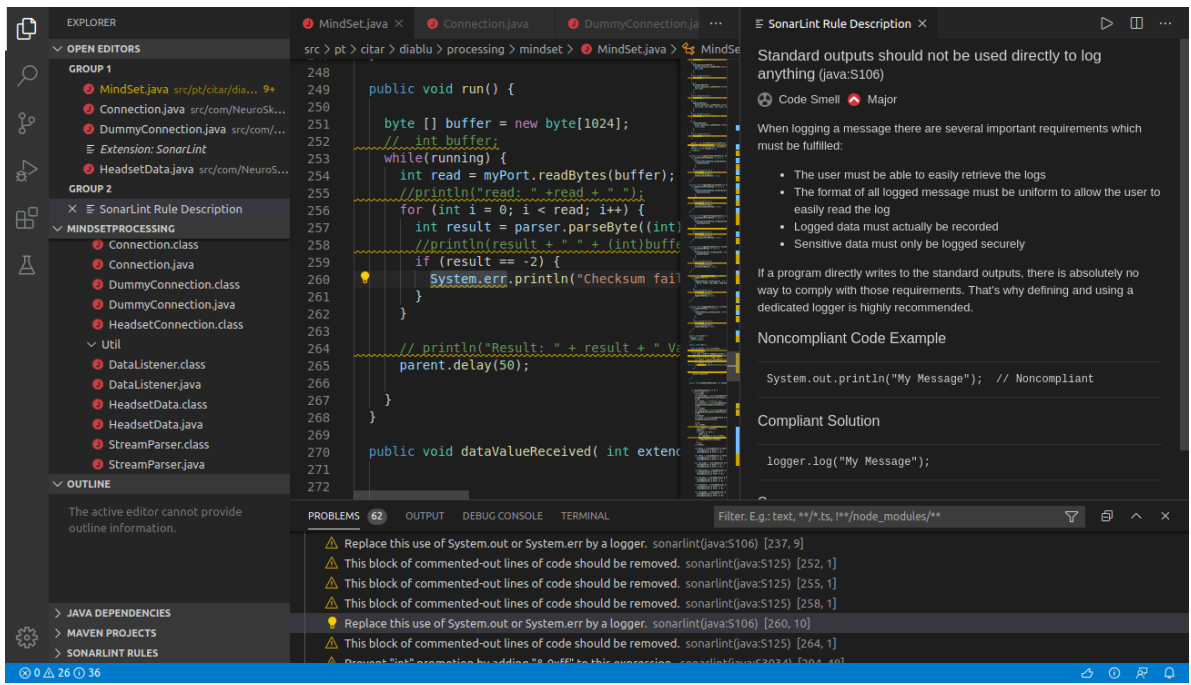

*Figure 74. Error message description.*

SonarLint allows bug detection as well as identifies common errors and shows possible vulnerabilities in the code as shown in Figure 75.

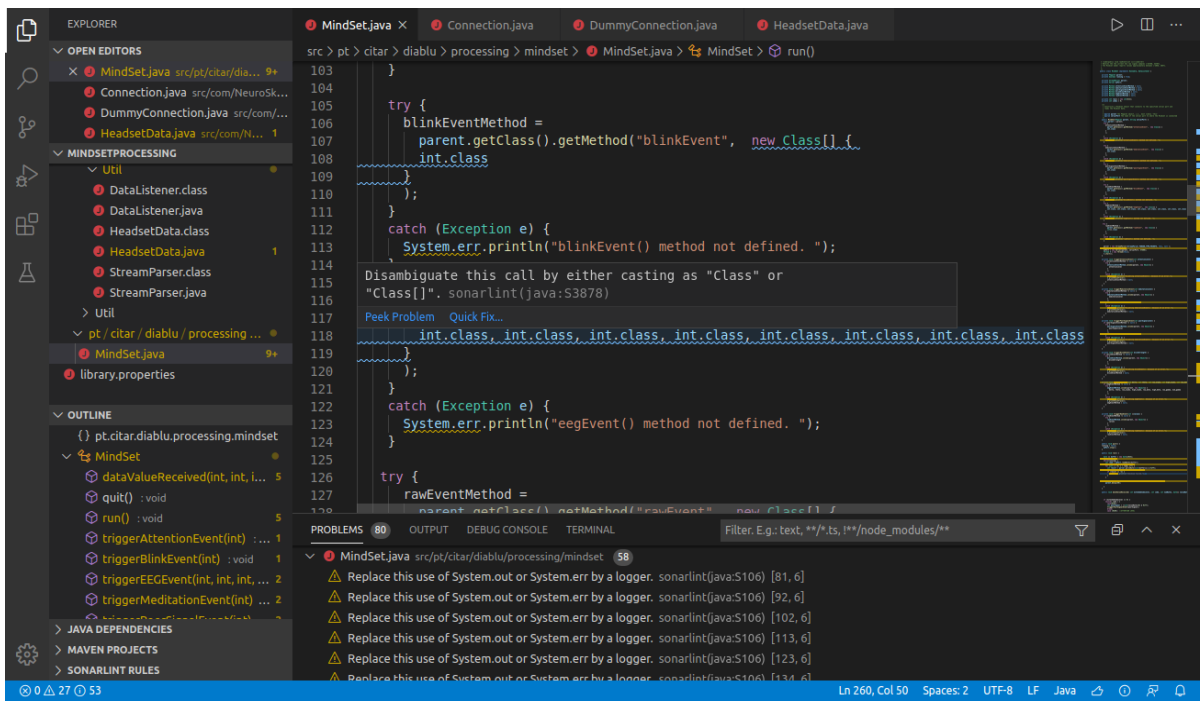

*Figure 75. Detection of an ambiguous instruction.*

The second tool used for static code analysis in the software development process was Checkstyle[7]. It allows analyzing if the code written in Java fulfills the programming rules and standards and helps the developers to adhere to a specific standard. Figure 76 shows the analyzer.

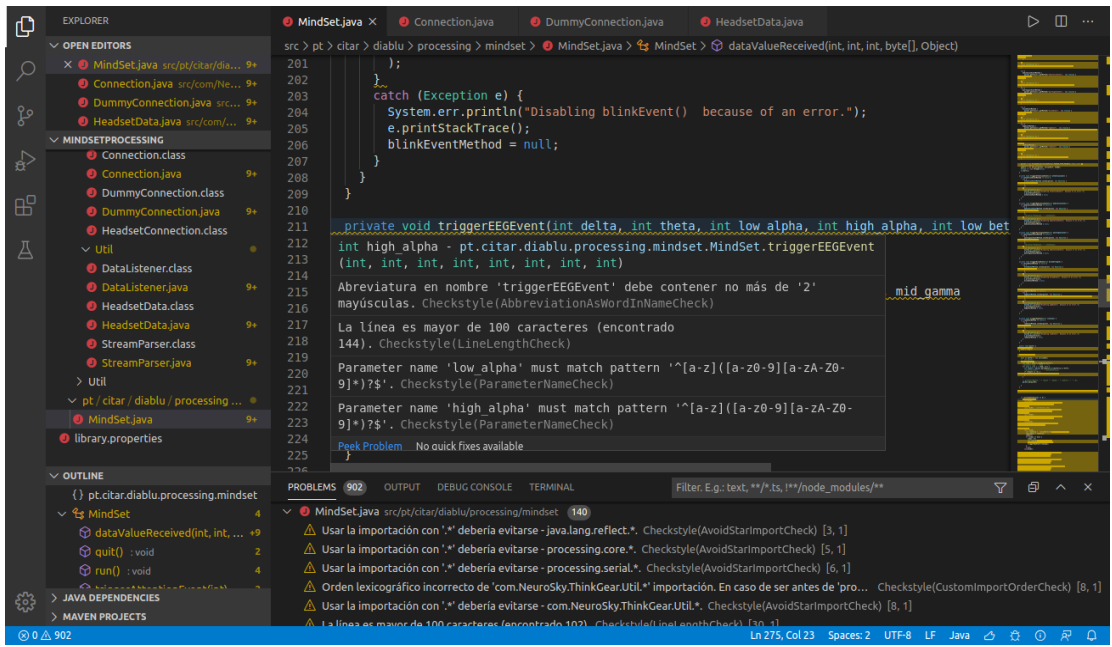

Figure 76. Checkstyle.

Checkstyle allows the developer to automate the checking process of code in Java. Besides, it is easily configurable with any programming standard, including the Sun Java standard and the Google standard, as shown in Figure 77.

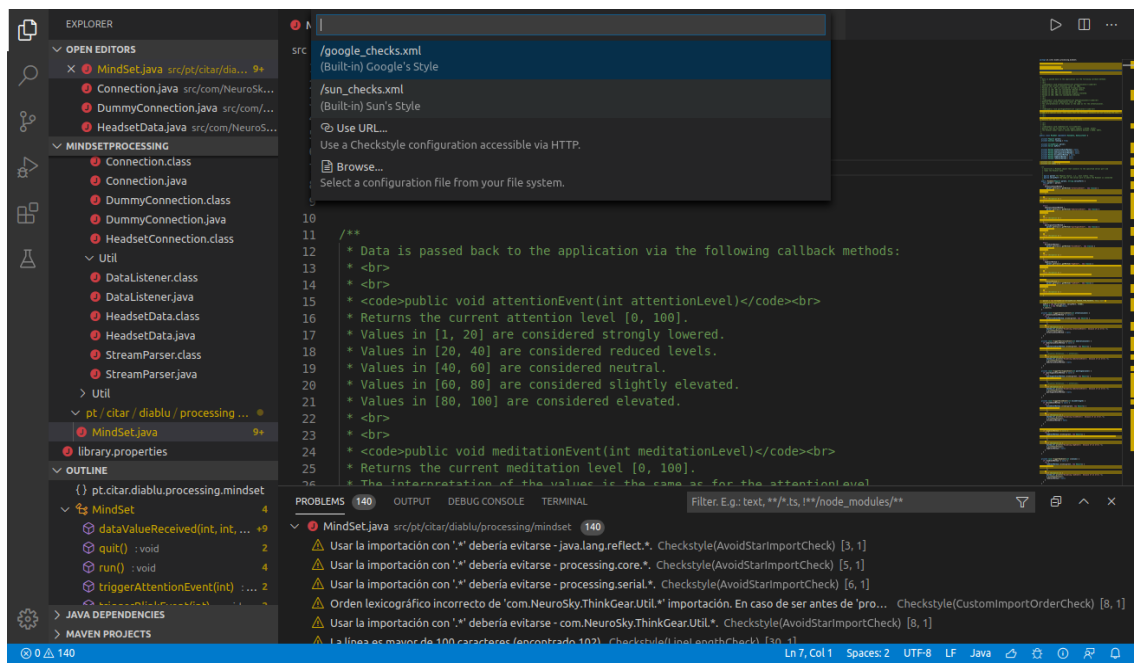

Figure 77. Programming standard configuration.

---

[7] https://checkstyle.sourceforge.io/

When selecting a different programming standard, Checkstyle automatically loads error messages and the highlighting color of the text, as shown in Figure 78.

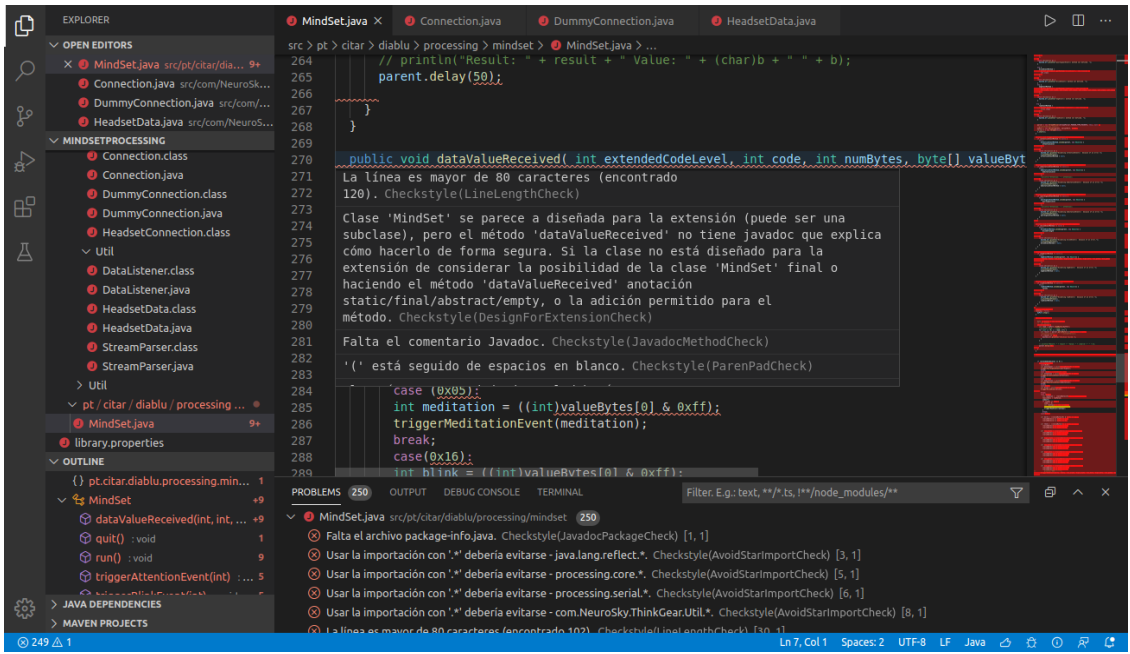

Figure 78. Checkstyle running the Java Sun Style.

The analysis that Checkstyle carries out on the code is limited to its presentation, so correctness or completeness is not confirmed. Checkstyle defines a set of available modules, each of which provides check rules at a configurable level. Some of the most common problems encountered are: The name of the attributes and methods, the number of functions in the parameters, the number of lines, spaces between characters, the use of imports, and scope modifiers among others, as shown in Figure 79.

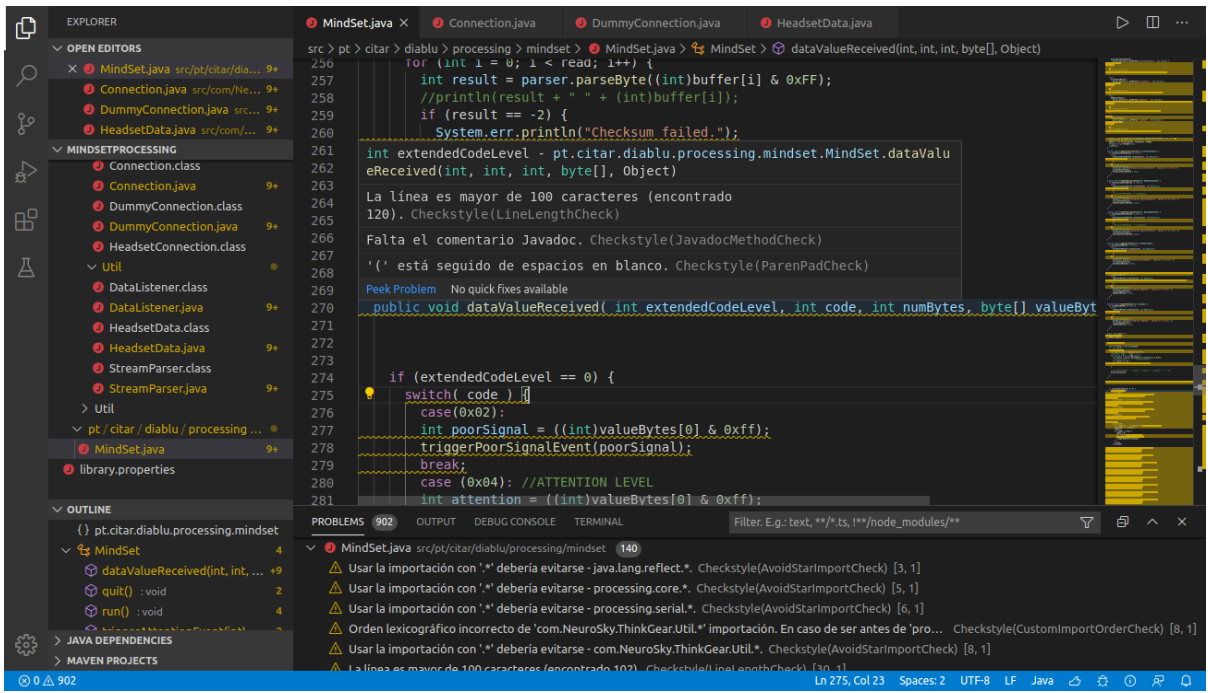

Figure 79. Problems found.

Therefore, SonarLint and Checkstyle are appropriate tools for static code analysis. Both are easily executable and configurable in Visual Studio Code.

**Results.**

In order to create a more stable and less error-prone software, both tools were used to detect and remove problems in the source code. When analyzing the code in its most stable version with SonarLint, 102 problems were found, as shown in Figure 80.

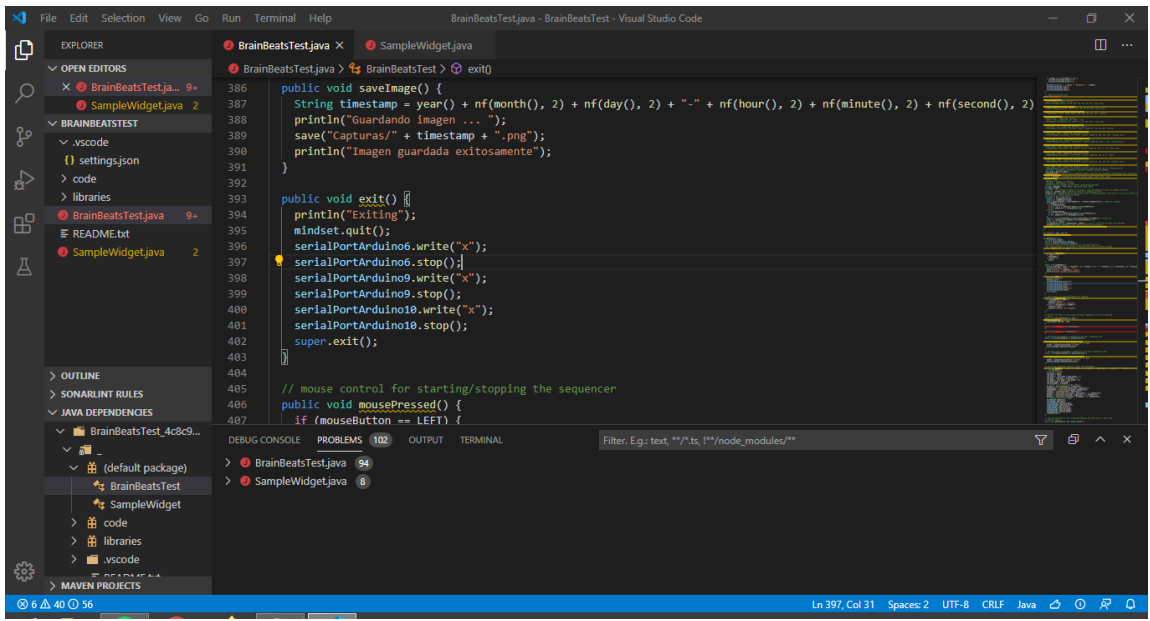

*Figure 80. SonarLint problem detection.*

The nature of the problems was of several kinds. Some of the most common were associated with cognitive complexity, which is a measurement that identifies how difficult it is to read and understand a unit of code intuitively. Methods with high cognitive complexity are difficult to maintain. Figure 81 shows one such method with high complexity.

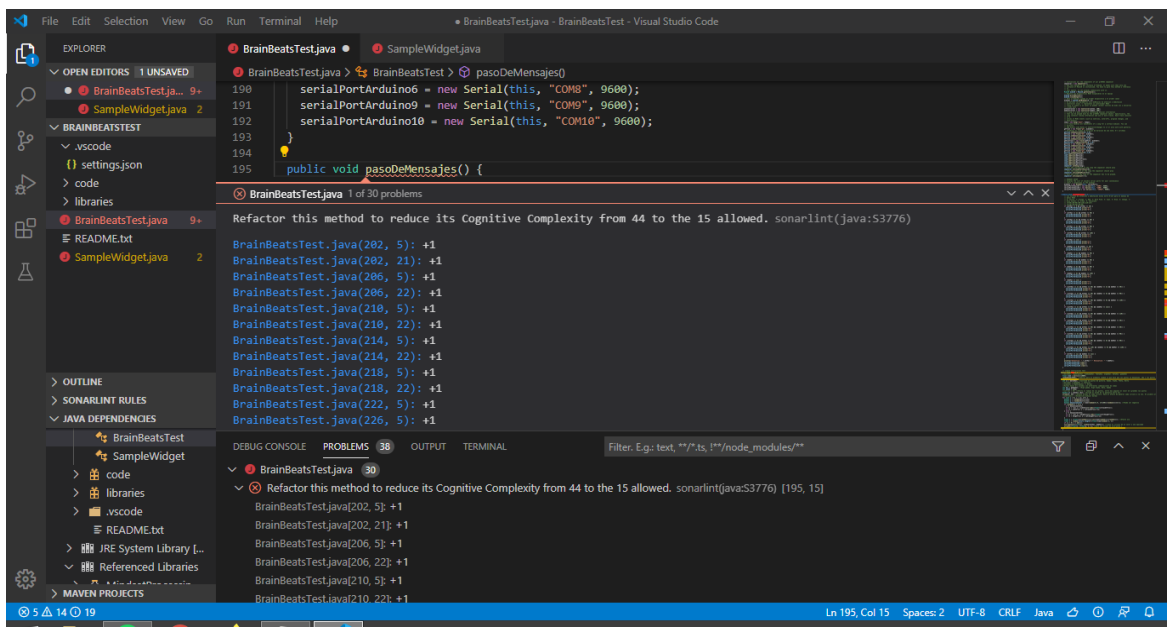

*Figure 81. Method with 44 complexity points out of 15 accepted.*

Another error found was related to programming practices. Therefore, in void-type methods that could be prone to errors, SonarLint proposes using *"@Override"* for two reasons: The first is because it causes a warning from the compiler if the method does not override anything, as in the case of a spelling error. The second is that it improves the readability of the source code by making it obvious that the methods are overridden. Figure 82 shows this error message.

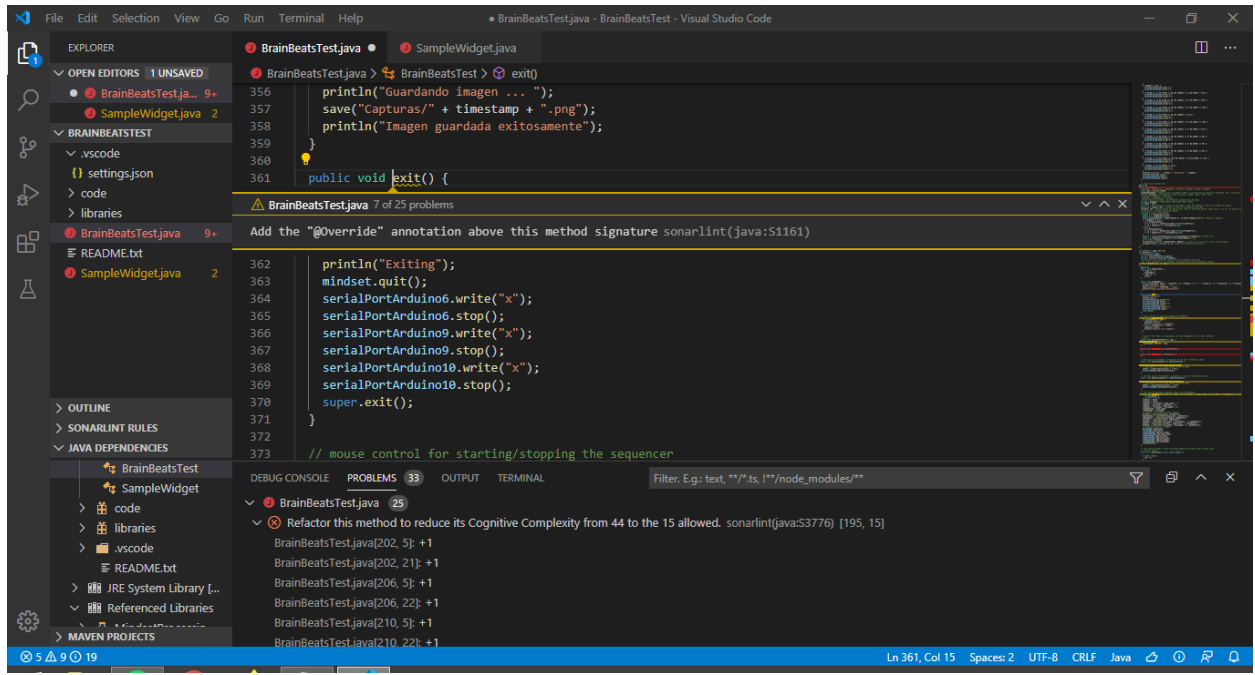

*Figure 82. Potential use of "@Override".*

SonarLint also analyzes when there are redundant or unnecessary instructions in the code. Figure 83 shows the detection of an unnecessary casting to a variable of the same type.

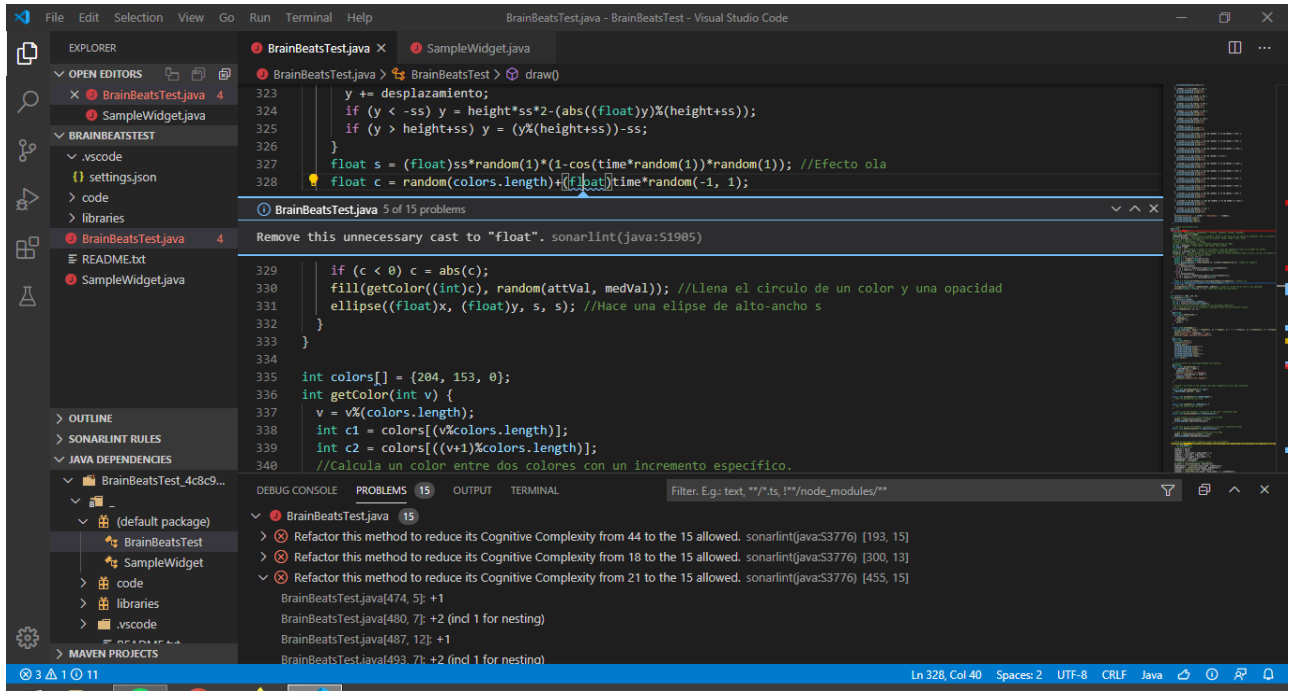

*Figure 83. Unnecessary casting.*

Figure 84 shows the detection of an error related to code efficiency. Here, an auxiliary variable was being used to calculate a value that could be returned directly. The detection of this error allows reducing the general complexity of the code, and the number of lines and calculations.

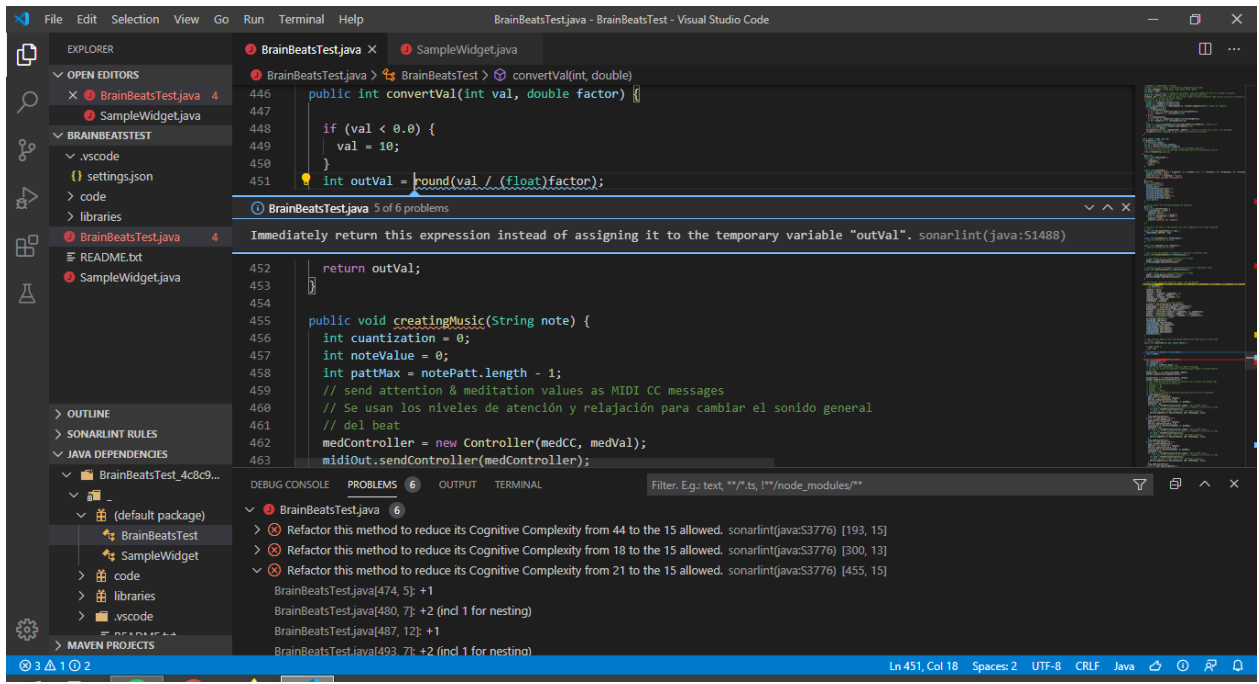

*Figure 84. Variable return.*

After making the necessary corrections and completely re-analyzing the source code with this tool, only 4 errors were left unresolved. These involved the reduction of essential lines of code into methods that required them. Therefore, it was decided to keep them to ensure the correct execution of the application. Figure 85 shows the new number of errors detected.

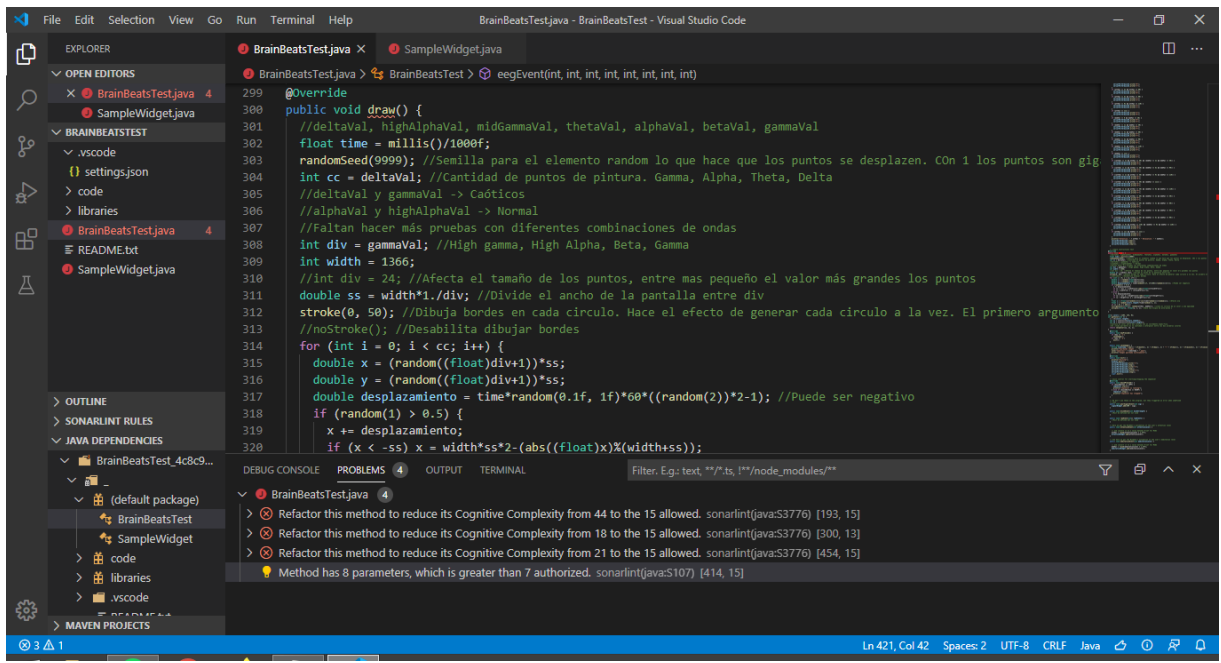

*Figure 85. New number of errors detected.*

Thus, the number of errors were significantly reduced from 102 errors to only 4, which is a 98% reduction. Similarly, when analyzing the code in its most stable version with Checkstyle 165 problems were found, as shown in Figure 86.

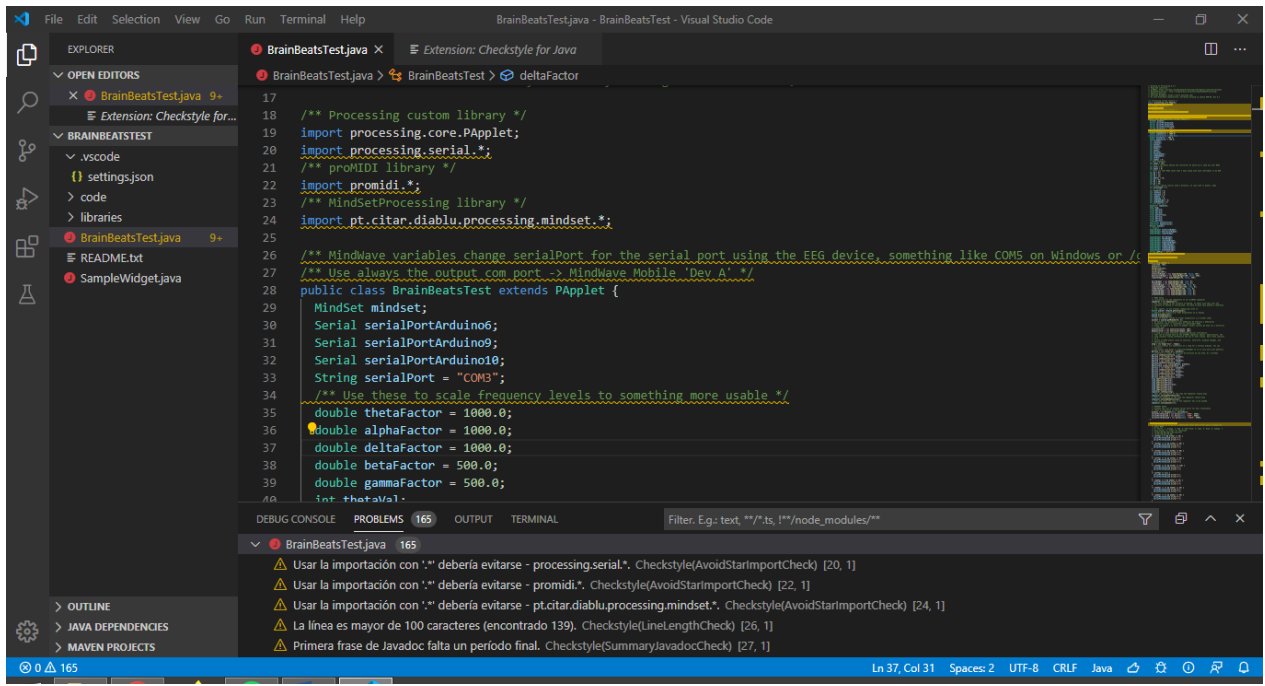

Figure 86. Errors found with Checkstyle.

Some of these problems were related to the number of characters in a comment, the use of keys in conditionals, and the lack of Javadoc comments, which provide a detailed description of the method or class to be used. The last-mentioned is shown in Figure 87.

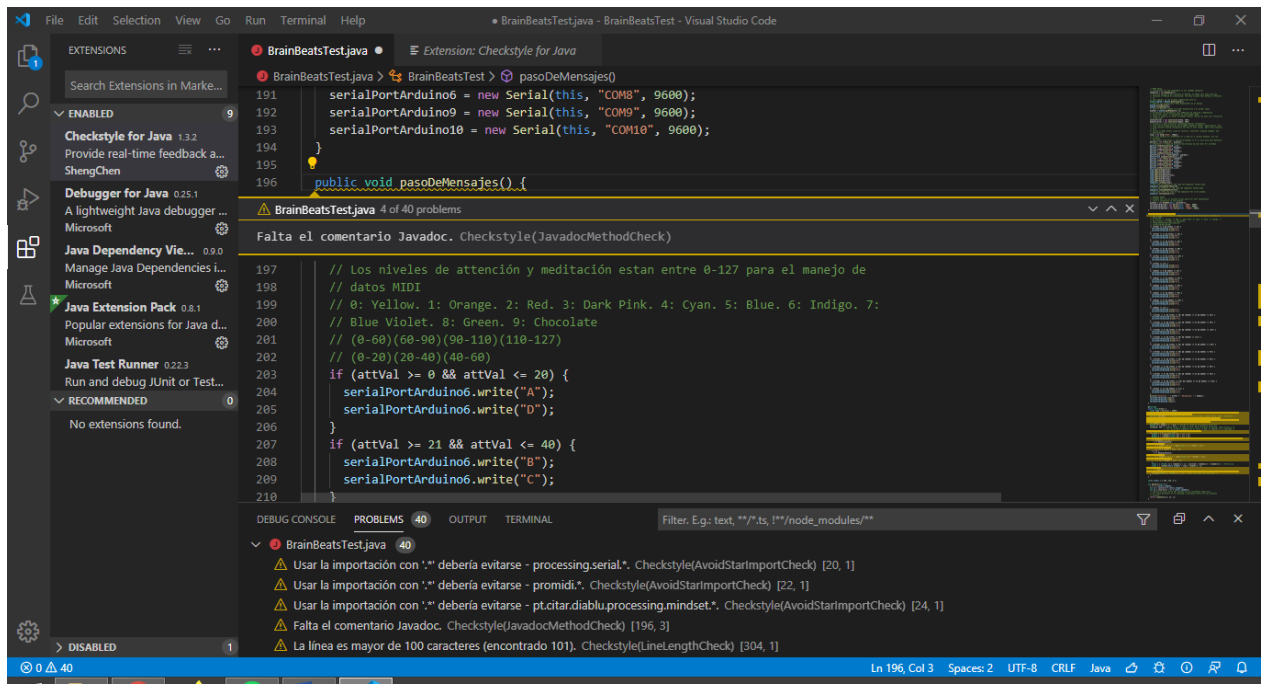

Figure 87. Javadoc.

After completing the analysis of the source code, the new number of errors found were 5, those being related to imports in the libraries used, and lines of code with many characters. The previous messages were ignored, since solving them would imply a malfunction in the application, which would lead to more errors. The new number of errors are shown in Figure 88.

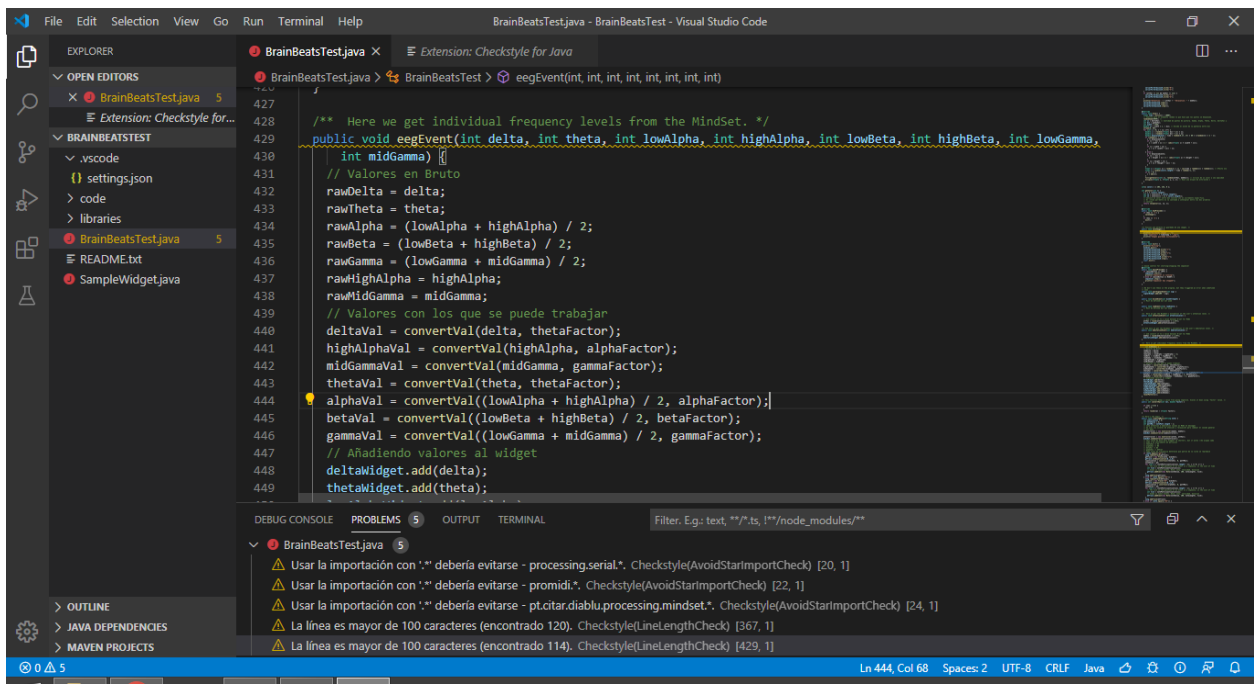

*Figure 88. New number of errors detected with Checkstyle.*

Thus, the number of errors decreased from 165 to 5 which is a 97% reduction in the number of errors when using Checkstyle and almost 98% when combining both tools. Thus, it can be seen that the use of static analyzers reduces the number of problems and potential failures in the code, helping developers code more stable, robust and less error-prone software, as shown in Figure 89.

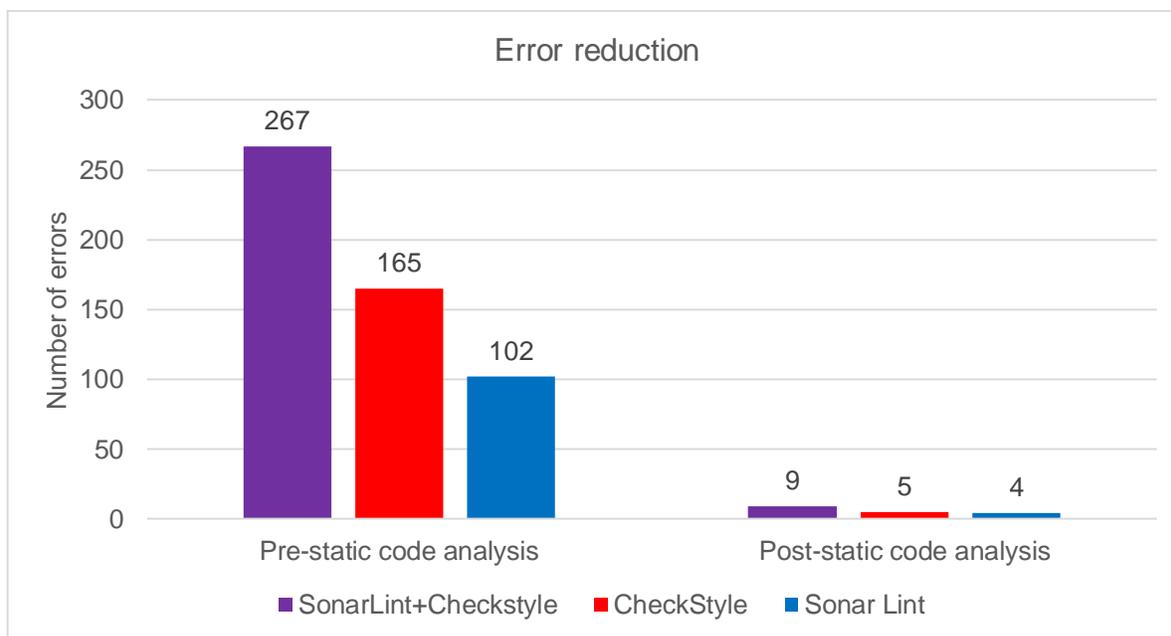

*Figure 89. Error reduction.*

### 6.3. Usability testing

To measure the user experience, a usability test that involves two techniques: Heuristic Evaluation and Direct Observation was designed.

The Heuristic evaluation was based on the 10 usability heuristics proposed by Nielsen (Nielsen, 2004) and validation criteria related to the characteristics of an Interactive Art Installation were adapted.

This test makes it possible to determine the minimum of necessary interface elements that the Artistic Installation must fulfill to improve the user's experience. It is noteworthy that in the literature review no heuristics adapted to generative art were found, and therefore, the design of this test is an interesting contribution to the academic community. The test design is available at the following link http://tiny.cc/UsabilityTest.

As for the direct observation, it is suggested to observe unobtrusively the process of interaction in real-time of a sample of users (sighted and visually impaired) to analyze the user experience as well as to take note of possible interaction problems that are not reflected in the heuristic evaluation.

The heuristic evaluation test, applied to the context of generative art, represents a contribution of great interest in this area, for this reason it was shared with the academic community. The test is available at the following link: http://tiny.cc/EvaluationHeuristics.

### 6.4. Results analysis

Next, the advantages and disadvantages of the Cerebrum Art Installation per module are described.

**Music Component.**

**Advantages:**

- **Use of MIDI data:** MIDI files are much more compact than digital audio files. Also, MIDI files may sound better than digital audio files if the MIDI sound source has high-quality. Using MIDI data allows creating well-defined and high-quality musical compositions.

- **Data sent to a DAW:** The MIDI data and musical structures created by user interaction can be sent to a DAW. In this way, a particular instrument can be removed from the song and/or a particular instrument can be changed by another just by selecting it, which increases the overall sound and its expressiveness. Also, the music generated can be recorded and printed as a musical sheet.

- **Use of brain waves:** The use of brain waves allows each user to create different and unique musical pieces. In each of them, their brain activity and mental states will be represented through musical data.

**Disadvantages:**

- **Dependency on the sound source:** MIDI data is device-dependent (the sounds produced by MIDI music files depend on the MIDI device used for playback). This is why, by using only the computer's output, the sound might have less quality. It can also differ according to the operating software.

**Generative Art Component.**

**Advantages:**

- **Use of design concepts:** Using design concepts allows the generated pieces to be displayed in a colorful and visually appealing way.

- **User Interaction:** One of the main advantages of Cerebrum is the fact that the spectator takes an active role in the creative process. In this regard, the user interaction not only captures the attention of the people present in the installation, but also takes advantage of the human component showing that every single interaction is different and produces diverse results.

- **Use of circles:** Using circles allows ensuring that the computational complexity of the displayed shapes won't be high. This allows using the available computational resources in adding characteristics, colors, and effects to the art pieces.

**Disadvantages:**

- **Limited to indoor spaces:** The installation facility is dark, only the brain data captured by the BCI device can bring color and start the interaction process. If the installation is set in an open-air space it would be harder to observe the art pieces. Also, If the user is located far away from the computer or the Bluetooth connection/range is diminished by interference with other devices, the art pieces won't be generated. For this reason, Cerebrum is thought and designed to be indoors.

- **Time-limited:** Using the BCI might cause fatigue in the users, which is why it is recommended not to use the device for more than 5 minutes. Another factor that could restrain the interaction is the fact that the BCI is non-rechargeable, which might limit its time if the battery is low.

To summarize, Cerebrum not only offers a new way of interaction but also adds an inclusive component, making art accessible to sighted and visually impaired users by adding an olfactory element. Nowadays, most art installations are designed to be enjoyed through visual and auditory senses, leaving aside users with different diversities. Also, unlike most art installations, in Cerebrum the users take an active role in the creative process.

In this regard, users are not limited to seeing what happens in the installation or walking around it but they become and create the installation. The installation lives with the people who interact with it. Table 7 shows a comparative analysis of Cerebrum and the art installations in section 2.2.3.

| Characteristics/ Art Installation | Projection type | Sensory type | Technology that allows Interaction | Bio signal | Audience role | Inclusive? |
|---|---|---|---|---|---|---|
| **Plasma Reflection** | None (TV) | Visual | Kinect | None | Spectator Creator | No |
| **As Above So Below** | Mapping | Visual | Kinect | None | Spectator | No |
| **The Act of Seeing** | Elevated | Visual Auditory | Electrode patches Arduino | EOG | Spectator | No |
| **Share your Unicorn** | Elevated | Visual | BCI device Drawing machine | EEG | Spectator Creator | No |
| **Cerebrum** | Back projection | Visual Auditory Olfactory | BCI device Air humidifier Arduino | EEG | Spectator Creator | Yes |

*Table 7. Art Installations comparison.*

# CHAPTER 7. CONCLUSIONS AND FUTURE WORK

The literature review regarding BCI showed that most of the research carried out is related to psychology and neuroscience. However, very few of them relate the use of BCI in contexts of artistic production and creation. The aforementioned made of this research work an innovative solution in the context of art and in particular in the areas of generative art and music. Its codification, development, and results show the integration of different areas of knowledge and disciplines, and it allows creating artistic pieces based on brain waves, mental states, and brain activity, creating an innovative bridge between human beings and art.

Thanks to the presence of Artificial Intelligence at the feature extraction stage and in the acquisition and processing of encephalographic signals it was possible to make the most out of the user interaction knowing that the time for this is limited. In this regard, the developed algorithms were able to respond to the user's mental states. In this way, not only the mental states prior to the interaction can be used and taken advantage of, but also the information received as a result of the interaction, thus becoming a system capable of taking advantage of the information received and then generating a multimodal artistic piece that responds to the information processed.

Using the SCRUM methodology in this research work allowed the software development process to be much more agile since there was a well-established Release Plan and each improvement in the deliverables redoubled the quality of the developed code. Besides, this framework allowed to have organized, quickly and concisely, results that have been published.

The experience acquired during the development of this research work allowed to evidence that the programming language Matlab offers robustness in the developed code, but neglected the visual component. The programming language ChucK is more oriented to the synthesis of sound in real-time and the creation of music, but leaves aside data processing. JavaScript allows integrating most of the components, however, this could avoid the decentralization and independence of each component from each other, as well as slowing down the browser and causing latency while exporting the processed data. In this regard it was necessary to apply the heuristic method of trial and error, until finding in Processing (Java) and the libraries mentioned in this document the development tools that allowed to incorporate the appropriate functionalities to create music and generative art based on brain activity.

The modularity of the algorithms allows the developed software to be used, adapted and extended in different areas, for example: 1) In the artistic context, since it presents a multimodal way of communication whose interaction involves the use of an EEG by means of producing an artistic piece, the software can be used in museums and interactive art installations, as those seen in section 2.2.3. 2) In the context of rehabilitation, under the supervision of a health professional, the software could be used as art therapy in the treatment of patients with Alzheimer's disease and other dementias, as it could help reduce agitation and similarly, to promote relaxation in patients. 3) In the context of artistic production, the software developed could be used so that patients with physical diversity or reduced mobility can create music and art on their own.

The design and execution of tests carried out in the BCI has shown that OpenViBE makes it easier to acquire data from BCI devices and to apply filters to EEG signals. In this sense, it was possible to determine which data could cause noise and thus omit it, and to identify which brain waves and mental states allow the algorithm to generate art pieces with greater naturalness and expressiveness.

Personally, doing BCI research allowed me to put into practice the skills obtained during my undergraduate studies, as well as it made me understand how important multidisciplinarity in research projects is. It also taught me how crucial it is to approach problems from different perspectives and points of view. In the same way, it encouraged me to continue doing research and scientific contributions.

**Future Work**

Some of the aspects to consider extending or improving the results of this undergraduate thesis are the following:

- To increase the expressiveness of the exported sound.

- To propose a haptic component that allows tactile feedback, especially for users in visual and auditory diversity.

- To carry out experiments with the use of organic forms in the visual component.

- To carry out usability tests to verify the satisfaction level and user experience.

Furthermore, it would be relevant to study how the generated artwork changes concerning the task performed and to the type of patient: For example, studying how activities such as listening to music, reading, solving arithmetical operations, or painting affect the final work. In this sense, there are many uses that the developed software offers in different contexts.